\documentclass[a4paper,11pt]{article}
\pdfoutput=1 

\usepackage{jheppub} 
\makeatletter
\DeclareRobustCommand*{\bfseries}{%
  \not@math@alphabet\bfseries\mathbf
  \fontseries\bfdefault\selectfont
  \boldmath
}
\makeatother

\usepackage{calc}
\usepackage{rotating}
\usepackage[english]{babel}
\usepackage{graphicx}
\usepackage{subfig}
\usepackage{float}
\usepackage{amsmath}
\usepackage{amssymb}
\usepackage{amsthm}
\usepackage{latexsym}
\usepackage{dcolumn}
\newcommand{\newc}{\newcommand*}

\long\def\begincomment#1\endcomment{%
        \begingroup\sf\baselineskip12pt#1\endgroup}

\newc{\etal}{\textrm{et al.}} 
\newc{\eg}{\textrm{e.g.}} 
\newc{\ie}{\textrm{i.e.}}
\newc{\etc}{\textrm{etc}}
\newc\vs{\textrm{vs.}}
\newc{\cl}{\rm {C.L.}}

\newc{\ev}{\ensuremath{\,\mathrm{eV}}}
\newc{\kev}{\ensuremath{\,\mathrm{keV}}}
\newc{\mev}{\ensuremath{\,\mathrm{MeV}}}
\newc{\gev}{\ensuremath{\,\mathrm{GeV}}}
\newc{\tev}{\ensuremath{\,\mathrm{TeV}}}
\newc{\MeV}{\mev} 
\newc{\TeV}{\tev}
\newc{\invpb}{\ensuremath{/\text{pb}}}
\newc{\invfb}{\ensuremath{\,\text{fb}^{-1}}}
\newc\nb{\ensuremath{\,\mathrm{nb}}} \newc\pb{\ensuremath{\,\mathrm{pb}}} \newc\fb{\ensuremath{\,\mathrm{fb}}}
\newc\pc{\ensuremath{\,\mathrm{pc}}}
\newc\kpc{\ensuremath{\,\mathrm{kpc}}}
\newc\mpc{\ensuremath{\,\mathrm{Mpc}}}
\newc\ps{\ensuremath{\,\mathrm{ps}}} 
\newc\cmeter{\ensuremath{\,\mathrm{cm}}} 
\newc\meter{\ensuremath{\,\mathrm{m}}} 
\newc\kmeter{\ensuremath{\,\mathrm{km}}}
\newc\second{\ensuremath{\,\mathrm{s}}}
\newc\msecond{\ensuremath{\,\mathrm{ms}}}
\newc\nsecond{\ensuremath{\,\mathrm{ns}}}
\newc\psecond{\ensuremath{\,\mathrm{ps}}}

\newc{\chisqmin}{\ensuremath{\chi^2_{\mathrm{min}}}}
\newc{\Delchisq}{\ensuremath{\Delta\chi^2}}
\newc{\chisq}{\ensuremath{\chi^2}}
\newc{\like}{\ensuremath{\mathcal{L}}}

\newc\lsim{\ensuremath{\mathrel{\rlap{\lower4pt\hbox{\hskip1pt$\sim$}}\raise1pt\hbox{$<$}}}}
\newc\gsim{\ensuremath{\mathrel{\rlap{\lower4pt\hbox{\hskip1pt$\sim$}}\raise1pt\hbox{$>$}}}}
\newc{\VEV}[1]{\ensuremath{\langle #1 \rangle}}
\newc{\dl}{\ensuremath{\stackrel{\leftarrow}{D}}}
\newc{\dr}{\ensuremath{\stackrel{\rightarrow}{D}}}

\newc{\bcenter}{\begin{center}}    \newc{\ecenter}{\end{center}}
\newc{\bfl}{\begin{flushleft}}    \newc{\efl}{\end{flushleft}}
\newc{\bfr}{\begin{flushright}}    \newc{\efr}{\end{flushright}}

\newc{\bi}{\begin{itemize}}
\newc{\ei}{\end{itemize}}
\newc{\bed}{\begin{description}}
\newc{\eed}{\end{description}}
\newc{\ben}{\begin{enumerate}}
\newc{\een}{\end{enumerate}}

\newc{\be}{\begin{equation}}
\newc{\ee}{\end{equation}}
\newc{\bea}{\begin{eqnarray}}
\newc{\eea}{\end{eqnarray}}
\newc{\ra}{\rightarrow}

\newc{\alphas}{\ensuremath{\alpha_s}}
\newc{\alphatwo}{\ensuremath{\alpha_2}}
\newc{\alphaone}{\ensuremath{\alpha_1}}
\newc{\alphai}[1]{\ensuremath{\alpha_{#1}}}
\newc{\alphaem}{\ensuremath{\alpha_{\mathrm{em}}}}
\newc{\alphaeff}{\ensuremath{\alpha_{\mathrm{eff}}}}
\newc{\sineff}{\ensuremath{\sin \theta_{\mathrm{eff}}}}
\newc{\sinsqeff}{\ensuremath{\sin^2 \theta_{\mathrm{eff}}}}
\newc{\dalphahad}{\ensuremath{\Delta \alpha_{\mathrm{had}}}}
\newc{\yt}{\ensuremath{h_t}} \newc{\yb}{\ensuremath{h_b}} \newc{\ytau}{\ensuremath{h_{\tau}}}
\newc\mz{\ensuremath{M_Z}} 
\newc\mw{\ensuremath{m_W}}
\newc\mZ{\mz}        \newc\mW{\mw}
\newc\mhsm{\ensuremath{ m_{H_{\mathrm{SM}}}}}
\newc{\mtop}{\ensuremath{ m_t}}               \newc{\mtpole}{\ensuremath{ M_t}}
\newc{\mbottom}{\ensuremath{ m_b}} 
\newc{\mtau}{\ensuremath{ m_{\tau}}}
\newc{\mt}{\mtpole}
\newc{\mb}{\mbottom} 
\newc{\rgg}{\ensuremath{R_{h}(\gamma\gamma)}}
\newc{\rzz}{\ensuremath{R_{h}(ZZ)}}
\newc{\rtwogg}{\ensuremath{R_{h_2}(\gamma\gamma)}}
\newc{\rtwozz}{\ensuremath{R_{h_2}(ZZ)}}
\newc{\ronegg}{\ensuremath{R_{h_1}(\gamma\gamma)}}
\newc{\ronezz}{\ensuremath{R_{h_1}(ZZ)}}
\newc{\rsiggg}{\ensuremath{R_{h_\textrm{sig}}(\gamma\gamma)}}
\newc{\rsigzz}{\ensuremath{R_{h_\textrm{sig}}(ZZ)}}
\newc{\llbar}{\ensuremath{\ell\bar{\ell}}}
\newc{\tauptaum}{\ensuremath{ \tau^+\tau^-}}
\newc{\qqbar}{\ensuremath{ q\bar{q}}} \newc{\ppbar}{\ensuremath{ p\bar{p}}}
\newc{\bbbar}{\ensuremath{ b\bar{b}}} \newc{\ttbar}{\ensuremath{ t\bar{t}}}
\newc{\ffbar}{\ensuremath{ f\bar{f}}} \newc{\tautaubar}{\ensuremath{ \tau\bar{\tau}}}
\newc{\mchi}{\ensuremath{m_{\chi}}}
\newc{\squark}{\ensuremath{\tilde{q}}}
\newc{\slepton}{\ensuremath{\tilde{l}}}
\newc{\gluino}{\ensuremath{\tilde{g}}} 
\newc{\mgluino}{\ensuremath{{m_{\gluino}}}}
\newc{\tone}{\ensuremath{{\tilde{t}_1}}}

\newc{\sthw}{\ensuremath{ \sin\theta_W}}              \newc{\cthw}{\ensuremath{\cos\theta_W}}
\newc{\tanthw}{\ensuremath{ \tan\theta_W}}              \newc{\cotthw}{\ensuremath{\cot\theta_W}}
\newc{\ssqthw}{\ensuremath{\sin^2 \theta_W}}
\newc{\msbar}{\ensuremath{\overline{MS}}} \newc{\drbar}{\ensuremath{\overline{DR}}}
\newc{\mtmtsmmsbar}{\ensuremath{ m_t(m_t)^{\msbar}_{{\mathrm{SM}}}}}
\newc{\mtmtsmdrbar}{\ensuremath{ m_t(m_t)^{\drbar}_{{\mathrm{SM}}}}}
\newc{\mtmtmssmdrbar}{\ensuremath{ m_t(m_t)^{\drbar}_{{\mathrm{SUSY}}}}}
\newc{\mbmbmsbar}{\ensuremath{ m_b(m_b)^{\msbar} }}
\newc{\mbmbsmmsbar}{\ensuremath{ m_b(m_b)^{\msbar}_{{\mathrm{SM}}}}}
\newc{\mbmzsmmsbar}{\ensuremath{ m_b(\mz)^{\msbar}_{{\mathrm{SM}}}}}
\newc{\mbmzsmdrbar}{\ensuremath{ m_b(\mz)^{\drbar}_{{\mathrm{SM}}}}}
\newc{\mbmzmssmdrbar}{\ensuremath{ m_b(\mz)^{\drbar}_{{\mathrm{SUSY}}}}}
\newc{\mtaumzsmmsbar}{\ensuremath{ m_{\tau}(\mz)^{\msbar}_{{\mathrm{SM}}}}}
\newc{\mtaumzsmdrbar}{\ensuremath{ m_{\tau}(\mz)^{\drbar}_{{\mathrm{SM}}}}}
\newc{\mtaumzmssmdrbar}{\ensuremath{ m_{\tau}(\mz)^{\drbar}_{{\mathrm{SUSY}}}}}
\newc{\alphasmzms}{\ensuremath{\alpha_s(M_Z)^{\overline{MS}}}}
\newc{\alphaimzms}[1]{\ensuremath{\alpha_{#1}(M_Z)^{\overline{MS}}}}
\newc{\alphaemmz}{\ensuremath{\alpha_{\mathrm{em}}(M_Z)^{\overline{MS}}}}

\newc{\mzero}{\ensuremath{{m_0}}}
\newc{\mhalf}{\ensuremath{ m_{1/2}}}
\newc{\tanb}{\ensuremath{\tan\beta}}
\newc{\azero}{\ensuremath{ A_0}}
\newc{\bzero}{\ensuremath{ B_0}}
\newc{\signmu}{\ensuremath{\rm{sgn}\,\mu}}
\newc{\mueff}{\ensuremath{\mu_{\rm{eff}}}}
\newc{\lam}{\ensuremath{{\lambda}}}
\newc{\kap}{\ensuremath{{\kappa}}}
\newc{\alam}{\ensuremath{{A_{\lambda}}}}
\newc{\akap}{\ensuremath{{A_{\kappa}}}}
\newc{\hs}{\ensuremath{ H_s}}      
\newc{\mhs}{\ensuremath{ m_{H_s}}} 
\newc{\mgut}{\ensuremath{ M_{\rm GUT}}}
\newc{\mplanck}{\ensuremath{ M_{\rm P}}}      \newc{\mpl}{\ensuremath{ M_{\rm Pl}}}
\newc{\msusy}{\ensuremath{ M_{\rm SUSY}}}      \newc{\ms}{\ensuremath{ M_{\rm S}}}
 \newc{\mhl}{\ensuremath{m_\hl}} 
 \newc{\mhone}{\ensuremath{m_{h_1}}} 
 \newc{\mhtwo}{\ensuremath{m_{h_2}}} 
 \newc{\mglu}{\ensuremath{m_{\tilde g}}} 
 \newc{\mul}{\ensuremath{m_{\tilde{u}_L}}} 
 \newc{\mtone}{\ensuremath{m_{\tilde{t}_1}}} 
 \newc{\ma}{\ensuremath{m_A}} 
 \newc{\maone}{\ensuremath{m_{a_1}}} 
 \newc{\matwo}{\ensuremath{m_{a_2}}}
 \newc{\hone}{\ensuremath{h_1}}
 \newc{\htwo}{\ensuremath{h_2}}
 \newc{\aone}{\ensuremath{a_1}}
 \newc{\atwo}{\ensuremath{a_2}}
 \newc{\mhu}{\ensuremath{ m_{H_u}}}       
 \newc{\mhd}{\ensuremath{ m_{H_d}}}
 \newc{\mhusq}{\ensuremath{ m_{H_u}^2}}       
 \newc{\mhdsq}{\ensuremath{ m_{H_d}^2}}
 \newc{\mhuew}{\ensuremath{ m^{\ast}_{H_u}}}       
 \newc{\mhdew}{\ensuremath{ m^{\ast}_{H_d}}}
 \newc{\mhuewsq}{\ensuremath{ m^{\ast\, 2}_{H_u}}}       
 \newc{\mhdewsq}{\ensuremath{ m^{\ast\, 2}_{H_d}}}
 \newc{\hu}{\ensuremath{ H_u}}       
 \newc{\hd}{\ensuremath{ H_d}}
 \newc{\barmhu}{\ensuremath{ \bar{m}_{H_u}}}
 \newc{\barmhd}{\ensuremath{ \bar{m}_{H_d}}}

 \newc{\mqthree}{\ensuremath{m_{\widetilde{Q}_3}^2}}
 \newc{\muthree}{\ensuremath{m_{\tilde{u}_3}^2}}
 \newc{\mdthree}{\ensuremath{m_{\tilde{d}_3}^2}}
 \newc{\mlthree}{\ensuremath{m_{\widetilde{L}_3}^2}}
 \newc{\methree}{\ensuremath{m_{\tilde{e}_3}^2}}
 \newc{\mqtwo}{\ensuremath{m_{\widetilde{Q}_2}^2}}
 \newc{\mutwo}{\ensuremath{m_{\tilde{u}_2}^2}}
 \newc{\mdtwo}{\ensuremath{m_{\tilde{d}_2}^2}}
 \newc{\mltwo}{\ensuremath{m_{\widetilde{L}_2}^2}}
 \newc{\metwo}{\ensuremath{m_{\tilde{e}_2}^2}}
 \newc{\mqone}{\ensuremath{m_{\widetilde{Q}_1}^2}}
 \newc{\muone}{\ensuremath{m_{\tilde{u}_1}^2}}
 \newc{\mdone}{\ensuremath{m_{\tilde{d}_1}^2}}
 \newc{\mlone}{\ensuremath{m_{\widetilde{L}_1}^2}}
 \newc{\meone}{\ensuremath{m_{\tilde{e}_1}^2}}
 \newc{\mone}{\ensuremath{M_1}}
 \newc{\monesq}{\ensuremath{M_1^2}}
 \newc{\mtwo}{\ensuremath{M_2}}
 \newc{\mtwosq}{\ensuremath{M_2^2}}
 \newc{\mthree}{\ensuremath{M_3}}
 \newc{\mthreesq}{\ensuremath{M_3^2}}
 \newc{\atau}{\ensuremath{{A_{\tau}}}}
 \newc{\at}{\ensuremath{{A_{t}}}}
 \newc{\ab}{\ensuremath{{A_{b}}}}
 \newc{\atausq}{\ensuremath{{A_{\tau}^2}}}
 \newc{\atsq}{\ensuremath{{A_{t}^2}}}
 \newc{\absq}{\ensuremath{{A_{b}^2}}}

 \newc{\dmzero}{\ensuremath{\Delta{_{m_0}}}}
 \newc{\dmhalf}{\ensuremath{\Delta{_{m_{1/2}}}}}
 \newc{\dmu}{\ensuremath{\Delta{_{\mu}}}}

 \newc{\pten}{\ensuremath{\psi_{10}}}
 \newc{\ffive}{\ensuremath{\phi_{5}}}
 \newc{\hfive}{\ensuremath{h_{5}}}
 \newc{\hbfive}{\ensuremath{h_{\bar{5}}}}
 \newc{\thet}{\ensuremath{\theta_{50}}}
 \newc{\thetb}{\ensuremath{\theta_{\,\overline{50}}}}
 \newc{\ptenhat}{\ensuremath{\hat{\psi}_{10}}}
 \newc{\ffivehat}{\ensuremath{\hat{\phi}_{5}}}
 \newc{\hfivehat}{\ensuremath{\hat{h}_{5}}}
 \newc{\hbfivehat}{\ensuremath{\hat{h}_{\bar{5}}}}
 \newc{\thethat}{\ensuremath{\hat{\theta}_{50}}}
 \newc{\thetbhat}{\ensuremath{\hat{\theta}_{\,\overline{50}}}}
 \newc{\si}{\ensuremath{\Sigma}}
 \newc{\mfive}{\ensuremath{m_5^2}}
 \newc{\mten}{\ensuremath{m_{10}^2}}
 \newc{\dfive}{\ensuremath{\Delta^2_5}}
 \newc{\dbfive}{\ensuremath{\Delta^2_{\bar{5}}}}
 \newc{\dfifty}{\ensuremath{\Delta^2_{50}}}
 \newc{\dfiftyb}{\ensuremath{\Delta^2_{\,\overline{50}}}}
 \newc{\msi}{\ensuremath{m_{\Sigma}^2}}
 \newc{\lamh}{\ensuremath{\lambda_{H}}}
 \newc{\lamhb}{\ensuremath{\lambda_{\bar{H}}}}
 \newc{\ah}{\ensuremath{A_{H}}}
 \newc{\ahb}{\ensuremath{A_{\bar{H}}}}
 \newc{\lams}{\ensuremath{\lambda_{S}}}
 \newc{\as}{\ensuremath{A_{S}}}
 \newc{\lamsig}{\ensuremath{\lambda_{\si}}}
 \newc{\asig}{\ensuremath{A_{\si}}}

 \newc{\msten}{\ensuremath{m_{16}^2}}
 \newc{\mhun}{\ensuremath{m_{126}^2}}
 \newc{\mhunb}{\ensuremath{m_{\bar{126}}^2}}
 \newc{\mthun}{\ensuremath{m_{210}^2}}
 \newc{\ahun}{\ensuremath{A_{\bar{126}}}}
 \newc{\yhun}{\ensuremath{Y_{\bar{126}}}}
 \newc{\aten}{\ensuremath{A_{10}}}
 \newc{\yten}{\ensuremath{Y_{10}}}
 \newc{\alone}{\ensuremath{A_{\lambda_1}}}
 \newc{\altwo}{\ensuremath{A_{\lambda_2}}}
 \newc{\althree}{\ensuremath{A_{\lambda_3}}}
 \newc{\althreeb}{\ensuremath{A_{\bar{\lambda_3}}}}
 \newc{\lone}{\ensuremath{\lambda_1}}
 \newc{\ltwo}{\ensuremath{\lambda_2}}
 \newc{\lthree}{\ensuremath{\lambda_3}}
 \newc{\lthreeb}{\ensuremath{\bar{\lambda_3}}}

\newc{\sigsip}{\ensuremath{\sigma^{\rm SI}_{p}}}	\newc{\sigsin}{\ensuremath{\sigma^{\rm SI}_{n}}}
\newc{\sigsdp}{\ensuremath{\sigma^{\rm SD}_{p}}}	\newc{\sigsdn}{\ensuremath{\sigma^{\rm SD}_{n}}}
\newc{\sigsi}{\ensuremath{\sigma^{\rm SI}}}	\newc{\sigsd}{\ensuremath{\sigma^{\rm SD}}}
\newc{\sigv}{\ensuremath{\sigma v}}
\newc{\abund}{\ensuremath{ \Omega h^2}}
\newc{\omegadm}{\ensuremath{ \Omega_{{\rm DM}}}}     \newc{\abunddm}{\ensuremath{ \Omega_{{\rm DM}} h^2}} 
\newc{\omegam}{\ensuremath{ \Omega_{{\rm m}}}}       \newc{\abundm}{\ensuremath{ \Omega_{{\rm m}} h^2}}
\newc{\omegab}{\ensuremath{ \Omega_{{\rm b}}}}	\newc{\abundb}{\ensuremath{ \Omega_{{\rm b}} h^2}}
\newc{\omegatot}{\ensuremath{ \Omega_{{\rm TOT}}}}
\newc{\omegacdm}{\ensuremath{ \Omega_{{\rm CDM}}}}   \newc{\abundcdm}{\ensuremath{ \Omega_{{\rm CDM}} h^2}}
\newc{\omegalambda}{\ensuremath{ \Omega_{\Lambda}}} \newc{\abundlambda}{\ensuremath{ \Omega_{\Lambda} h^2}}
\newc{\omegarad}{\ensuremath{ \Omega_{{\rm rad}}}}  \newc{\abundrad}{\ensuremath{ \Omega_{{\rm rad}} h^2}}
\newc{\rhocrit}{\ensuremath{ \rho_{\rm crit}}}
\newc{\rhochi}{\ensuremath{ \rho_{\chi}}}
\newc{\abunchi}{\ensuremath{\Omega_\chi h^2}}
\newc{\abundlsp}{\ensuremath{\Omega_{\rm LSP}h^2}}
\newcommand*{\abundchi}{\ensuremath{\Omega_\chi h^2}}

\newc{\amu}{\ensuremath{ a_{\mu}}}        \newc{\amususy}{\ensuremath{ a_{\mu}^{\mathrm{SUSY}}}}
\newc{\amuexpt}{\ensuremath{ a_{\mu}^{\mathrm{expt}}}}        \newc{\amusm}{\ensuremath{ a_{\mu}^{\mathrm{SM}}}}
\newc\deltaamu{\ensuremath{\Delta a_{\mu}}} \newc{\deltaamususy}{\ensuremath{\delta a_{\mu}^{\mathrm{SUSY}}}}
\newc\gmtwo{\ensuremath{ (g-2)_{\mu}}} 
\newc{\deltagmtwomususy}{\ensuremath{\delta\left(g-2\right)_{\mu}^{\mathrm{SUSY}}}}
\newc{\deltagmtwomu}{\ensuremath{\delta\left(g-2\right)_{\mu}}}
\newc\BR{\ensuremath{\rm BR}}
\newc\bsgamma{\ensuremath{ b\rightarrow s \gamma }}
\newc\bxsgamma{\ensuremath{\overline{B}\rightarrow X_{s}\gamma}}
\newc\brbsgamma{\ensuremath{\BR\left(\bsgamma\right)}}
\newc\brbxsgamma{\ensuremath{\BR\left(\bxsgamma\right)}}
\newc\bsmumu{\ensuremath{B_s\to\mu^+\mu^-}}
\newc\brbsmumu{\ensuremath{\BR\left(B_s\to\mu^+\mu^-\right)}}
\newc\bdmmumu{\ensuremath{\overline{B}_d\to\mu^+\mu^-}}
\newc\bbbarmix{\ensuremath{\overline{B}_s\mbox{-}B_s}}      
\newc\delmbs{\ensuremath{\Delta M_{B_s}}}
\newc{\butaunu}{\ensuremath{B_u \rightarrow \tau \nu}}
\newc{\brbutaunu}{\ensuremath{\BR\left(B_u \rightarrow \tau \nu\right)}}

\newcommand*{\reffig}[1]{Fig.~\ref{#1}}

     \newcommand*{\refsec}[1]{Sec.~\ref{#1}}

\newcommand*{\charone}{\ensuremath{\chi^{\pm}_1}}

\newcommand*{\stau}{\ensuremath{\tilde{\tau}}}

\newcommand*{\mstopone}{\ensuremath{m_{\tilde{t}_1}}}
\newcommand*{\mstoptwo}{\ensuremath{m_{\tilde{t}_2}}}

\newcommand*{\alphaT}{\ensuremath{\alpha_T}}

\newcommand*{\alphaTelefb}{\ensuremath{\cms\ \alphaT\ 11.7\invfb}}
\newcommand*{\razor}{\textrm{razor}}
\newcommand*{\razorfourfb}{\ensuremath{\cms\ \razor\ 4.4\invfb}}

\newcommand*{\cms}{\text{CMS}}

\newcommand*{\xenononet}{\text{XENON-1T}}

\let\oldcite\cite
\renewcommand*{\cite}{~\oldcite}

\newcommand*{\hl}{\ensuremath{h}}

\restylefloat{figure}

\title{What next for the CMSSM and the NUHM: improved prospects for superpartner and dark matter detection}

\author[1]{Leszek Roszkowski,\note{On leave of absence from the University of Sheffield, U.K.}}
\author{Enrico Maria Sessolo}
\author{and Andrew J.~Williams}

\affiliation{National Centre for Nuclear Research,\\
  Ho{\. z}a 69, 00-681 Warsaw, Poland} 

\emailAdd{L.Roszkowski@sheffield.ac.uk}
\emailAdd{Enrico-Maria.Sessolo@fuw.edu.pl}
\emailAdd{Andrew.Williams@fuw.edu.pl}

\abstract{We present an updated analysis of the CMSSM and the NUHM using the
  latest experimental data and numerical tools.  We map out favored
  regions of Bayesian posterior probability in light of data from the LHC,
  flavor observables, the relic density and dark matter searches.  We
  present some updated features with respect to our previous analyses:
  we include the effects of corrections to the light Higgs mass beyond
  the 2-loop order using $\tt FeynHiggs~2.10.0$; we include in the
  likelihood the latest limits from direct searches for squarks and
  gluinos at ATLAS with $\sim 20\invfb$; the latest constraints on the
  spin-independent scattering cross section of the neutralino from LUX
  are applied taking into account uncertainties in the nuclear form
  factors.  We find that in the CMSSM the posterior distribution now tends
  to favor smaller values of \msusy\ than in the previous analyses.
  As a consequence, the statistical weight of the $A$-resonance region
  increases to about 30\% of the total probability, with interesting
  new prospects for the 14\tev\ run at the LHC. The most favored
  region, on the other hand, still features multi-TeV squarks and
  gluinos, and $\sim1\tev$ higgsino dark matter whose detection
  prospects by current and one-tonne detectors look very promising. The same region is
  predominant in the NUHM, although the $A$-resonance region is also present there as well as a new solution, 
  of neutralino-stau coannihilation through the channel
  $\tilde{\tau}\tilde{\tau}\rightarrow hh$ at very large $\mu$.  
  We derive the expected sensitivity of the future CTA experiment to $\sim1\tev$ higgsino
  dark matter for both models and show that the prospects for probing
  both models are realistically good.  We comment on the
  complementarity of this search to planned direct detection one-tonne
  experiments.}

\begin{document}
\maketitle
\flushbottom

\section{\label{sec:intro}Introduction}

The recent discovery of a Higgs boson at the
LHC\cite{Aad:2012tfa,Chatrchyan:2012ufa} raised widespread excitement
in the particle physics community and spurred a lot of activity to
interpret the new discovery in the context of the Standard Model (SM) and models of new
physics.  In particular, the mass of the newly discovered particle,
$\mhl\simeq 126\gev$, is well within (albeit on the upper side) the
predictions of low-scale supersymmetry (SUSY). In fact, since in SUSY
the quartic coupling of the scalar potential is related to the gauge
couplings of the electroweak (EW) sector, the lightest Higgs mass can
deviate from the masses of the $W$ and $Z$ bosons only through
radiative corrections so that its value is effectively bounded to be less
than about 135\gev.

The implications of the Higgs discovery for the parameter space of the
popular Constrained Minimal Supersymmetric Standard Model (CMSSM)\cite{Kane:1993td}
and Non-Universal Higgs Model (NUHM), have been intensely investigated (see, e.g.,\cite{Baer:2011ab,Kadastik:2011aa,Cao:2011sn,Ellis:2012aa,Baer:2012uya,Bechtle:2012zk,Balazs:2013qva,
Fowlie:2012im,Akula:2012kk,Buchmueller:2012hv,Strege:2012bt,Cabrera:2012vu,Kowalska:2013hha,Dighe:2013wfa,Cohen:2013kna} 
for some of the papers that followed the Higgs discovery). 
Many of those studies, including our own ones, explored statistical combinations of the constraints from the 
Higgs measurements at the LHC with other pieces of experimental information: the measurement of the dark matter relic density\cite{Ade:2013zuv}; 
a number of EW precision observables; 
measurements of rare-decay branching ratios like \brbxsgamma\cite{bsgamma}, \brbutaunu\cite{Adachi:2012mm}, or the 
recent measurement of a SM-like \brbsmumu\ at the
LHC\cite{Aaij:2013aka,Chatrchyan:2013bka}; the anomalous magnetic
moment of the muon\cite{Bennett:2006fi,Miller:2007kk}, etc.
Additional constraints came from direct SUSY searches at the LHC. For instance, 
in our previous CMSSM and NUHM analyses\cite{Fowlie:2012im,Kowalska:2013hha} the likelihood 
function included limits from the \razorfourfb\ analysis at 7\tev\cite{Chatrchyan:2012uea} and the \alphaTelefb\ analysis at 8\tev\cite{Chatrchyan:2013lya}, 
which were obtained by simulating the SUSY signal and detector response
and comparing them to the observed and background yields in different channels given by the experimental collaboration. 

It is important to point out that, besides the Higgs boson mass measurement and LHC direct bounds, 
the constraint showing by far the strongest impact on the parameter space of the Minimal Supersymmetric SM (MSSM) is the relic density. 
It is measured very precisely and its value tends to be too large in broad regions of the parameter space. 
The mechanisms for reducing the relic abundance are quite general and insensitive to the particular pattern of scalar mass unification with the exception, 
obviously, of solutions requiring the presence of light sleptons, which in the CMSSM/NUHM are excluded by direct LHC limits on the squarks. 
In other words, while featuring a limited number of free parameters that make them more predictive than general phenomenological parametrizations 
defined at the scale of the lightest SUSY partners, the CMSSM and the NUHM produce solutions 
to the relic density that are present and play an important role in more general models. 
From this point of view, the CMSSM ansatz and its most immediate extension, the NUHM, are very useful 
frameworks to investigate the predictions for dark matter in constrained frameworks 
within the MSSM with gaugino unification at the GUT scale.
 
The picture that emerged particularly in the analyses\cite{Cabrera:2012vu,Kowalska:2013hha} for the CMSSM and\cite{Strege:2012bt,Kowalska:2013hha} for the NUHM, 
is that the relatively large value of the Higgs mass, 
the SM-like nature of its couplings to the other SM particles, the
measured values of the flavor physics observables in great agreement with the SM,
and the non-observation of SUSY particles below $\sim1-1.5\tev$ at the LHC, 
can all be easily accommodated in an extended region of the parameter space characterized
by squarks and gluinos in the multi-TeV regime and the heavy Higgs sector effectively decoupled. 
In this region, the relic density can naturally assume the value measured by PLANCK, as the lightest SUSY particle (LSP) 
is an almost pure higgsino neutralino with a mass $\mchi\simeq 1\tev$.\footnote{While other regions of the parameter space, 
characterized by bino dark matter and such that the relic abundance is saturated through mechanisms of $A$-resonance or stau co-annihilation,
are consistent with the constraints at the 1 or $2\sigma$ level, the $\sim1\tev$ higgsino region shows by far the best agreement with
all of them with the exception of \deltagmtwomu, which favors low
\msusy\ below ATLAS/CMS bounds in unified models.} The existence of
the $\sim1\tev$ higgsino solution for DM in the MSSM has been long known\cite{Profumo:2004at,ArkaniHamed:2006mb}
but in the framework of unified SUSY was first pointed out in a pre-LHC study of the NUHM\cite{Roszkowski:2009sm}.

Interestingly, parameter space regions with sparticles in the
multi-TeV regime and $\sim1\tev$ higgsino dark matter in agreement
with the value of the relic density were shown to be very favored also
in scans of the phenomenological MSSM\cite{Fowlie:2013oua}, in the
Next-to-Minimal Supersymmetric SM\cite{Kaminska:2013mya}, and in a
variety of models with non-universal boundary conditions at the GUT
scale\cite{Kowalska:2014hza}. Incidentally, it was shown
in\cite{Kowalska:2014hza} that for some of these models focus
point-like mechanisms significantly increase the naturalness of the
$\sim1\tev$ higgsino region with respect to the CMSSM, without
affecting dark matter properties and prospects for detection.  As a
matter of fact, the best prospects for detection of the $\sim 1\tev$
higgsino region indeed come from dark matter direct detection
experiments, particularly at 1-tonne detectors, as the spin-independent neutralino-proton
cross section, \sigsip, is well within the projected sensitivities of
the currently running and future 1-tonne experiments.

Also very interestingly, recent studies of the sensitivity of the Cherenkov Telescope Array (CTA)\cite{Acharya:2013sxa} 
show the largest projected reach in the region characterized by dark matter mass and annihilation cross section typical
of the $\sim1\tev$ higgsino region\cite{Doro:2012xx,Pierre:2014tra}, thus opening up the enticing possibility of complementary detection
for these scenarios.

On the other hand, the statistical analyses of
Refs.\cite{Bechtle:2012zk,Balazs:2013qva,Fowlie:2012im,Akula:2012kk,Buchmueller:2012hv,Strege:2012bt,Cabrera:2012vu,Kowalska:2013hha}
were based on the calculation of the Higgs mass performed at
two loops, implemented in the most popular SUSY spectrum
calculators\cite{Allanach:2001kg,Djouadi:2002ze,Porod:2003um} or in
earlier versions of $\tt
FeynHiggs$\cite{Heinemeyer:1998np,Heinemeyer:1998yj,Degrassi:2002fi,Frank:2006yh}.
Significant effort in the direction of improving the
theoretical precision of the Higgs mass calculation in the
MSSM\cite{Martin:2007pg,Harlander:2008ju,Kant:2010tf,Hahn:2013ria,Draper:2013oza} has recently prompted
some groups to update their previous
analyses\cite{Feng:2013tvd,Buchmueller:2013psa,Buchmueller:2013rsa}.
Reference\cite{Buchmueller:2013rsa} in particular shows the first
global statistical analysis to incorporate these recent developments
in a frequentist approach.

In this paper, we update the global Bayesian analyses of the CMSSM and the NUHM previously produced by our group, 
by including the following new elements:\smallskip

$\bullet$ We calculate the Higgs mass with $\tt FeynHiggs~2.10.0$\cite{Hahn:2013ria}, 
which incorporates results beyond two loops with a resummation of leading and 
sub-leading logarithms in the top/stop sector.

$\bullet$ We update the likelihood map for direct SUSY searches with the most constraining limits from ATLAS with 20\invfb\ at 8\tev.

$\bullet$ We include the recent constraints from the LUX experiment\cite{Akerib:2013tjd} in the likelihood function.

$\bullet$ We add an analysis of the prospect for CTA to independently
explore the favored regions.\smallskip

We will focus in particular on the properties of the regions  that present the highest 
posterior probability, favored by the value of the dark matter relic density. 
We will show that the bulk of the posterior still lies on the $\sim 1\tev$ higgsino region, 
but in the CMSSM its statistical significance is not as overwhelmingly predominant as previously shown. 
We will touch on the prospects for detection at the LHC and in
future colliders, but will focus particularly on 
dark matter searches like the above-mentioned 1-tonne detectors and the CTA. 

The paper is organized as follows. In \refsec{sec:constraints} we describe the statistical setup,
the constraints included in the likelihood function, the parameter prior ranges and distributions, and the numerical 
tools used in our scans. In \refsec{sec:CMSSM} we present the results of the Bayesian analysis of the CMSSM, including prospects at
collider and dark matter experiments. In \refsec{sec:NUHM} we present equivalent results for the NUHM.
We give our summary and conclusions in \refsec{sec:summary}.

\section{Scanning methodology and experimental constraints}\label{sec:constraints}

Our goal is to determine the regions in the parameter space of the CMSSM and the NUHM that are favored by all of the experimental data available. 
We follow a Bayesian approach outlined in\cite{Fowlie:2011mb,Roszkowski:2012uf,Fowlie:2012im,Kowalska:2013hha} and 
map out the $68\%$ and $95\%$ credible regions in two-dimensional (2D) projections of the marginalized posterior probability 
density function (pdf) and/or the one-dimensional (1D) marginalized pdf of some interesting parameters. 
The posterior pdf (or, simply, the posterior), $p(m|d)$, is given by Bayes' Theorem,
\begin{equation}
p(m|d) = \frac{p(d|m) \pi(m)}{p(d)}\,,
\end{equation}
where $m$ is the set of model parameters, $\pi(m)$ is prior probability distribution of the parameters $m$,  $p(d|m) \equiv {\cal L}(m)$ is 
the probability of obtaining the experimental data $d$ given the model parameters, known as the likelihood function, 
and $p(d)$ is a normalization factor called the evidence, which is required for the comparison of different models. 
In this framework the likelihood function encodes all of our information from experimental constraints and their associated uncertainties.

We can consider marginalized posterior distributions of subsets of the parameters $m$ by integrating over the remaining parameters. 
The posterior distribution of a subset of parameters $\psi_{1...r}$ from the full set of parameters $m_{1...N}$ is then given by
\begin{equation}
p(\psi_{1..r}|d) = \int  p(m|d) d^{N-r}m\,.
\end{equation}
This leads to a natural prescription for dealing with nuisance parameters, since these can be included as parameters of the model 
with suitable prior probability distributions and then marginalized over to obtain posterior distributions of the parameters of interest. 

\bigskip
We construct the likelihood function from the experimental data. 
We account for positive measurements with a gaussian likelihood function and combine experimental and theoretical errors in quadrature. 
A summary of the experimental constraints used in this analysis is given in Table~\ref{tab:exp_constraints}. 
In addition, we adopt a specific procedure for incorporating the constraints from Higgs boson searches at the LHC, 
direct searches for SUSY, and the constraints from dark matter direct detection at LUX. 

\begin{table}[t]
\begin{center}
\begin{tabular}{|c|c|c|c|c|}
\hline
Constraint & Mean & Exp. Error & Th. Error & Ref. \\
\hline
Higgs sector & See text. & See text. & See text. & \cite{Bechtle:2013xfa,Bechtle:2008jh,Bechtle:2011sb,Bechtle:2013wla} \\
\hline
Direct SUSY searches & See text. & See text. & See text. & \cite{Drees:2013wra,Barr:2003rg,Cheng:2008hk,Cacciari:2005hq,Cacciari:2008gp,Cacciari:2011ma,deFavereau:2013fsa,Lester:1999tx,Read:2002hq} \\
\hline
\sigsip\ & See text. & See text. & See text. & \cite{Akerib:2013tjd}\\
\hline
\abunchi\ & 0.1199 & 0.0027 & 10\% & \cite{Ade:2013zuv}\\
\hline
\sinsqeff\ & 0.23155 & 0.00015 & 0.00015 &  \cite{Beringer:1900zz} \\
\hline
$\deltagmtwomu\times 10^{10}$ & 28.7 & 8.0 & 1.0 & \cite{Bennett:2006fi,Miller:2007kk} \\
\hline
$\brbxsgamma\times 10^4$ & 3.43 & 0.22 & 0.21 & \cite{bsgamma} \\
\hline
$\brbutaunu \times 10^4$ & 0.72 & 0.27 & 0.38 & \cite{Adachi:2012mm} \\
\hline
\delmbs\ & 17.719~ps$^{-1}$ & 0.043~ps$^{-1}$ & 2.400~ps$^{-1}$ & \cite{Beringer:1900zz} \\
\hline
$M_W$ & $80.385\gev$ & $0.015\gev$ & $0.015\gev$ & \cite{Beringer:1900zz} \\
\hline
$\brbsmumu \times 10^9$ & 2.9 & $0.7$ & 10\% & \cite{Aaij:2013aka,Chatrchyan:2013bka} \\
\hline
\end{tabular}
\caption{The experimental constraints used in this study.}
\label{tab:exp_constraints}
\end{center}
\end{table}%

The discovery of a SM-like Higgs boson with $\mhl\simeq126\gev$ presents a challenging constraint. 
The experimental collaborations have published the Higgs signal rates in several channels and we make use of this 
information by interfacing with the public code $\tt HiggsSignals\ v1.0.0$\cite{Bechtle:2013xfa}.  
We supply $\tt HiggsSignals$ with the Higgs boson production cross section and branching ratios calculated 
by $\tt FeynHiggs$\cite{Heinemeyer:1998np,Heinemeyer:1998yj,Degrassi:2002fi,Frank:2006yh} and use the calculated $\chi^2$ result in the likelihood function. 
An accurate prediction for the lightest Higgs boson is important due to the small width of the observed experimental signal. 
Recently\cite{Hahn:2013ria} $\tt FeynHiggs\ v2.10.0$ has incorporated results beyond two loops with a resummation of leading and sub-leading logarithms 
in the top/stop sector. We include these in our scan. 
We fix the uncertainty in the mass of the lightest Higgs to 2\gev, as a conservative estimate 
of the remaining sources of theory error. As well as the observed signal at $\sim126\gev$, Higgs searches 
in other mass ranges may constrain the heavy Higgs bosons. We use 
$\tt HiggsBounds\ v4.1.0$\cite{Bechtle:2008jh,Bechtle:2011sb,Bechtle:2013wla} to reject points excluded at 95\%~\cl\ by those searches.

\begin{figure}[t]
\centering
\subfloat[]{%
\label{fig:a}%
\includegraphics[width=0.40\textwidth]{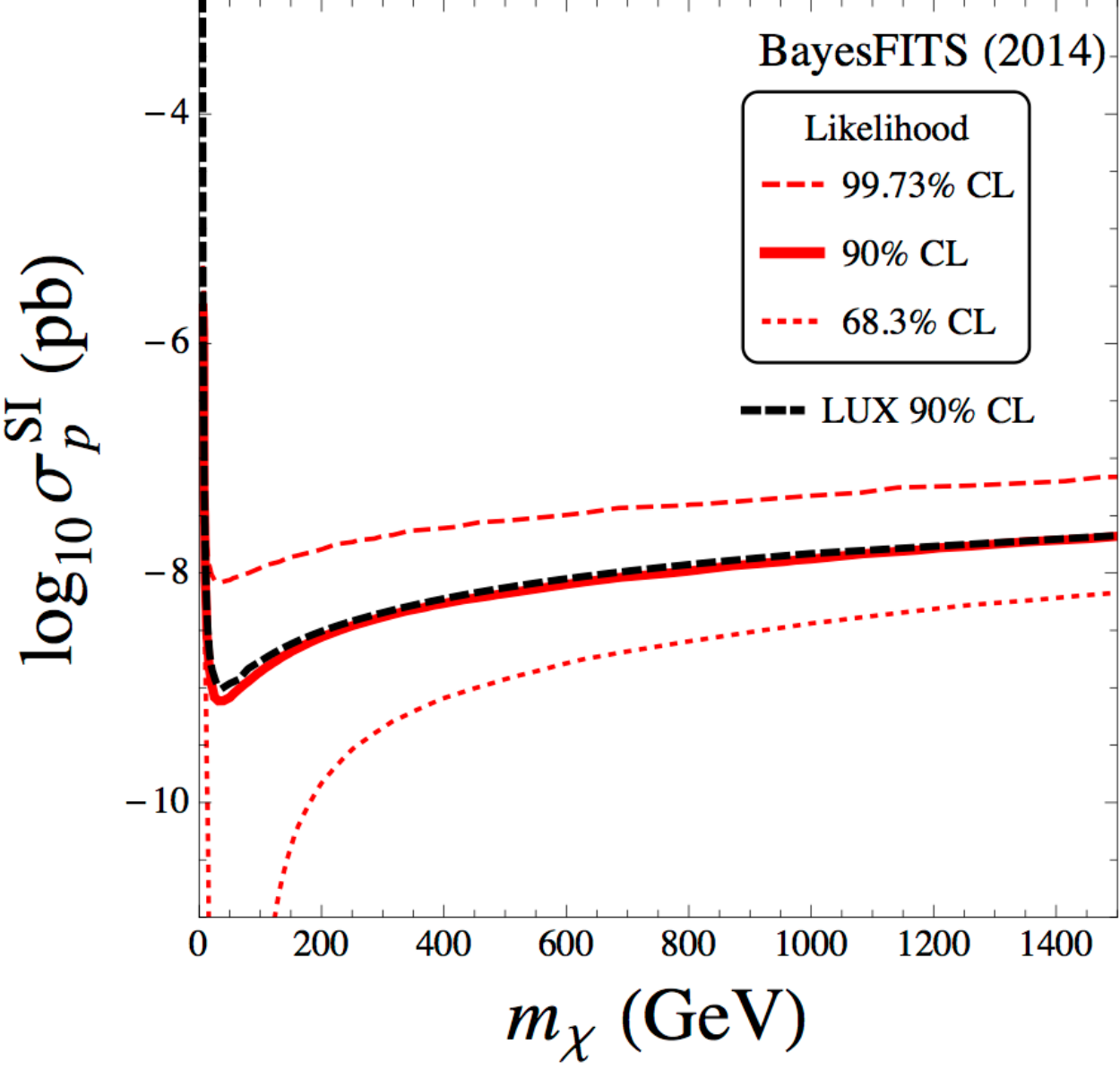}
}%
\hspace{0.07\textwidth}
\subfloat[]{%
\label{fig:b}%
\includegraphics[width=0.411\textwidth]{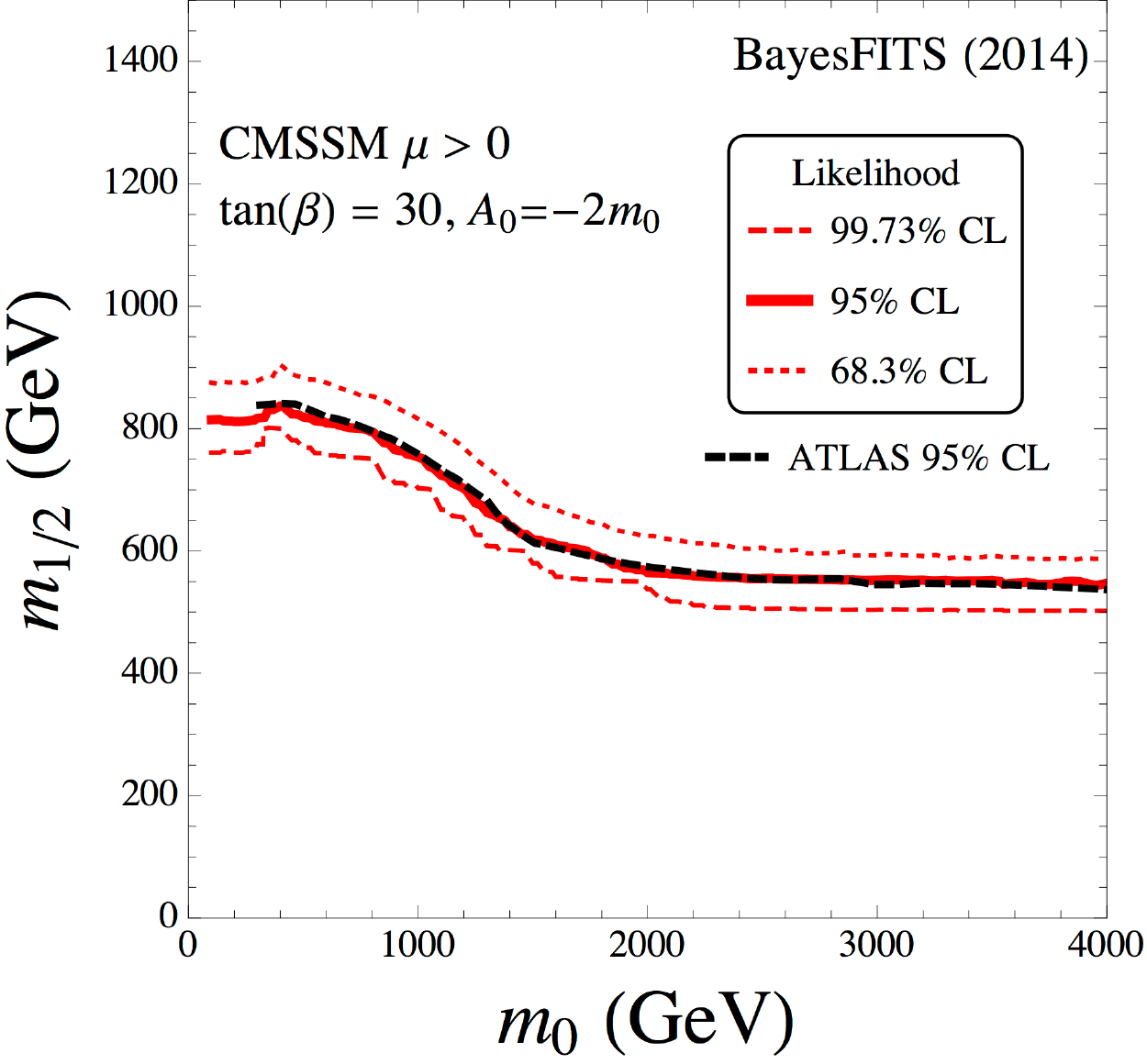}
}%
\caption{\footnotesize \protect\subref{fig:a} The 68.3\%~C.L. (red dotted line), 90\%~C.L. (red solid line) and 99.7\%~C.L. (red dashed line) 
exclusion bounds given by our likelihood map for dark matter direct detection experiments, compared with the 90\%~C.L. limit 
published by LUX\cite{Akerib:2013tjd} (black dashed line). \protect\subref{fig:b} The 68.3\%~C.L. (red dotted line), 95\%~C.L. (red solid line) and 99.7\%~C.L. 
(red dashed line) exclusion bounds for our 
combination of the ATLAS searches compared to the original 95\%~C.L. contour from ATLAS (black dashed line).}
\label{fig:likemaps}
\end{figure}

The contribution to the likelihood arising from the results of the LUX
experiment\cite{Akerib:2013tjd} is derived as was explained
in\cite{Kowalska:2014hza}, i.e., by applying the procedure developed
in Ref.\cite{Cheung:2012xb} for the likelihood of
XENON100\cite{Aprile:2012nq} to the data from LUX.  We assume that the
number of observed events follows a Poisson distribution centered on
the predicted signal plus background.  A likelihood map in the (\mchi,
\sigsip) plane is generated by simulating signal events in $\tt
micrOMEGAs$\cite{Belanger:2013oya} and marginalizing over the
uncertainty in the expected number of background events.  In
\reffig{fig:likemaps}\subref{fig:a} we plot the 68.3\%, 90\%, and
99.7\%~C.L. exclusion bounds obtained with our procedure.  The
dashed black line gives the official 90\%~C.L. exclusion bound.  In
our scans, we also account for uncertainties in the predicted elastic
scattering cross section\cite{Ellis:2008hf,Young:2013nn} by including
the nuclear form factors $\sigma_s$ and $\Sigma_{\pi N}$ as nuisance
parameters.

We finally account for the direct SUSY searches at the LHC by updating the method developed in\cite{Fowlie:2012im,Kowalska:2013hha}. 
We generate a grid in the (\mzero, \mhalf) plane at 50-GeV intervals. 
At each point we generate squark- and gluino-production events using $\tt Madgraph$\cite{Alwall:2011uj} and produce the parton shower in 
$\tt pythia$\cite{Sjostrand:2007gs}. 
The cross sections are calculated using $\tt nll-fast$\cite{Beenakker:1996ch,Kulesza:2008jb,Kulesza:2009kq,Beenakker:2009ha,Beenakker:2011fu} 
to include the next-to-leading order and next-to-leading log contributions. 
We evaluate the expected number of events in a given signal region for the searches considered using $\tt CheckMATE$
\cite{Drees:2013wra,Barr:2003rg,Cheng:2008hk,Cacciari:2005hq,Cacciari:2008gp,Cacciari:2011ma,deFavereau:2013fsa,Lester:1999tx,Read:2002hq}. 
$\tt CheckMATE$ includes a number of validated SUSY searches and includes an advanced tuning of the fast detector simulation. 
We calculate a likelihood for each search from the product of Poisson distributions for each signal region. 
We account for the uncertainties in the background rate by marginalizing over the background rate with a gaussian distribution. 
When calculating the likelihood, we consider the two searches that give the strongest limits in the CMSSM: a 0 lepton 2--6 jets 
ATLAS search\cite{ATLAS-CONF-2013-047} and a 0--1 lepton 3 $b$-jets ATLAS search\cite{ATLAS-CONF-2013-061}.  
We scale the total squark and gluino production rate by a small constant factor to match the limit achieved by the experimental analyses in order to account for the remaining differences in efficiencies due to the fast detector simulation. 
To combine the results of the two ATLAS searches we evaluate at each point which of the two searches 
has the largest expected exclusion and then use that search to calculate the likelihood for that point. 
The combination of the two searches compared to the official 95\%~C.L. line from ATLAS 
(where we have combined the lines from the two different searches) is shown in \reffig{fig:likemaps}\subref{fig:b}. 
The agreement is very good across the entire range of \mzero\ and \mhalf\,.

\bigskip

In this study, we also estimate the sensitivity of the Cherenkov Telescope Array (CTA)\cite{Acharya:2013sxa} for the favored parameter space 
of the CMSSM and the NUHM. The results will be shown in \refsec{sec:dmcmssm} and \refsec{sec:dmnuhm}, respectively. 

The CTA project will build the next generation air Cherenkov
telescope observatory. For dark matter masses greater than $\sim
100\gev$ CTA is expected to significantly exceed current
experimental limits for dark matter annihilations such as those from
HESS\cite{Abazajian:2011ak} and Fermi-LAT\cite{Ackermann:2013yva}.
CTA may even probe cross sections below the canonical thermal relic
value for some final states\cite{Pierre:2014tra}.

Reference\cite{Pierre:2014tra} estimated the future sensitivity of CTA to dark-matter annihilation in the Galactic Center (GC) for the final states 
$b\bar{b}$, $\mu^+\mu^-$ and $\tau^+\tau^-$ assuming 500 hours of observation time. 
In order to directly apply their limits to more generic neutralino annihilations, whose final states 
also include gauge bosons, $Z Z$ and $W^+ W^-$, we infer from\cite{Pierre:2014tra} the 95\%~C.L. limit on the expected flux of signal photons per $J$-factor, 
$N_{\gamma,95}$, by convolving the photon flux with the effective area given in\cite{Bernlohr:2012we} 
in a single energy bin between $30\gev$ and the dark matter mass:
\begin{equation}
N_{\gamma,95} =  t_{\text{obs}} \frac{\langle \sigma v \rangle_{95}^{(b\bar{b},\mu^+\mu^-,\tau^+\tau^-)}}{8 \pi m_\chi^2} N_{\gamma,\text{obs}}\,J_{\textrm{fact}}\,, 
\end{equation} 
where $t_{\text{obs}}$ is the observation time, $\langle \sigma v \rangle_{95}$ 
is the projected 95\%~C.L. limit for each final state, and $N_{\gamma,\text{obs}}$ is given by
\begin{equation}
N_{\gamma,\text{obs}}=\int_{30\gev}^{\mchi}\frac{dN_{\gamma}(E)}{dE}A_{\text{eff}}(E)dE\,,
\end{equation}
where $A_{\text{eff}}$ is the effective area, $dN_{\gamma}(E)/dE$ is the energy spectrum per annihilation 
and for simplicity we have neglected the effect of finite energy resolution.
For validation, one must make sure that the $N_{\gamma,95}$ obtained from the different final-state bounds of\cite{Pierre:2014tra} 
are all in very good agreement.
 
\begin{figure}[t]
\centering
\includegraphics[width=0.45\textwidth]{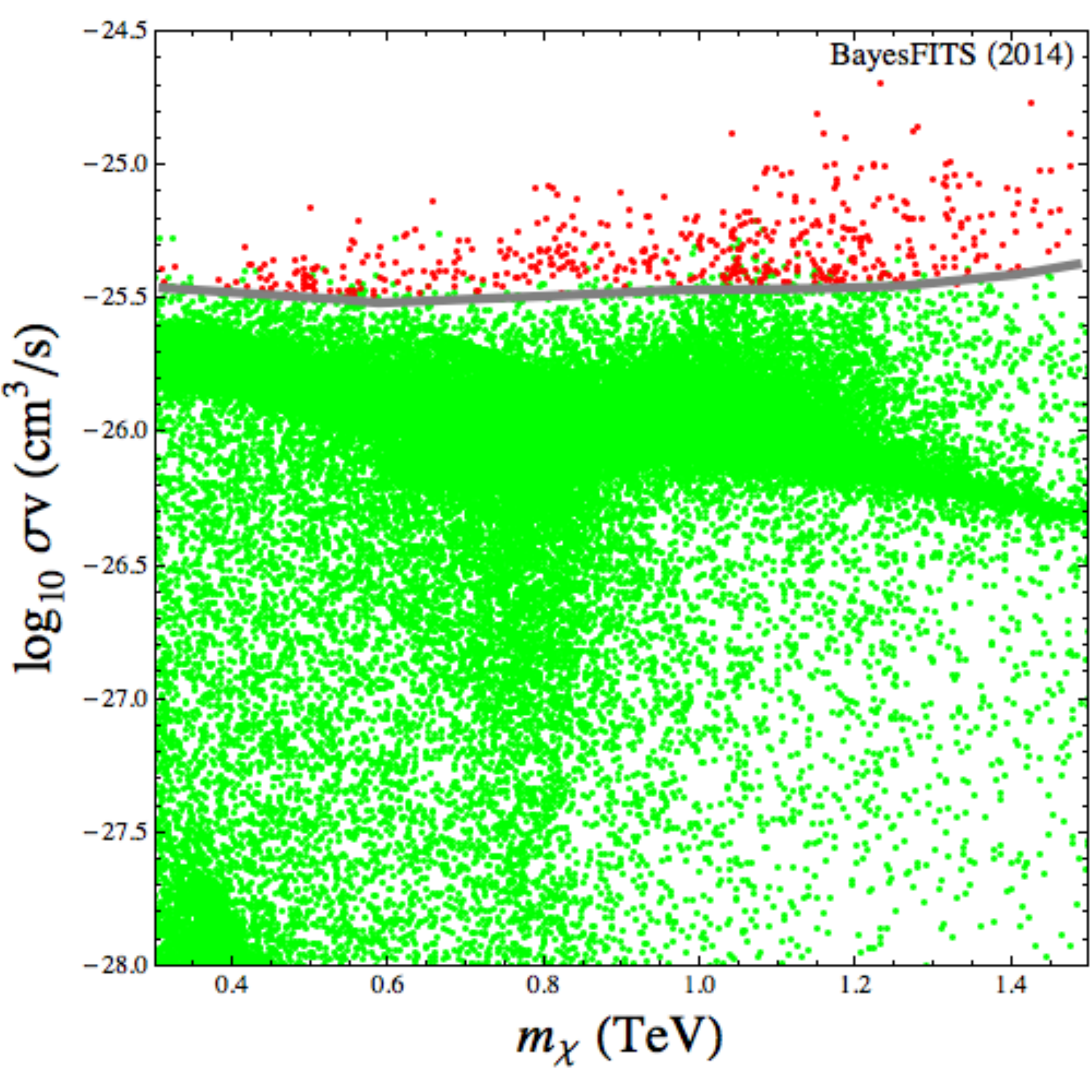}
   \caption{\footnotesize Our indicative 95\%~C.L. projected sensitivity in the (\mchi, $\sigma v$) plane of the MSSM
   for 500~h of observation at CTA is shown as a gray solid line. 
   The red (dark) points are excluded at the 95\%~C.L. through direct comparison with $N_{\gamma,95}$, which we derived from the projected bounds given in\cite{Pierre:2014tra} for different annihilation final states. 
   The gray line marks the maximum cross section bound, which for each point depends on the individual final states.}
   \label{fig:cmssm_mx_sigmav_cta}
\end{figure}

Once $N_{\gamma,95}$ is derived, the photon flux is calculated point by point in our scans using $\tt micrOMEGAs$.
In \reffig{fig:cmssm_mx_sigmav_cta} we show the approximate 95\%~C.L. limit in the plane of the annihilation cross section 
times velocity at zero momentum, $\sigma v\equiv\lim_{p\rightarrow 0}\langle\sigma v\rangle$, 
versus the neutralino mass, \mchi. The limit is extracted by testing each point against the expected $N_{\gamma,95}$, 
and showing them in red if they are excluded.
The large statistical sample shown in \reffig{fig:cmssm_mx_sigmav_cta} includes the scans performed for this study
and those used in\cite{Fowlie:2013oua}. 

Note that the limit is derived for the photon flux, so that there is no corresponding clear-cut limit in $\sigma v$.
Nevertheless, we show with a gray solid line the maximum extent that can be probed with this method, and
we will apply the obtained limit to the figures in \refsec{sec:dmcmssm} and \refsec{sec:dmnuhm}.

\bigskip

The scans in this study are performed with the BayesFITS package\cite{Fowlie:2012im,Kowalska:2012gs,Kowalska:2013hha,Fowlie:2013oua}, 
which interfaces several publicly available tools to 
direct the scanning procedure and calculate physical observables. The sampling is performed by MultiNest\cite{Feroz:2008xx} with 4000 and 10000 live points 
for the CMSSM and NUHM respectively. The evidence tolerance is set to 0.5 and the sampling efficiency to 0.8 for all the scans. 
We use $\tt SoftSusy \,v.3.3.9$\cite{Allanach:2001kg} to calculate the mass spectrum. 
As was explained above, this is passed via the SUSY LesHouches Accord format to $\tt FeynHiggs\ v.2.10.0$ 
to calculate the higher-order corrections to the Higgs mass. 
$\tt FeynHiggs$ is interfaced with $\tt HiggsSignals$ and $\tt HiggsBounds$ 
to evaluate the constraints on the Higgs sector. $\tt SuperISO\ v.3.3$\cite{Mahmoudi:2008tp} 
is used to calculate \brbxsgamma, \brbsmumu,\footnote{We use the \textit{non} time-averaged output of $\tt SuperISO$, as explained in\cite{Fowlie:2013oua}. 
It is numerically closer to the time-averaged SM prediction of Ref.\cite{Bobeth:2013uxa} 
than the code's own time-averaged calculation.} \brbutaunu, and \deltagmtwomu. 
The observables $M_W$, \sinsqeff, \delmbs\ are calculated using $\tt FeynHiggs$. 
The dark matter observables \abundchi, \sigsip, and $\sigma v$ are computed using $\tt micrOMEGAs\ v.3.5.5$\cite{Belanger:2013oya}. 

The prior distributions of the model and nuisance parameters for the CMSSM are given in Table~\ref{tab:cmssmprior}. Additionally the parameters scanned in the NUHM are indicated with an asterisk. The sign of $\mu$ is fixed for each scan.
Note that for $\tanb<3$ it becomes very difficult to obtain at the same time EWSB, 
the correct value of the relic density,  or the correct Higgs mass,
as the parameters \mhu, \mhd, \tanb, and the one-loop tadpole corrections must be fine-tuned very precisely.
Note also that for the scans with $\mu<0$ we do not include \deltagmtwomu\ in the likelihood
as it is known to be poorly fitted.
We use logarithmic priors for the universal mass parameters $m_0$ and $m_{1/2}$ to reduce volume effects. 

\begin{table}[t]
\begin{center}
\begin{tabular}{|c|c|c|c|}
\hline
\footnotesize Parameter & \footnotesize Description & \footnotesize Range & \footnotesize Distribution \\
\hline
\mzero\ & \footnotesize Universal scalar mass & $0.1,20$ & \footnotesize Log \\
\hline
\mhalf & \footnotesize Universal gaugino mass & $0.1,10$ & \footnotesize Log \\
\hline
\azero\ & \footnotesize Universal trilinear coupling & $-20,20$ & \footnotesize Linear \\
\hline
\tanb\ & \footnotesize Ratio of the Higgs vevs & $3,62$ & \footnotesize Linear \\
\hline
\signmu\ & \footnotesize Sign of the Higgs/higgsino mass parameter & $+1$ or $-1$ &  \\
\hline
$\mhdsq/\sqrt{|\mhdsq|}\,^{(\ast)}$ & \footnotesize Signed GUT-scale soft mass of $H_d$ & $-20,20$ & \footnotesize Linear \\
\hline
$\mhusq/\sqrt{|\mhusq|}\,^{(\ast)}$ & \footnotesize Signed GUT-scale soft mass of $H_u$ & $-10,10$ & \footnotesize Linear  \\
\hline
\hline
\footnotesize Nuisance parameter & \footnotesize Description & \footnotesize Central value & \footnotesize Distribution \\
\hline
$M_t$ & \footnotesize Top quark pole mass & $173.34 \pm 0.76\gev$\cite{ATLAS:2014wva} & \footnotesize Gaussian \\
\hline
\mbmbmsbar\ & \footnotesize Bottom quark mass & $4.18 \pm 0.03\gev$\cite{Beringer:1900zz} & \footnotesize Gaussian\\
\hline
\alphasmzms\ & \footnotesize Strong coupling &  $0.1185 \pm 0.0006$\cite{Beringer:1900zz} & \footnotesize Gaussian\\
\hline
$1/\alphaemmz$ & \footnotesize Reciprocal of electromagnetic coupling & $127.944 \pm 0.014 $\cite{Beringer:1900zz} & \footnotesize Gaussian \\
\hline
$\Sigma_{\pi N}$ & \footnotesize Nucleon sigma term & $ 34 \pm 2\mev$\cite{Belanger:2013oya} & \footnotesize Gaussian\\ 
\hline
$\sigma_s$ & \footnotesize Strange sigma commutator & $42 \pm 5\mev$\cite{Belanger:2013oya} & \footnotesize Gaussian\\
\hline
\end{tabular}
\footnotesize $(\ast)$ These quantities are independently scanned in the NUHM analysis.
\caption{Prior distributions of the CMSSM and nuisance parameters used in the scans. All dimensionful parameters are given in\tev\ unless indicated otherwise.}
\label{tab:cmssmprior}
\end{center}
\end{table}%

\section{Results in the CMSSM}\label{sec:CMSSM}

\subsection{Posterior distributions and prospects for collider searches}

\begin{figure}[t] 
   \centering
    \subfloat[]{
     \label{fig:a}
    \includegraphics[width=0.47\textwidth]{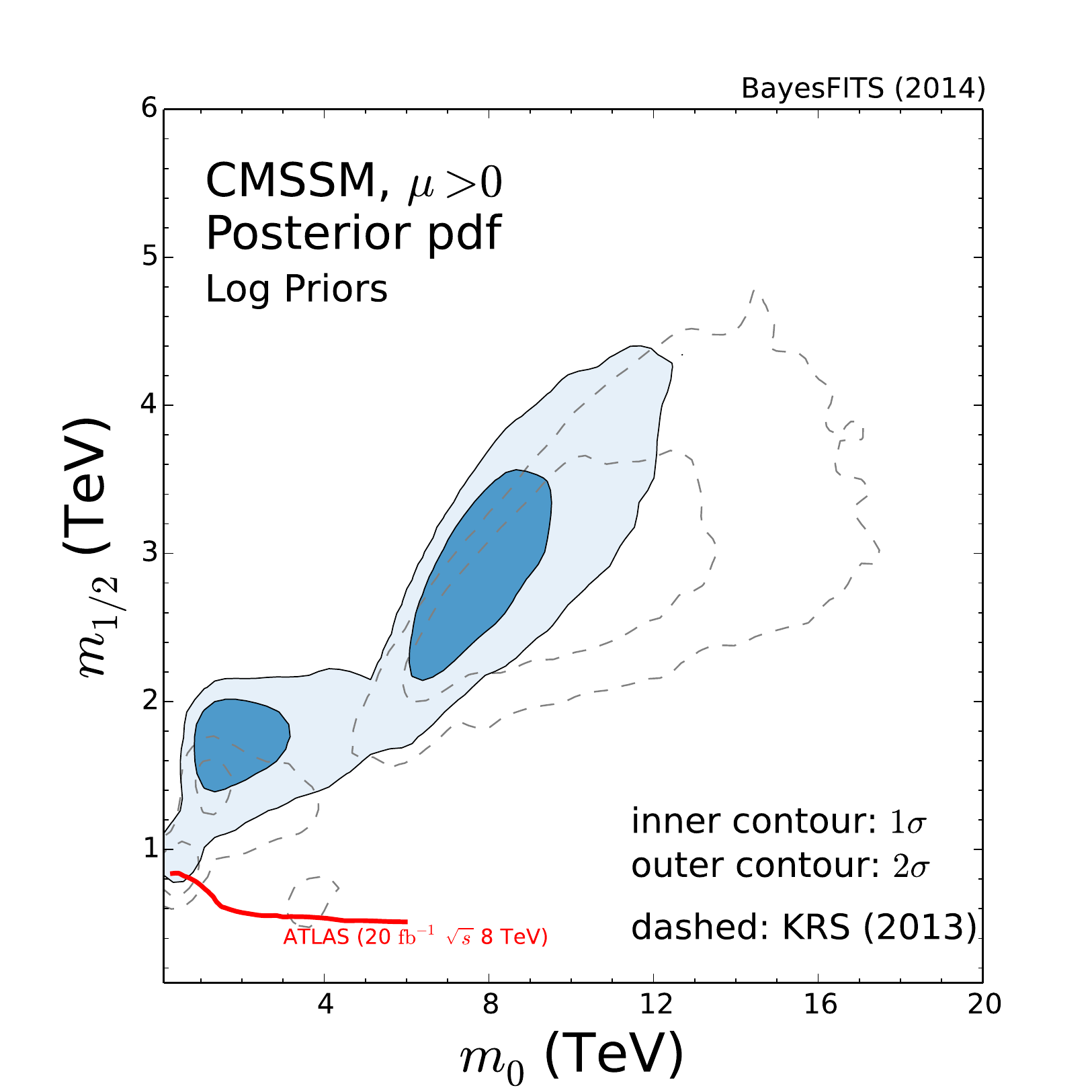} 
}
\hspace{0.01\textwidth}
          \subfloat[]{
   \label{fig:b}
   \includegraphics[width=0.47\textwidth]{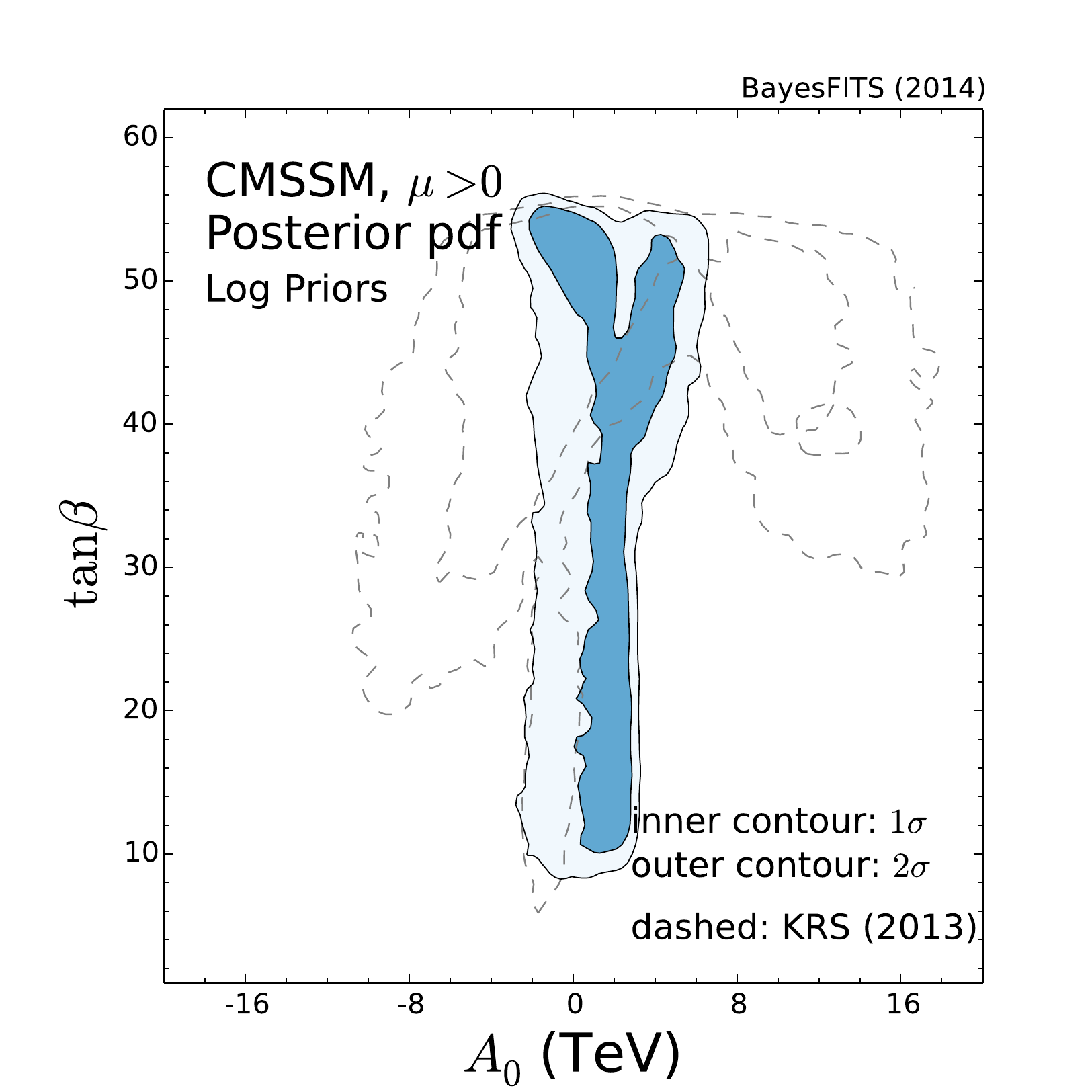} 
   }
\\
\subfloat[]{%
\label{fig:c}%
\includegraphics[width=0.47\textwidth]{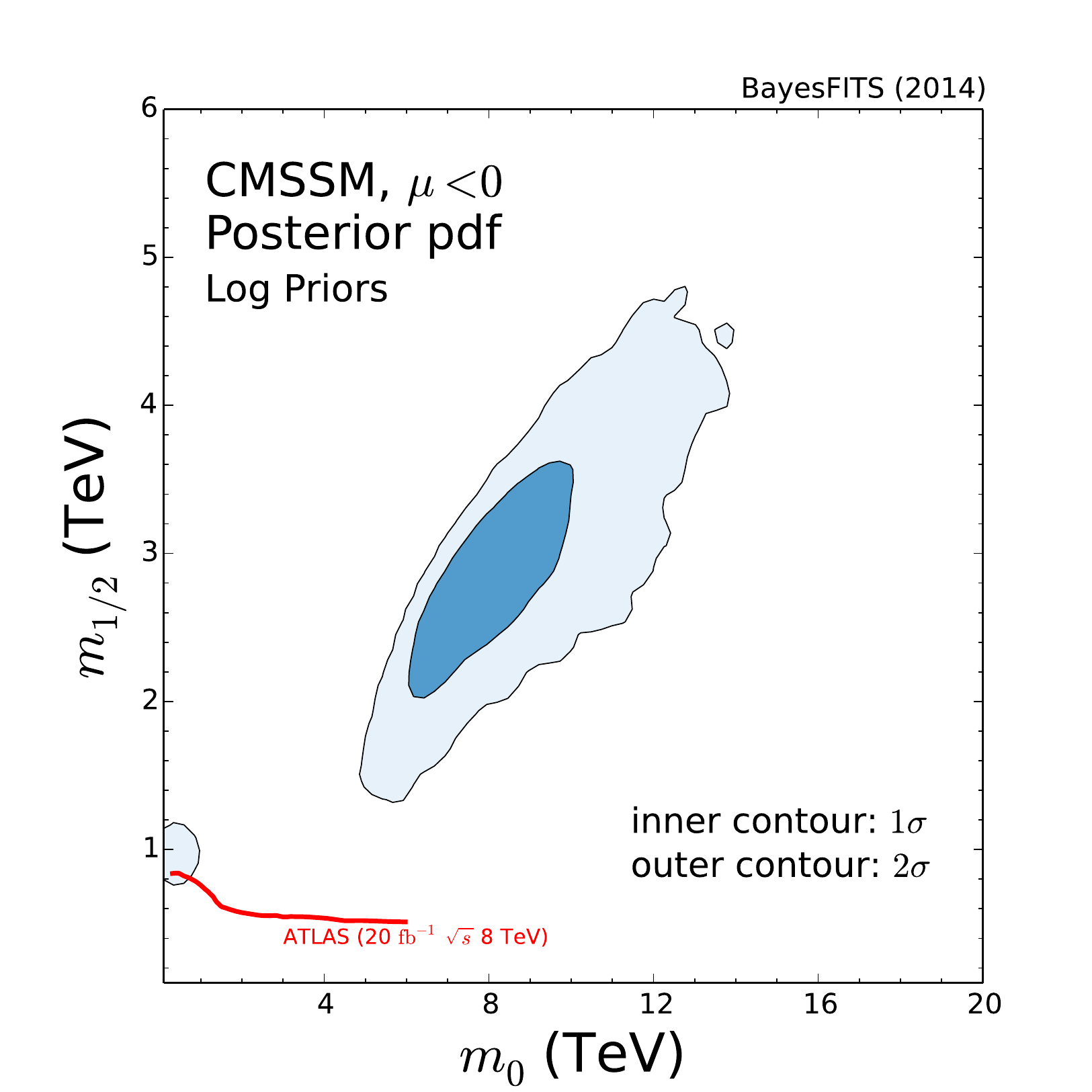}
}%
\hspace{0.01\textwidth}
\subfloat[]{%
\label{fig:d}%
\includegraphics[width=0.47\textwidth]{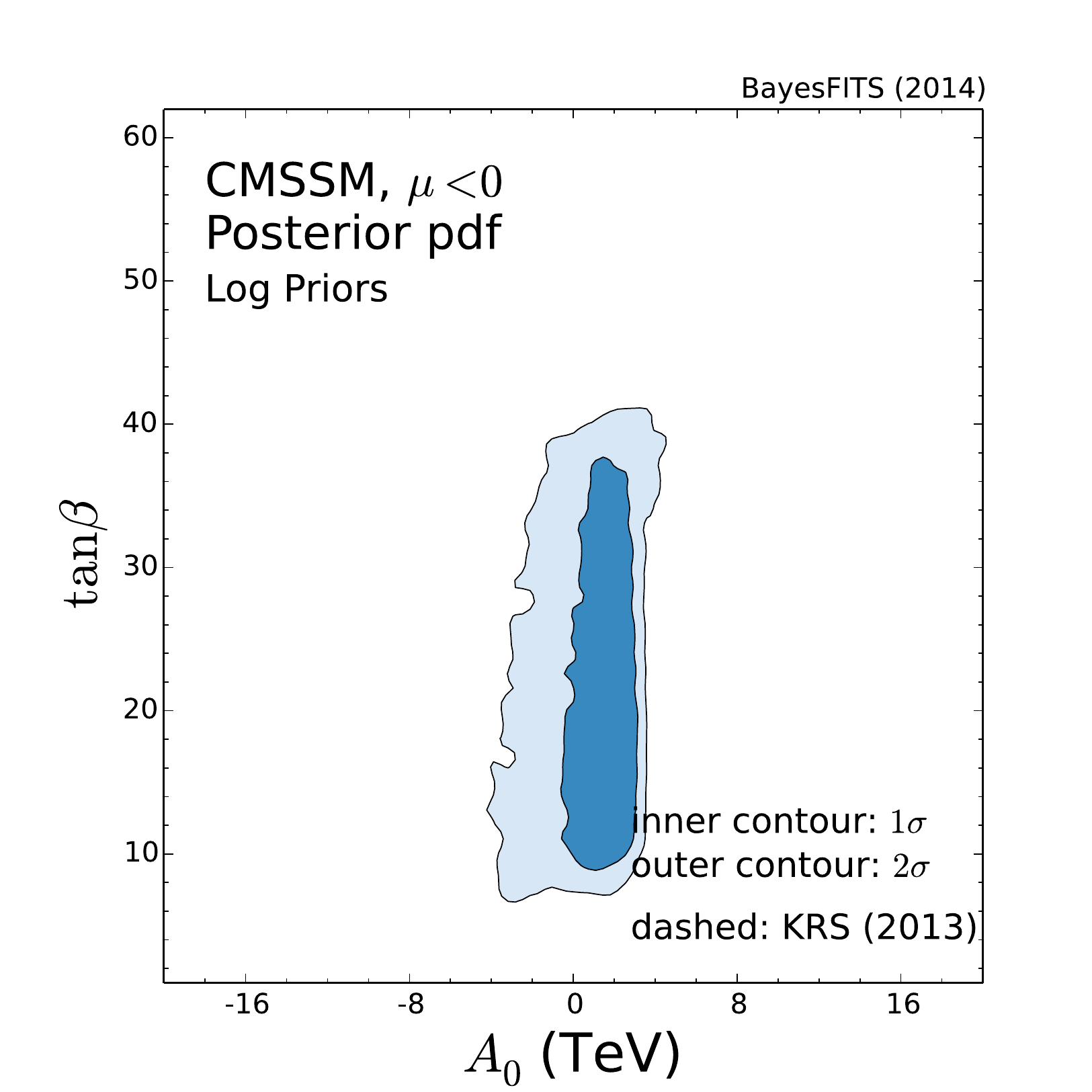}
}%
\caption{\footnotesize Marginalized 2D posterior distribution for the CMSSM in \protect\subref{fig:a} the (\mzero, \mhalf) plane 
   for $\mu>0$, \protect\subref{fig:b} the (\azero, \tanb) plane for $\mu>0$, \protect\subref{fig:c} the (\mzero, \mhalf) plane 
   for $\mu<0$, and \protect\subref{fig:d} the (\azero, \tanb) plane for $\mu<0$. 
   The 68\% credible regions are shown in dark blue and the 95\% credible regions in light blue. 
   For comparison we show the 68\% and 95\% credible regions of\cite{Kowalska:2013hha} (KRS (2013) hereafter) 
   encapsulated by thin gray dashed lines. The ATLAS 95\%~C.L. exclusion line is shown as a red solid line for reference.}
   \label{fig:cmssm_m0_m12_A0_tanb}
\end{figure}

In \reffig{fig:cmssm_m0_m12_A0_tanb}\subref{fig:a} we show the 68\% and 95\% credible regions of the marginalized 2D pdf in the 
(\mzero, \mhalf) plane of the CMSSM, for $\mu>0$. The gray dashed contours mark the previous 68\% and 95\% regions obtained in\cite{Kowalska:2013hha}, 
which we present for comparison to highlight the impact of the new 
more precise calculation of \mhl\  and the new improved constraints.   

As has been long standing practice, in the CMSSM the modes of the posterior pdf are 
identified according to the respective mechanisms to satisfy the relic density constraint.
The little, round, 95\% credibility region just above the ATLAS line at low \mzero\ is
the stau-coannihilation region\cite{Ellis:1998kh}; the large region
immediately above it, for
$\mzero\lesssim 5\tev$ and $\mhalf\gsim 1.2\tev$ is the $A$-resonance region\cite{Drees:1992am}; the remaining
mode for $\mzero> 5\tev$ and $\mhalf\gsim 1.5\tev$ is the above-mentioned
$\sim 1\tev$ higgsino region.

The different shape of the present pdf relative to the one given
in\cite{Kowalska:2013hha} is due to the new higher-order determination
of the Higgs mass. Over the whole parameter space, given equivalent
$\msusy= (\mstopone\mstoptwo)^{1/2}$, the value of the Higgs mass has
increased by about 2\gev.  In practice, this means that in the $\sim
1\tev$ higgsino region, where the Higgs mass constraint is always
more easily satisfied thanks to the large scalar masses, the favored
\mzero\ values are now limited to less than 12\tev, above which the
Higgs mass becomes too heavy. In the $A$-resonance and
stau-coannihilation regions, the Higgs mass value has increased from
an average 122--123\gev\ to values closer to 126\gev, thus improving
the \chisq. As a consequence, the statistical weight of the
$A$-resonance region has now much increased, with new, improved
 prospects for collider phenomenology, as we shall see later.

In spite of higher mass values for the Higgs boson, the
stau-coannihilation region is now becoming disfavored by increasing
tension with direct SUSY limits at the LHC, and it will be most likely
probed in its entirety in the 14\tev\ run. Note also that, with
respect to\cite{Kowalska:2013hha}, the focus point
region\cite{Chan:1997bi,Feng:1999mn,Feng:1999zg}, which is shown in
\reffig{fig:cmssm_m0_m12_A0_tanb}\subref{fig:a} as a 95\% dashed
contour just above the ATLAS line at $\mzero\simeq 4\tev$, is now
disfavored.  Interestingly enough, the new estimate for the Higgs mass
in the focus point region is now much closer to the experimental value
than before, but the region is disfavored by the LUX results: as is
well known, the neutralino there is a mixed composition of bino and
higgsino with a mass of $\mchi\lesssim 600\gev$.  Its large \sigsip\
is now excluded by LUX at 90\%~C.L. We will come back to this point in
\refsec{sec:dmcmssm}.

In \reffig{fig:cmssm_m0_m12_A0_tanb}\subref{fig:b} we show the credible regions of the 2D pdf in the 
(\azero, \tanb) plane of the CMSSM, for $\mu>0$.
The three favored modes of the posterior are not as clearly separated as in \reffig{fig:cmssm_m0_m12_A0_tanb}\subref{fig:a},
or as they were in the (\azero, \tanb) plane shown in\cite{Kowalska:2013hha}. 
One can now recognize an elongated 68\% credible region for $0\lesssim\azero\lesssim2\tev$ and
$8\lesssim\tanb\lesssim 45$, which belongs to the $\sim 1\tev$ higgsino region and extends to slightly
larger, positive \azero\ values for $\tanb\gsim 45$. An elongated 95\% credible area is adjacent to it to the left:
it encompasses part of the $\sim 1\tev$ higgsino region, and the stau-coannihilation region. 
Finally, a roundish 68\% and 95\% credibility region at $-3\tev\lesssim\azero\lesssim 2\tev$
and large \tanb\ is the $A$-resonance region. 

The most striking difference with the results of\cite{Kowalska:2013hha} is that now the favored \azero\ range is much reduced,
from a maximum span of approximately 28--30\tev, covering almost the full parameter space, to
the present span of about 10\tev, centered around zero.
By comparing the solid contours in color with the previous ones in dashed lines, 
one can see that the two modes that were previously predominant at large $|\azero|$
in the $\sim 1\tev$ higgsino region
are not as favored in the present scan and do not appear at 95\% credibility.
This is because in\cite{Kowalska:2013hha} one needed solutions with larger \msusy\ to fit the Higgs mass 
measurement. When \mzero\ becomes very large, electroweak-symmetry breaking (EWSB) can only be obtained with large trilinear couplings,
which enhance the RGE running of \mhusq\,.
However, as we said above, the correct value of the Higgs mass can now be obtained with 
an \msusy\ on average smaller, even in the $\sim 1\tev$ higgsino region.
Thus, the scanning program can more naturally find solutions with a
reduced $|\azero|$.   

In \reffig{fig:cmssm_m0_m12_A0_tanb}\subref{fig:c} we show the 68\%
and 95\% credible regions of the 2D pdf in the (\mzero, \mhalf) plane
of the CMSSM for $\mu<0$. Two of the modes of the posterior are the
same as in the $\mu>0$ case, with the exception of the $A$-resonance
region, which does not appear for negative $\mu$ because the
conditions for EWSB are not met there.\footnote{To be precise, the
  $A$-resonance region is absent not because it is disfavored by any
  particular constraint, but because at large \tanb\ $\tt SoftSusy$
  easily incurs negative values of the $\overline{DR}$
  pseudoscalar mass squared, $\ma^2(\overline{DR})$, during the
  iterative procedure of RGE running. The matter was solved in earlier
  versions of the program by switching to the pole mass at each
  iteration while looking for convergence. It is debatable whether
  such a solution is enough to produce the correct EWSB. Thus,
  following the default setting in recent versions of $\tt SoftSusy$,
  we have decided not to include the points of the $A$-resonance
  region for $\mu<0$. We also abstain from showing a direct comparison
  with the results of\cite{Kowalska:2013hha} in this case, as they
  were obtained with an earlier modified version of $\tt SoftSusy$.}
In \reffig{fig:cmssm_m0_m12_A0_tanb}\subref{fig:d} we show the
corresponding 2D pdf in the (\azero, \tanb) plane.

%
%
%
%
%
%
%

\begin{figure}[t]
\centering
\subfloat[]{
\label{fig:a}
\includegraphics[width=0.47\textwidth]{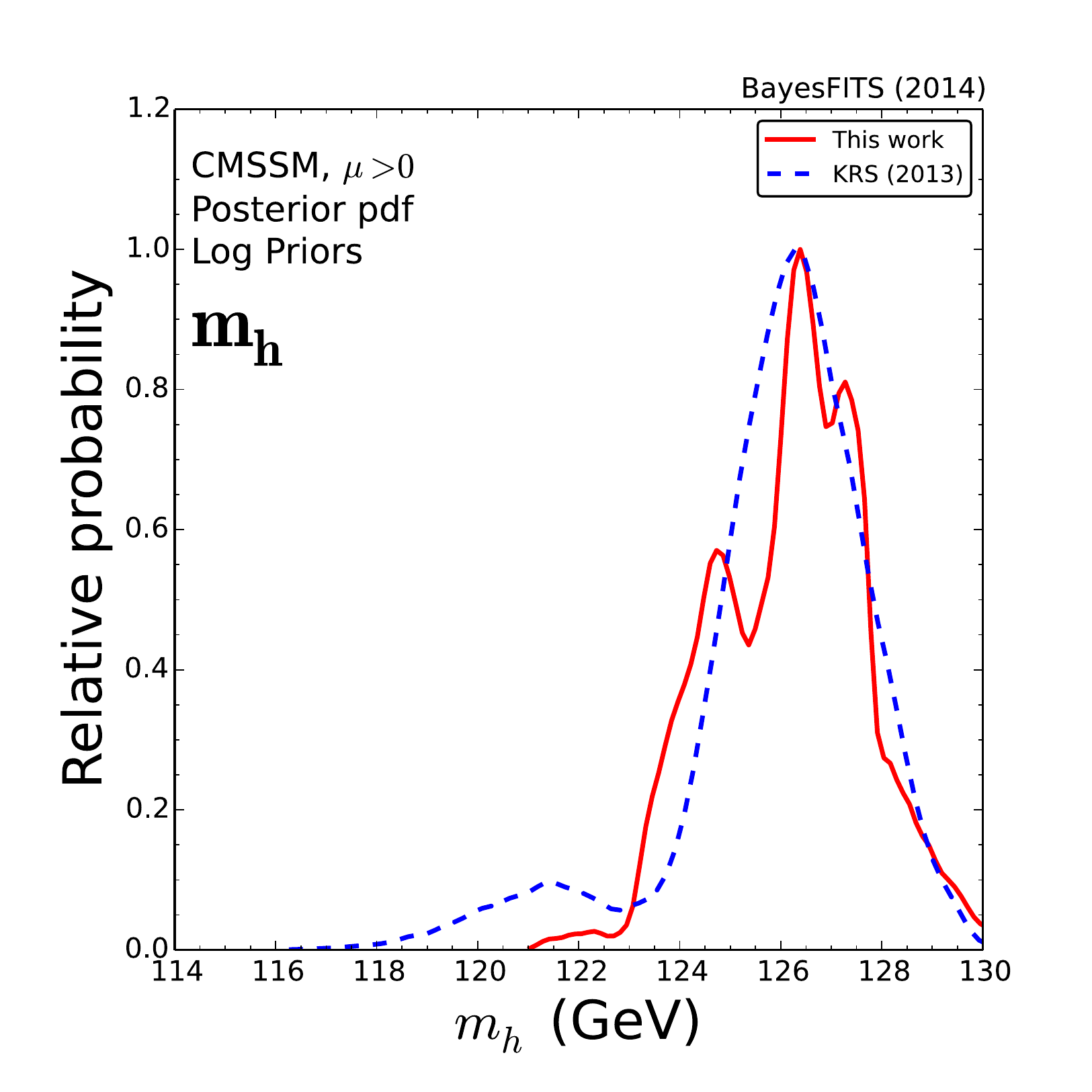}
}
\hspace{0.01\textwidth}
          \subfloat[]{
   \label{fig:b}
   \includegraphics[width=0.47\textwidth]{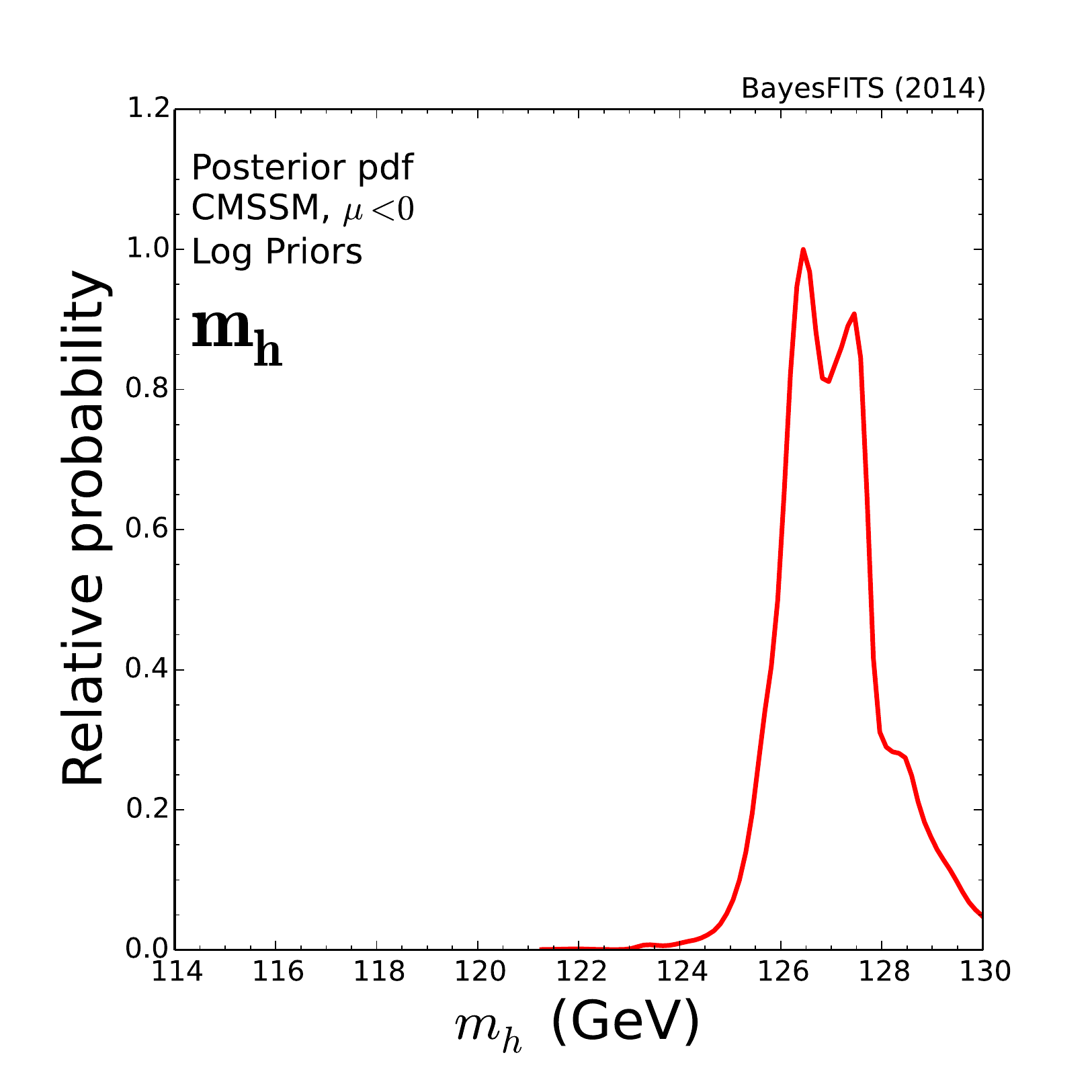} 
   }
\caption{\footnotesize Marginalized 1D pdf of $m_h$ for the CMSSM with \protect\subref{fig:a} $\mu > 0$ and 
\protect\subref{fig:b} $\mu<0$. Red solid lines is this work, blue dashed line represents the distribution obtained in KRS 2013.}
\label{fig:cmssm_mh}
\end{figure}

In \reffig{fig:cmssm_mh}\subref{fig:a} we show in solid red the marginalized 1D posterior distribution for the lightest Higgs
mass. We also show with a dashed blue line the distribution obtained in\cite{Kowalska:2013hha}, to facilitate comparison.
The posterior is in good agreement with the previous study, in which the Higgs-mass likelihood was 
a simple gaussian function centered about the measured CMS value, with experimental and theoretical uncertainties added in quadrature.    
The present likelihood function is instead determined 
by the sum of the \chisq\ contributions of the individual Higgs searches, encoded in the $\tt HiggsSignals$ program.
Note that the two main peaks in the distribution, characteristic of the $A$-resonance region about 124--125\gev, and 
of the $\sim 1\tev$ higgsino region for larger \mhl, are now much more in agreement with each other than in the previous
study. This is the effect of including the higher order Higgs corrections: there are virtually no regions left in the parameter space for which
constraints other than the Higgs searches push the posterior distribution toward $\mhl<123\gev$.  

\begin{figure}[t]
\centering
\subfloat[]{%
\label{fig:a}%
\includegraphics[width=0.36\textwidth]{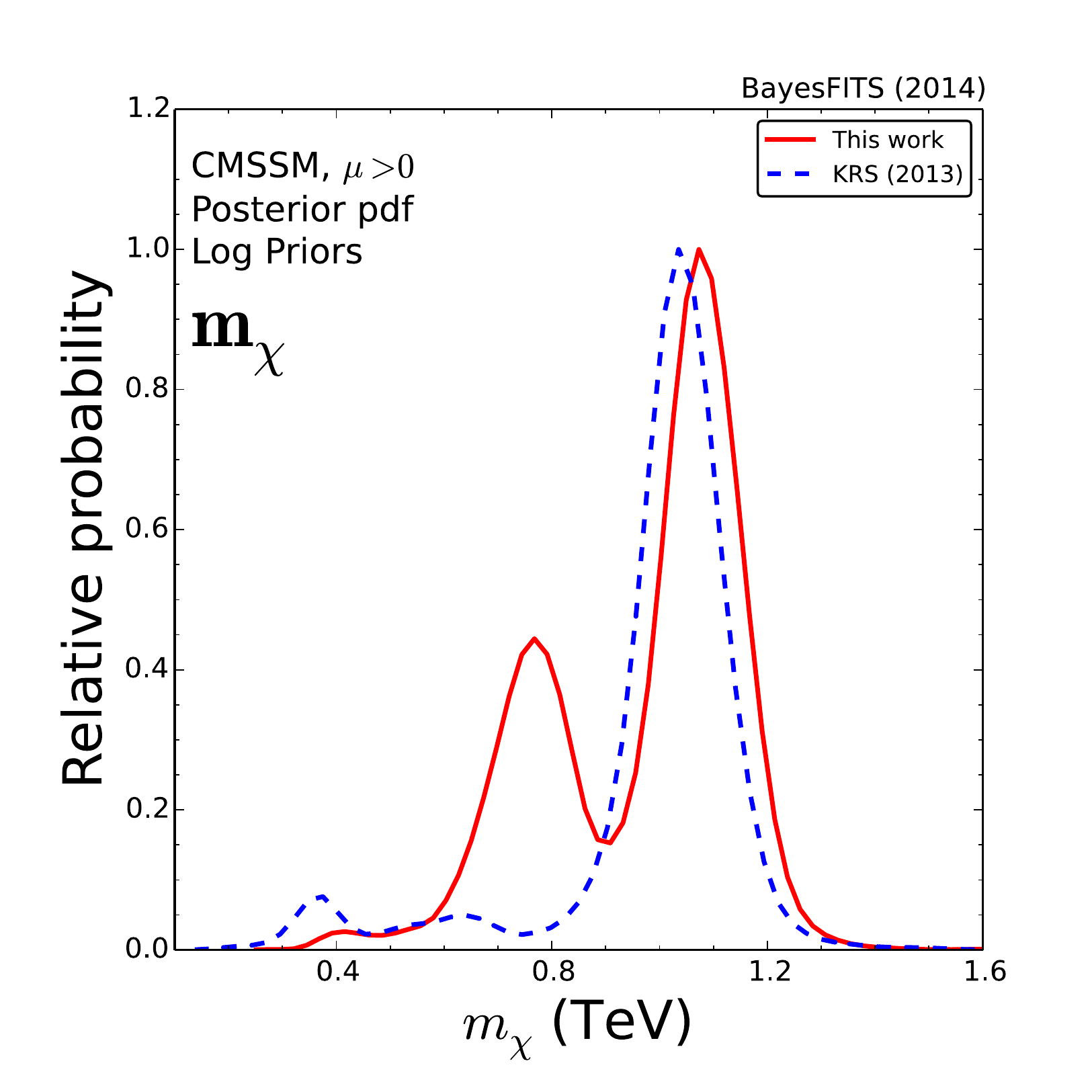}
}%
\hspace{0.07\textwidth}
\subfloat[]{%
\label{fig:b}%
\includegraphics[width=0.36\textwidth]{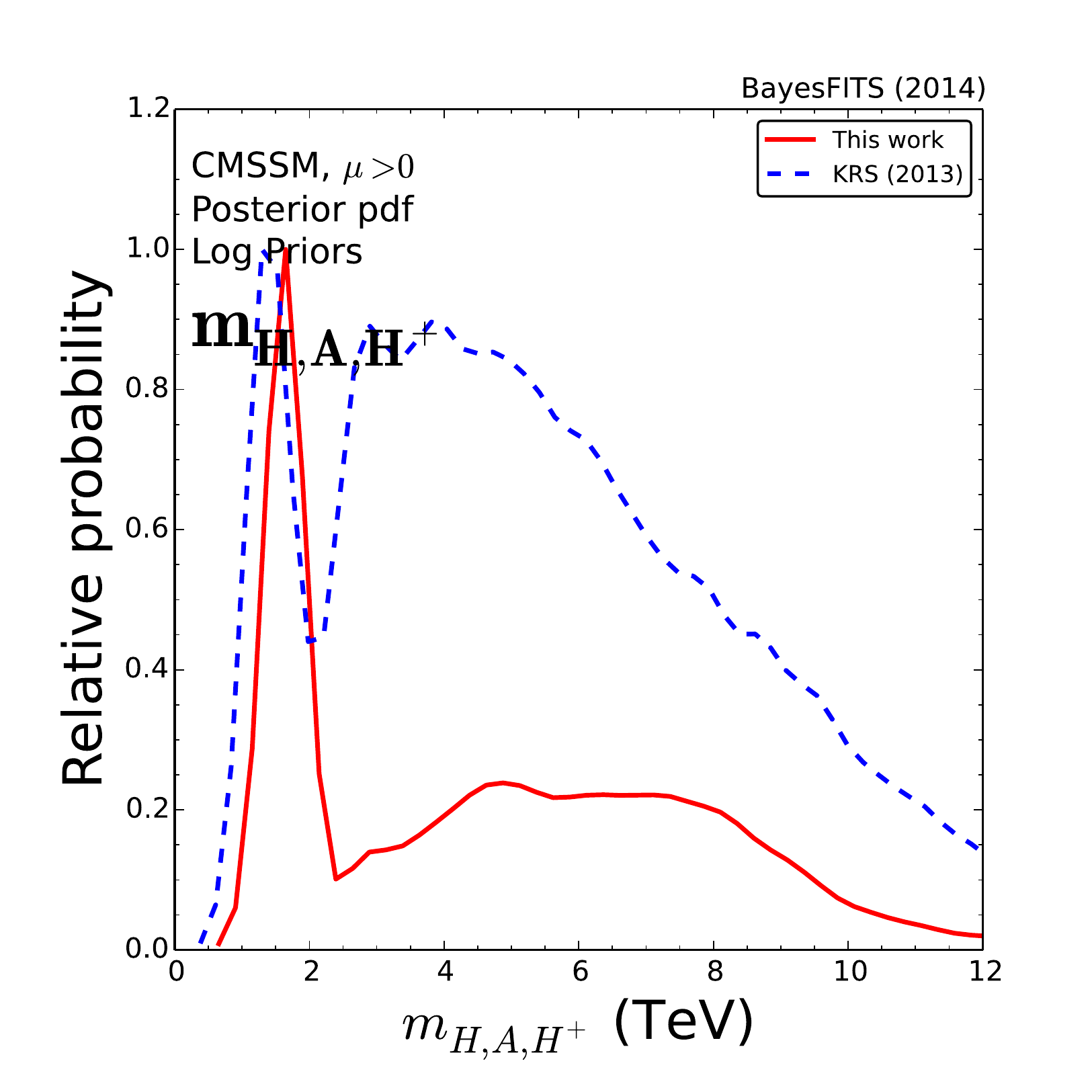}
}%

\subfloat[]{%
\label{fig:c}%
\includegraphics[width=0.36\textwidth]{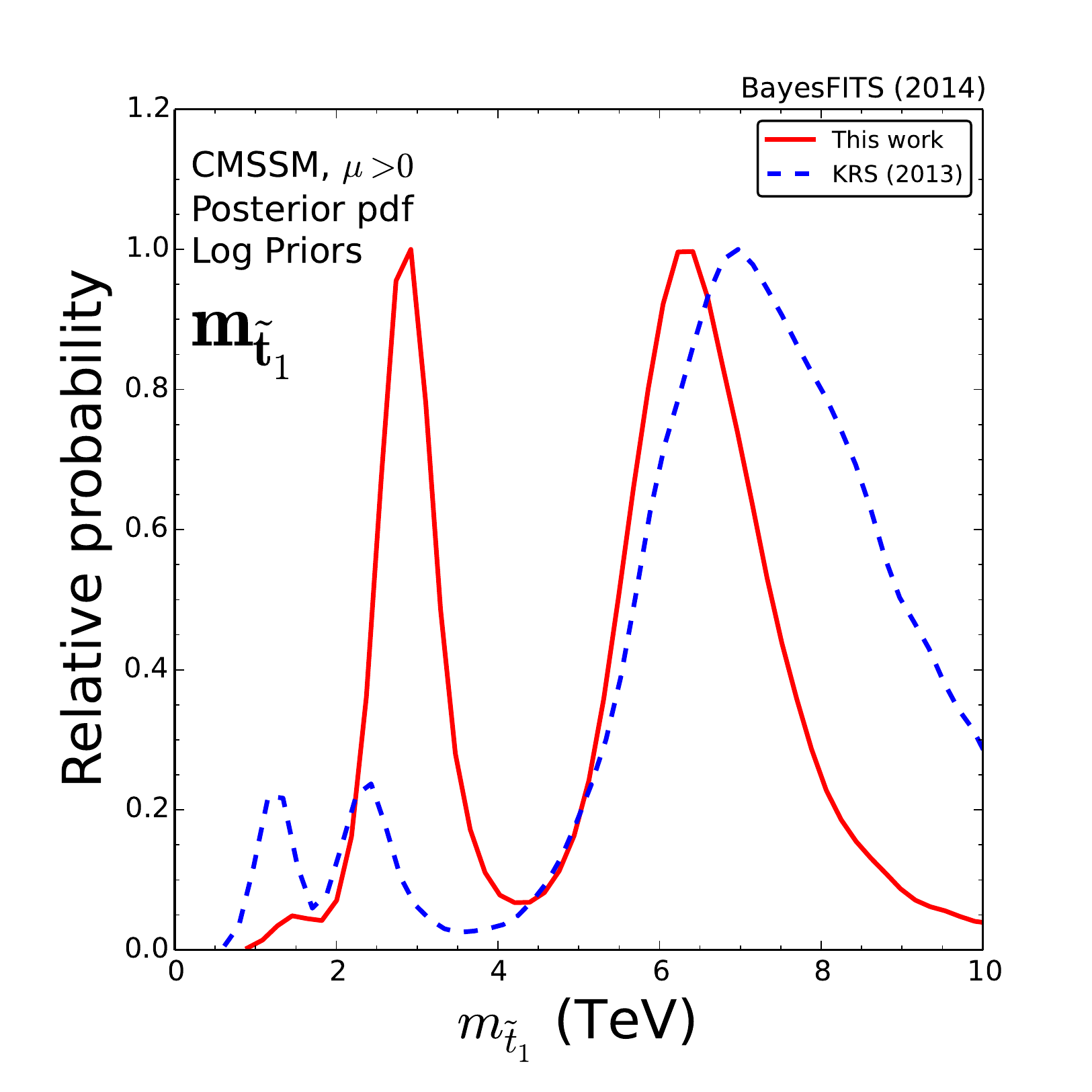}
}%
\hspace{0.07\textwidth}
\subfloat[]{%
\label{fig:d}%
\includegraphics[width=0.36\textwidth]{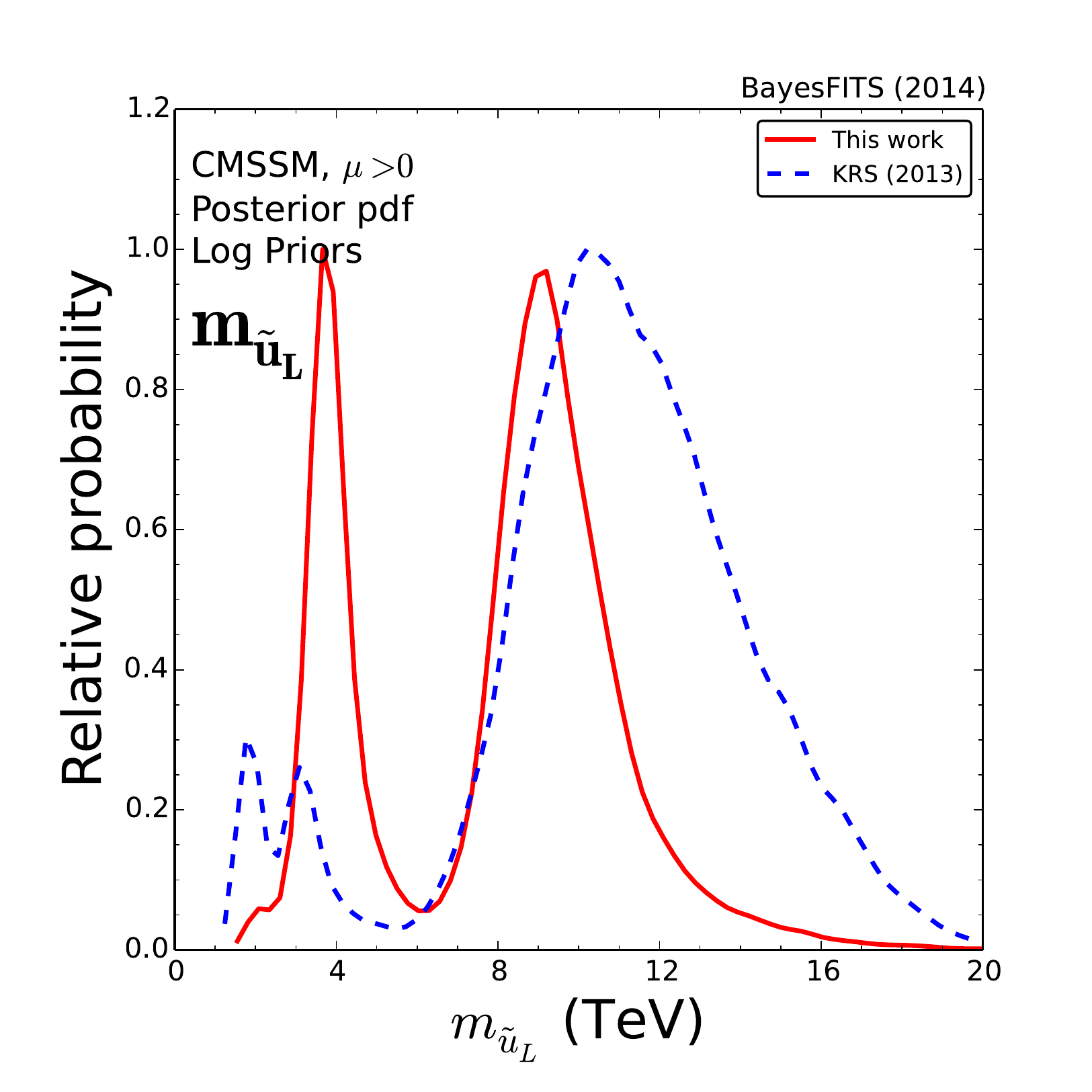}
}%

\subfloat[]{%
\label{fig:e}%
\includegraphics[width=0.36\textwidth]{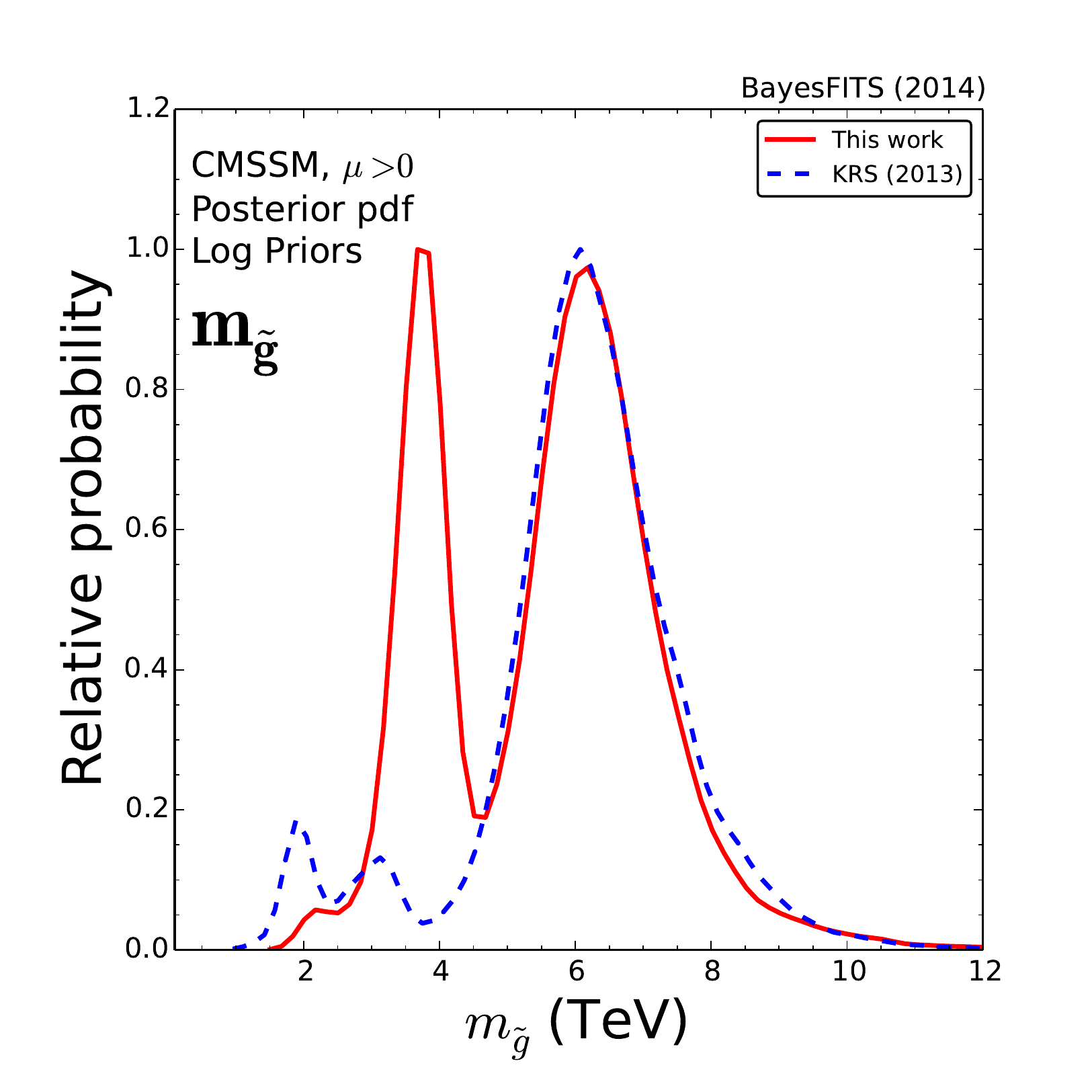}
}%
\hspace{0.07\textwidth}
\subfloat[]{%
\label{fig:f}%
\includegraphics[width=0.36\textwidth]{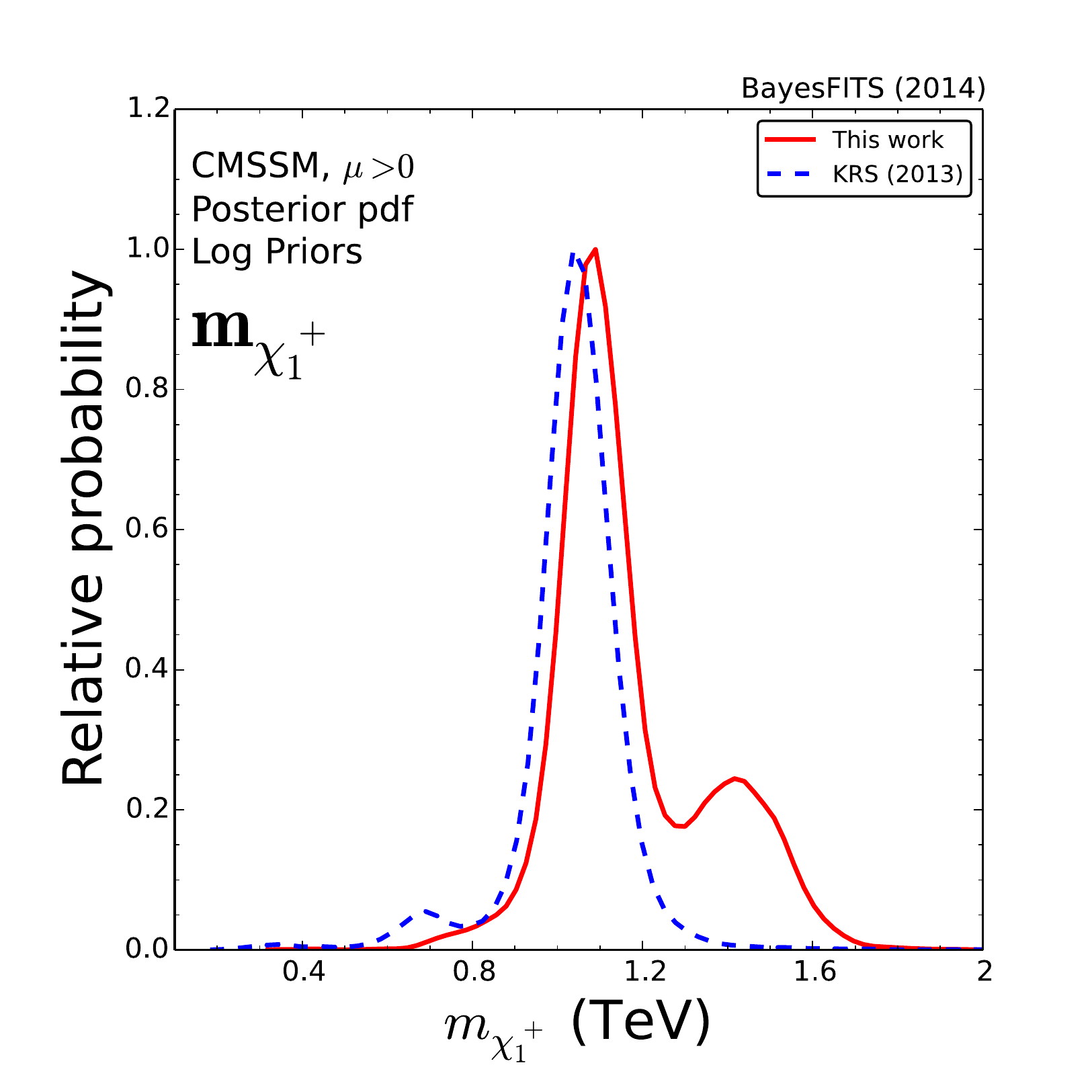}
}%

\caption{\footnotesize Marginalized 1D pdf for the heavy Higgs bosons and a selection of superpartner masses in the the CMSSM with $\mu > 0$. 
Red solid line is this work, blue dashed lines represent distributions obtained in KRS 2013.}  
\label{fig:cmssm_1D}
\end{figure}

The equivalent distribution for the case with $\mu<0$ is shown in \reffig{fig:cmssm_mh}\subref{fig:b}.  
In \reffig{fig:cmssm_1D} we show the marginalized 1D posteriors for the heavy Higgs bosons and a selection of superpartner masses for $\mu>0$. 
Again, the dashed blue line in the figures shows the distribution for the scan of Ref.\cite{Kowalska:2013hha}.

In \reffig{fig:cmssm_1D}\subref{fig:a} we show the pdf for the mass of
the lightest neutralino.  The posterior presents three main
modes, corresponding to the regions defined in
\reffig{fig:cmssm_m0_m12_A0_tanb}\subref{fig:a}.  From the left to the
right one can can see: the stau-coannihilation region, with
$0.3\tev\lesssim\mchi\lesssim 0.6\tev$, subtending approximately 1\%
of the total probability; the $A$-resonance region,
$0.6\tev\lesssim\mchi\lesssim 0.9\tev$, with a probability of
$\sim30\%$; and the $\sim 1\tev$ higgsino region for $\mchi\gsim
0.9\tev$, with $\sim 69\%$ probability.
The inclusion of higher order corrections to the Higgs mass has caused the shift of a substantial fraction of posterior probability to the 
$A$-resonance region. As the dashed blue line shows,
the $\sim 1\tev$ higgsino region was strongly favored in\cite{Kowalska:2013hha}, featuring $\sim94\%$ of the total probability, while 
the remainder was split in approximately equal parts between the other two regions.  
Note also that the average \mchi\ in the $\sim 1\tev$ higgsino region is slightly larger than in\cite{Kowalska:2013hha},   
as the PLANCK-measured value of the relic density is a little larger than the one measured by WMAP\cite{Komatsu:2010fb}.

The 1D pdf for the heavy Higgs masses is shown in \reffig{fig:cmssm_1D}\subref{fig:b}. 
As could be expected, the distribution features a sharp peak in the $A$-resonance region, 
for $\ma\approx 2\mchi\lesssim 2.5\tev$. The $\sim 1\tev$ higgsino region is instead characterized by a much broader range of 
values, extending up to $\sim 15\tev$. 
We will show below that a large fraction of the $A$-resonance regions is within the reach of Higgs searches
at the LHC 14\tev\ run. 

The 1D posterior distributions for the mass of the lightest stop, $\mstopone$, and that of the left-handed first and second 
generation squarks, generically indicated with $m_{\tilde{u}_L}$, are shown in Figs.~\ref{fig:cmssm_1D}\subref{fig:c} and \ref{fig:cmssm_1D}\subref{fig:d}, respectively.
While the lightest squark masses in the figures, typical of the stau-coannihilation region, 
are now disfavored by the present limits from direct SUSY searches at the LHC, 
the increased relevance of the $A$-resonance region, which features
in general lighter squarks than the $\sim 1\tev$ higgsino region, 
leads to a moderate optimism for squark detection in future-generation colliders.
In particular, by comparing the solid red and dashed blue lines  
in, e.g., Figs.~\ref{fig:cmssm_1D}\subref{fig:c} and \ref{fig:cmssm_1D}\subref{fig:a}, one can see that in\cite{Kowalska:2013hha}
there was a $\sim30\%$ of probability favoring $\mstopone\gsim 8\tev$ (with $\mchi\gsim 1\tev$) that has now been virtually erased.
On the other hand, the parameter space featuring $\mstopone\lesssim 4\tev$ (with $\mchi\lesssim 0.8\tev$)
is now favored by roughly the same odds.    

Equivalently, one can see in \reffig{fig:cmssm_1D}\subref{fig:e}, where we show the 1D pdf for the gluino mass \mglu,
that the increased relevance of the $A$-resonance region leads to the emergence of a new peak in probability at $\mglu\simeq 3-4\tev$.     

Finally, we show in \reffig{fig:cmssm_1D}\subref{fig:f} the 1D posterior for the lightest chargino mass,
$m_{\charone}$. The bulk of the probability is subtended by the $\sim 1\tev$ higgsino region, 
in which the lightest chargino is also higgsino-like and almost degenerate with the neutralino,
in the range $0.9\tev\lesssim m_{\charone}\lesssim 1.3\tev$. 
To the left and to the right of this larger mode, one can see the 
modes for the stau-coannihilation and $A$-resonance region, respectively. In both of them, the neutralino is bino-like, the chargino wino-like,
and one finds $m_{\charone}\approx 2\mchi$.  

An ATLAS study\cite{ATL-PHYS-PUB-2013-011} of the sensitivity reach for direct SUSY searches at the LHC 14\tev\ run 
showed that the chances of probing in this way squark masses typical of the $A$-resonance region are scant, to say the least.
For example, with 3000\invfb\ of integrated luminosity, 0- and 1-lepton searches for third generation squarks have the potential, 
when combined, to exclude at the 95\%~C.L. simplified models with stop next-to-LSP (NLSP) up to $\mstopone\simeq 1.4-1.5\tev$ for 
$\mchi\lesssim 0.6\tev$\,.   
However, it was shown in\cite{Cohen:2013xda} that if the same luminosity 
were obtained at a future
33~TeV proton collider, one could 
exclude at the 95\%~C.L. simplified models with gluino NLSP up to 
$\mglu\simeq 5\tev$ for $\mchi\lesssim 2\tev$, or first-two generation 
squark NLSP up to $m_{\tilde{u}_L}\simeq 3.5\tev$ for $\mchi\lesssim 1\tev$.

The reach for the CMSSM is going to be reduced with respect to the 
simplified models, due to complex decay chains that include intermediate charginos and neutralinos.
Nevertheless, the same paper\cite{Cohen:2013xda} also showed that 3000\invfb\ in a 100~TeV machine 
can extend the sensitivity  for gluino NLSP to $\mglu\simeq 12\tev$ for $\mchi\lesssim 4\tev$, or first-two generation 
squark NLSP up to $m_{\tilde{u}_L}\simeq 8\tev$ for $\mchi\lesssim 2\tev$.

Obviously, it seems that the $\sim 1\tev$ higgsino region of the CMSSM remains for the most part beyond 
the direct reach of conceivable future colliders.
Note, however, that it was shown in\cite{Kowalska:2014hza} that this is not necessarily the case when 
the assumptions of GUT-scale universality are relaxed.
 
For $\mu < 0$ the 1D distributions of particle masses share identical features with the positive $\mu$ case with the exception that 
the peaks associated with the $A$-resonance region described above are absent. 
Since this generally leaves only posterior probability in the region of large superpartner masses far beyond the reach of the LHC we do not show these distributions. 
 
\begin{figure}[t] 
   \centering
    \subfloat[]{
     \label{fig:a}
    \includegraphics[width=0.47\textwidth]{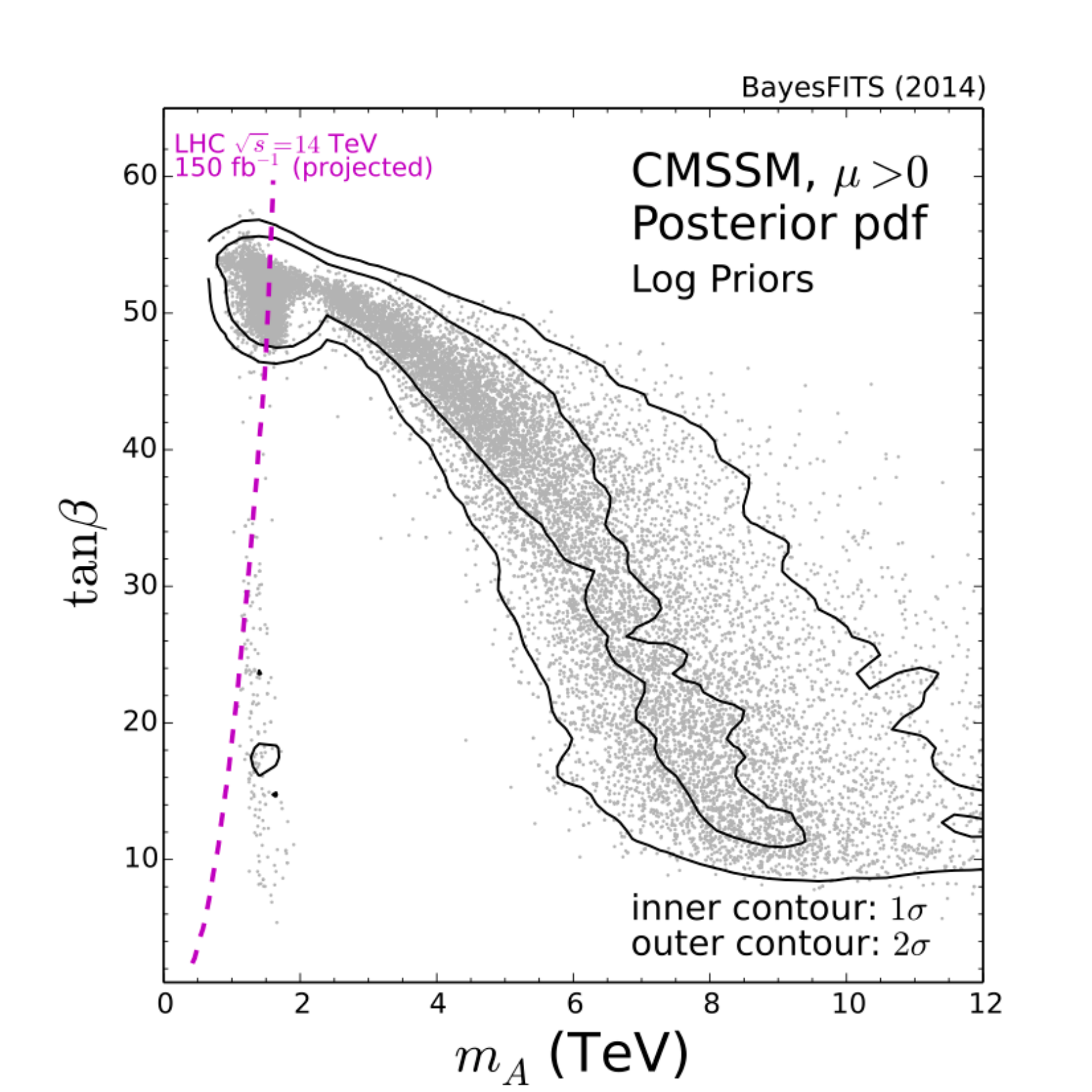} 
}
\hspace{0.01\textwidth}
          \subfloat[]{
   \label{fig:b}
   \includegraphics[width=0.47\textwidth]{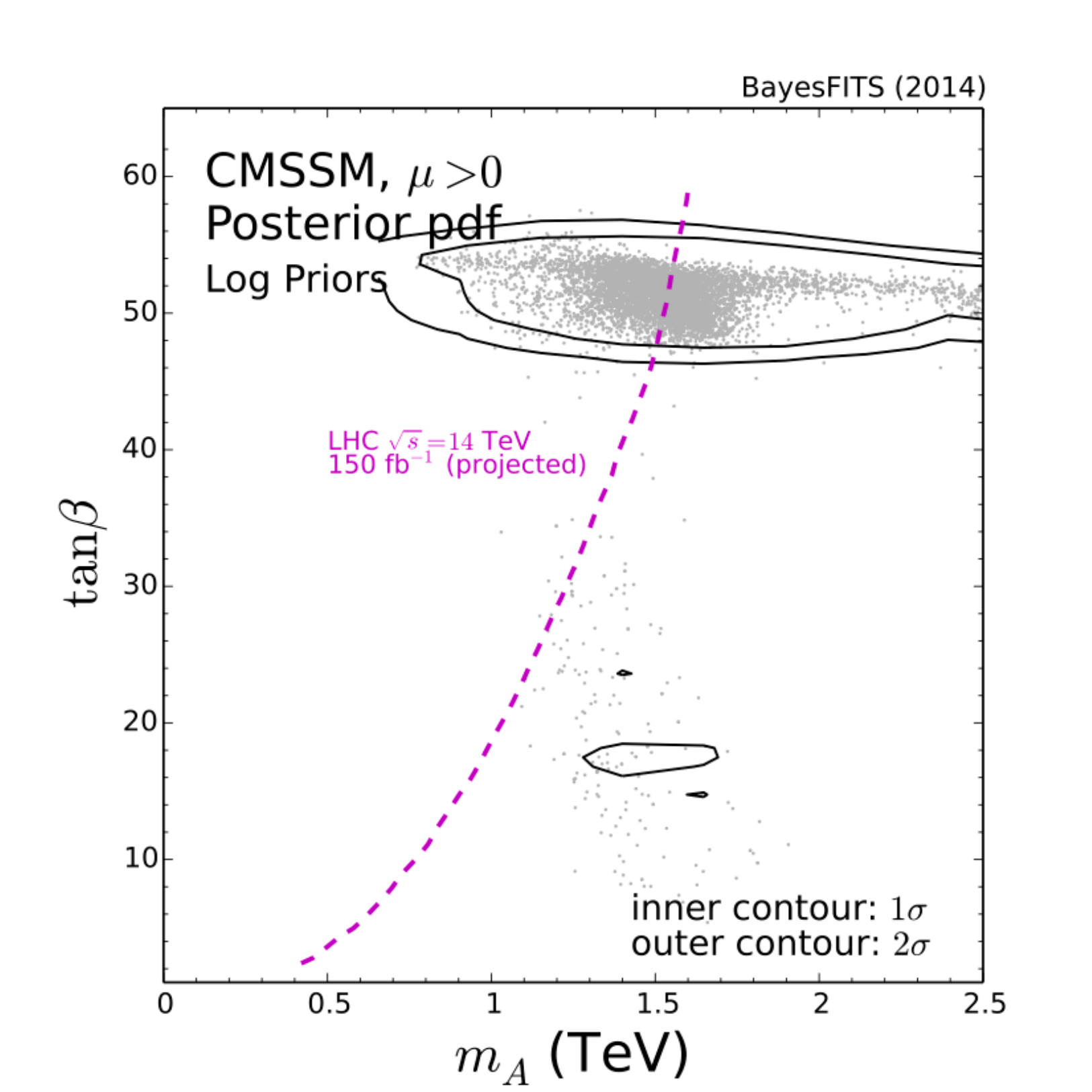} 
   }
\caption{\footnotesize \protect\subref{fig:a} Marginalized 2D posterior distribution for the CMSSM with $\mu > 0$ in the (\ma, \tanb) plane. 
   \protect\subref{fig:b} A zoomed-in fraction of the same. The inner contours give the 68\% credible regions and the outer ones the 95\% credible regions.
   A sample of points (in gray) drawn from the posterior distribution is shown for clarity. 
   The magenta dashed lines show the expected reach at LHC 14\tev\ with $\sim 150\invfb$ 
   estimated in\cite{Arbey:2013jla}.}
   \label{fig:cmssm_mA_tanb}
\end{figure}

In \reffig{fig:cmssm_mA_tanb}\subref{fig:a} we show the marginalized
2D pdf projected to the (\ma, \tanb) plane.  In the $A$-resonance
region, characterized by $\ma< 2.5\tev$ and $\tanb>45$, the posterior
appears to be well localized.  In the CMSSM, this limited range of
\ma\ and \tanb\ values is a proper feature of the $A$-resonance
region, as was extensively explained
in\cite{Kowalska:2013hha}, namely large \tanb\ value and neutralino
not excessively heavy must conspire to produce a region of resonance,
$\ma\approx2\,\mchi$, and a cross section large enough to yield the
relic density measured at PLANCK.  One can see that \ma\ and \tanb\
are not as tightly constrained for the two remaining modes, so that
the posterior spreads over a broad range of values. In particular, the
stau-coannihilation region is localized in \ma, $\ma\simeq1.5-2\tev$, but not in \tanb; the $\sim
1\tev$ higgsino region covers instead the majority of the parameter
space with $\ma\gsim2.5\tev$, $\tanb\gsim 5$.  To make this feature
more visible we superimposed in the figure a sample of points drawn
from the posterior distribution.

It was shown in\cite{Kowalska:2013hha} that, since the points of the $A$-resonance region are well localized 
in the (\ma, \tanb) plane, the whole region can be easily tested through precise measurement of \brbsmumu, which is proportional to
$\tan^6\beta/\ma^4$. Here, we show that the same features also make a large fraction of 
the points testable at the LHC through Higgs searches in the $\tau^+\tau^-$ channel.

In \reffig{fig:cmssm_mA_tanb}\subref{fig:b} we show a zoomed-in fraction of the (\ma, \tanb) plane.
The limits from direct $A\rightarrow\tau^+\tau^-$ searches at CMS with $\sim 17\invfb$\cite{CMS-PAS-HIG-12-050} are included 
in $\tt HiggsBounds$. In the $M_h^{\textrm{max}}$ scenario\cite{Carena:2002qg}, they exclude $m_A$ up to approximately 800\gev\ 
for $\tanb\simeq 50$. 
However, it was explained in\cite{Djouadi:2013vqa} that they are robust against radiative corrections to
the MSSM Higgs, and can be considered almost scenario-independent.
Ref.\cite{Arbey:2013jla} estimated the sensitivity reach of the LHC 14\tev\ run with $\sim 150\invfb$ in the $\tau^+\tau^-$ channel
for the heavy MSSM Higgs bosons. We show the expected 95\%~C.L. reach as a 
magenta dashed line in Figs.~\ref{fig:cmssm_mA_tanb}\subref{fig:a} and \ref{fig:cmssm_mA_tanb}\subref{fig:b}.
Approximately 50\% of the points in the $A$-resonance region fall within the expected sensitivity.

\subsection{Prospects for dark matter detection}\label{sec:dmcmssm}

\begin{figure}[t]
\centering
\subfloat[]{%
\label{fig:a}%
\includegraphics[width=0.47\textwidth]{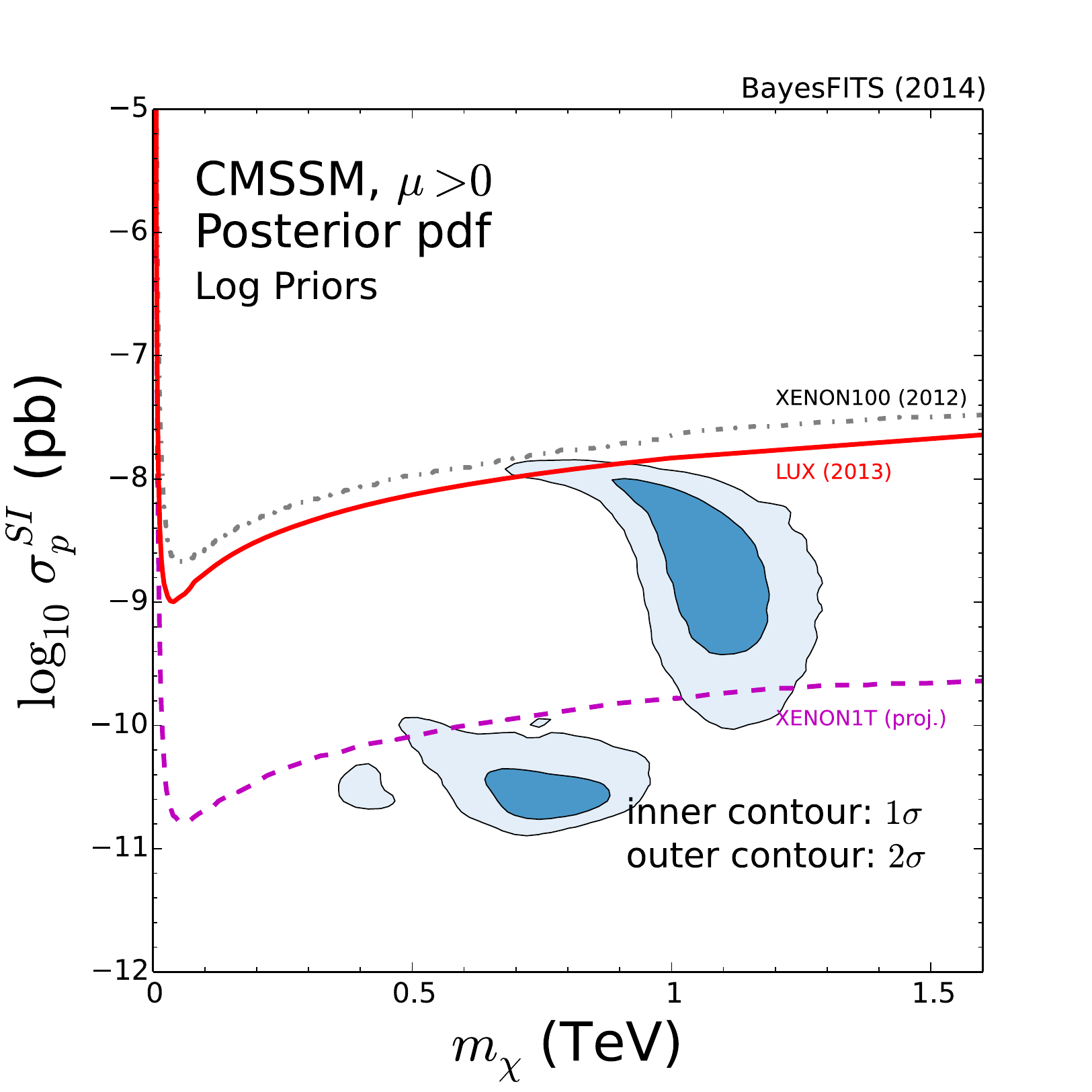}
}%
\hspace{0.01\textwidth}
\subfloat[]{%
\label{fig:b}%
\includegraphics[width=0.47\textwidth]{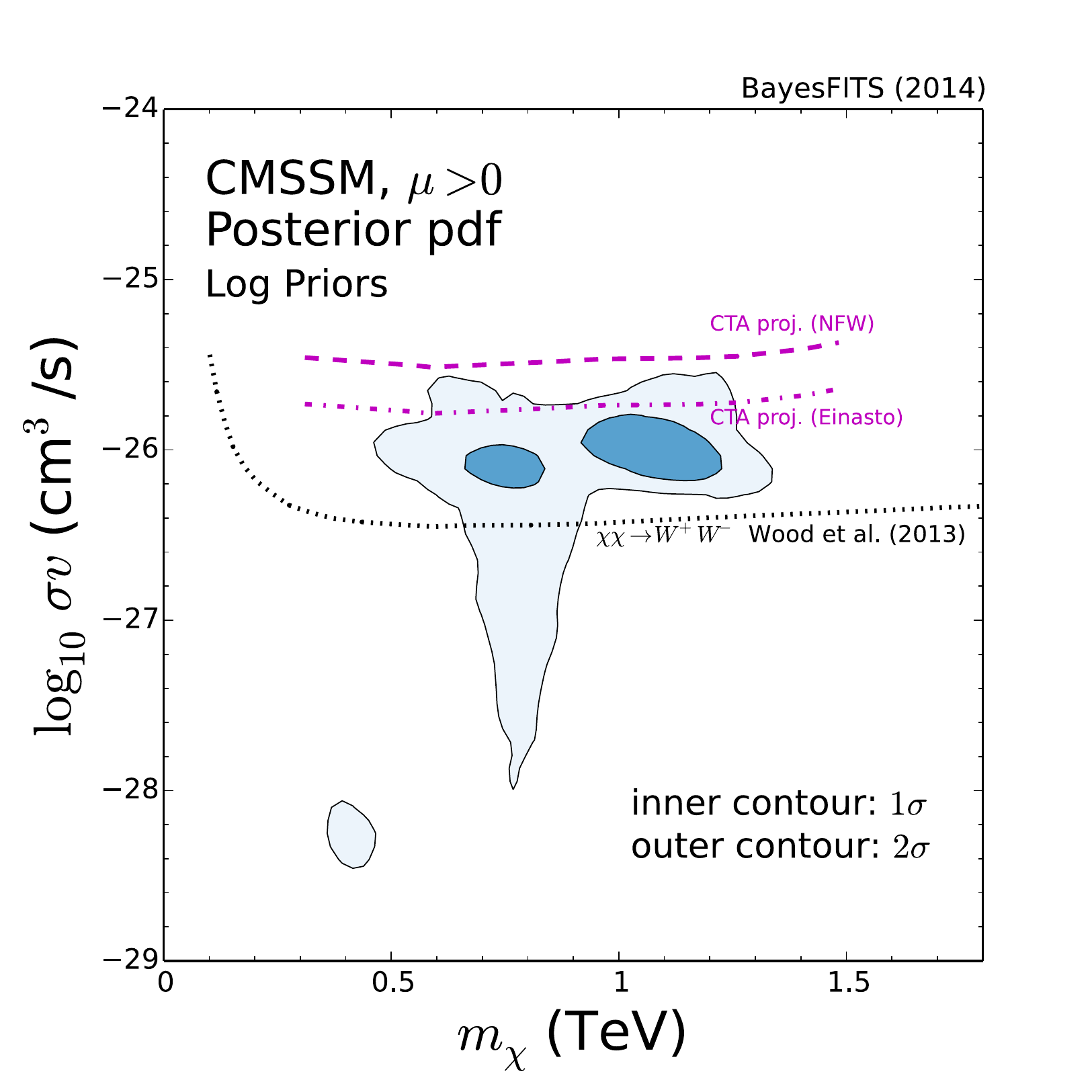}
}%
\caption{\footnotesize \protect\subref{fig:a} Marginalized 2D posterior distribution for the CMSSM with $\mu > 0$ in the (\mchi, \sigsip) plane. 
The red solid line shows the 90\%~C.L. upper bound as given by LUX, here included in the likelihood function. The gray dot-dashed line
shows the 2012 XENON100 90\%~C.L. bound\cite{Aprile:2012nq} and the magenta dashed line shows projected sensitivity for 2017 at \xenononet\cite{Aprile:2012zx}. 
\protect\subref{fig:b} Marginalized 2D posterior distribution for the CMSSM with $\mu > 0$ in the (\mchi, $\sigma v$) plane.
The magenta dashed line shows the expected sensitivity of CTA under the assumptions of\cite{Pierre:2014tra} for a NFW halo profile. 
The magenta dot-dashed line shows the corresponding sensitivity with Einasto profile.
The dotted black line shows the projected sensitivity of the CTA expansion considered in\cite{Wood:2013taa}.}
\label{fig:cmssm_mx_sigmap}
\end{figure}

In \reffig{fig:cmssm_mx_sigmap}\subref{fig:a} we show the 2D posterior
distribution in the (\mchi, \sigsip) plane for $\mu>0$.  The different
regions are well separated and can be identified from left to right as
the stau-coannihilation, $A$-resonance and $\sim 1\tev$ higgsino
regions.  We show the current best upper limit from LUX 90\%~\cl\
as a red solid line, the previous one from
XENON100 as a gray dot-dashed line, and the
projected sensitivity of \xenononet\cite{Aprile:2012zx} as a magenta dashed line.  The
bino-like neutralino typical of the stau-coannihilation and
$A$-resonance regions has a suppressed coupling to the nucleus, so
that both regions lie well below the current LUX bound and it is very
unlikely they will be tested, even with the improved sensitivity of
\xenononet.  In contrast, the $\sim 1\tev$ higgsino region lies almost
entirely within the projected \xenononet\ sensitivity.  The entire
68\% and nearly all of the 95\% credibility region have the potential
to be probed in the next few years, encompassing about 70\% of the
points in the scan.  This makes dark matter direct detection searches
the predominant tool for exploration of the CMSSM.
   
In the CMSSM the largest cross section values, $\sigsip\gsim 10^{-8}\pb$, are obtained in the focus point region. 
One can see the beginning of the horizontal branch joining the higgsino and focus point regions, at $\mchi\simeq 0.7-0.8\tev$.
The effect of the LUX limit in the likelihood is visible, as the credibility region is cut off rapidly after crossing the 90\%~\cl\ bound, shown in red. 
In contrast to\cite{Kowalska:2013hha}, this causes the focus point region to be disfavored by the scan.
In the $\mu < 0$ scenario we obtain the same results albeit with the absence of the $A$-resonance region. 
The sign of the $\mu$ parameter has little impact on \sigsip\ for the neutralino and the $\sim 1\tev$ higgsino region with $\mu < 0$ 
can also be entirely probed by \xenononet.

In \reffig{fig:cmssm_mx_sigmap}\subref{fig:b} we show the 2D posterior distribution in the (\mchi, \sigv) plane.
The node at $\sigv\lesssim 10^{-28}\textrm{ cm}^3/\textrm{s}$ is the stau-coannihilation region, 
which has a much reduced \sigv\ in the present day due to the absence of co-annihilations with the stau NLSP, 
which are instead only present in the early Universe.
The $A$-resonance and $\sim1\tev$ higgsino regions are visible at larger \sigv, from left to right, respectively.
The $A$-resonance region is characterized by a broad range of cross section values, with a deep funnel at 95\% credibility
that extends down to $\sigv\simeq 10^{-28}\textrm{ cm}^3/\textrm{s}$.
This corresponds to a large resonant effect in the early Universe when the neutralinos are distributed thermally, 
but the present value of \sigv\ is small since the colliding neutralinos have insufficient energy to produce the pseudoscalar on shell
(see, e.g., Appendix~B in\cite{Fowlie:2013oua}).
\sigv\ is reduced by orders of magnitude in this funnel and is effectively impossible to probe via indirect detection.

As was the case for direct detection, the $\sim 1\tev$ higgsino region
presents a particularly promising target for indirect detection since
the annihilation cross section is restricted to a small range close to
the thermal value compared to other dark matter candidates in the MSSM
which can have much lower annihilation cross sections.  The magenta
dashed and dot-dashed lines in
\reffig{fig:cmssm_mx_sigmap}\subref{fig:b} show the expected
sensitivity of CTA derived in \refsec{sec:constraints} under the
assumptions of the NFW\cite{Navarro:1995iw} and Einasto\cite{Einasto}
halo profile, respectively.  The dotted line shows the projected
sensitivity of CTA to the $W^+W^-$ final state, as obtained in
Ref.\cite{Wood:2013taa} under a more optimistic setup including more
telescopes in the array.  As previously discussed, limits on single
final states may not automatically exclude all points in the posterior
since the neutralino will usually annihilate into several different
final states depending on its mass and composition.  The $W^+W^-$
final state does however provide a good approximation in the $\sim
1\tev$ higgsino region and points in this region lying above the
$W^+W^-$ line have the potential to be constrained.  We note that
current limits from the Cherenkov telescope array
HESS\cite{Abazajian:2011ak} are approximately an order of magnitude
larger than the projected CTA sensitivity and so do not constrain the
posterior regions found.

One can see that the derived CTA limits lie just above above the 68\%
credibility posterior region for either choice of dark matter profile,
but significant improvements to the limits are possible by improving
the experimental setup.  It should be noted that a factor of five
improvement in the model independent limit on the number of observed
gamma rays would be sensitive to the entire $\sim 1 \tev$ higgsino
region and to the bulk of the $A$-resonance region, thus the vast
majority of the favored points in the CMSSM.

The higgsino and stau-coannihilation regions in the negative $\mu$ case share the same properties as the positive $\mu$ case and 
have the same prospects for detection, we therefore do not show their distribution here. 

\section{Results in the NUHM}\label{sec:NUHM}

\subsection{Posterior distributions and prospects for collider searches}

We proceed now to the analysis of the NUHM. The parameters \mzero,
\mhalf, \azero, and \tanb\ were scanned in the same ranges as in the
CMSSM. The parameters \mhdsq\ and \mhusq\ were allowed to assume
negative values at the GUT scale; see Table~\ref{tab:cmssmprior}. We
limit ourself to the case with $\mu>0$, which, as was seen for the
CMSSM, presents a greater number of solutions. One must keep in mind
that the solutions with $\mu<0$ can generally be mapped to a subset of
the ones we present in here, without novel phenomenological features.

\begin{figure}[t]
\centering
\subfloat[]{%
\label{fig:a}%
\includegraphics[width=0.47\textwidth]{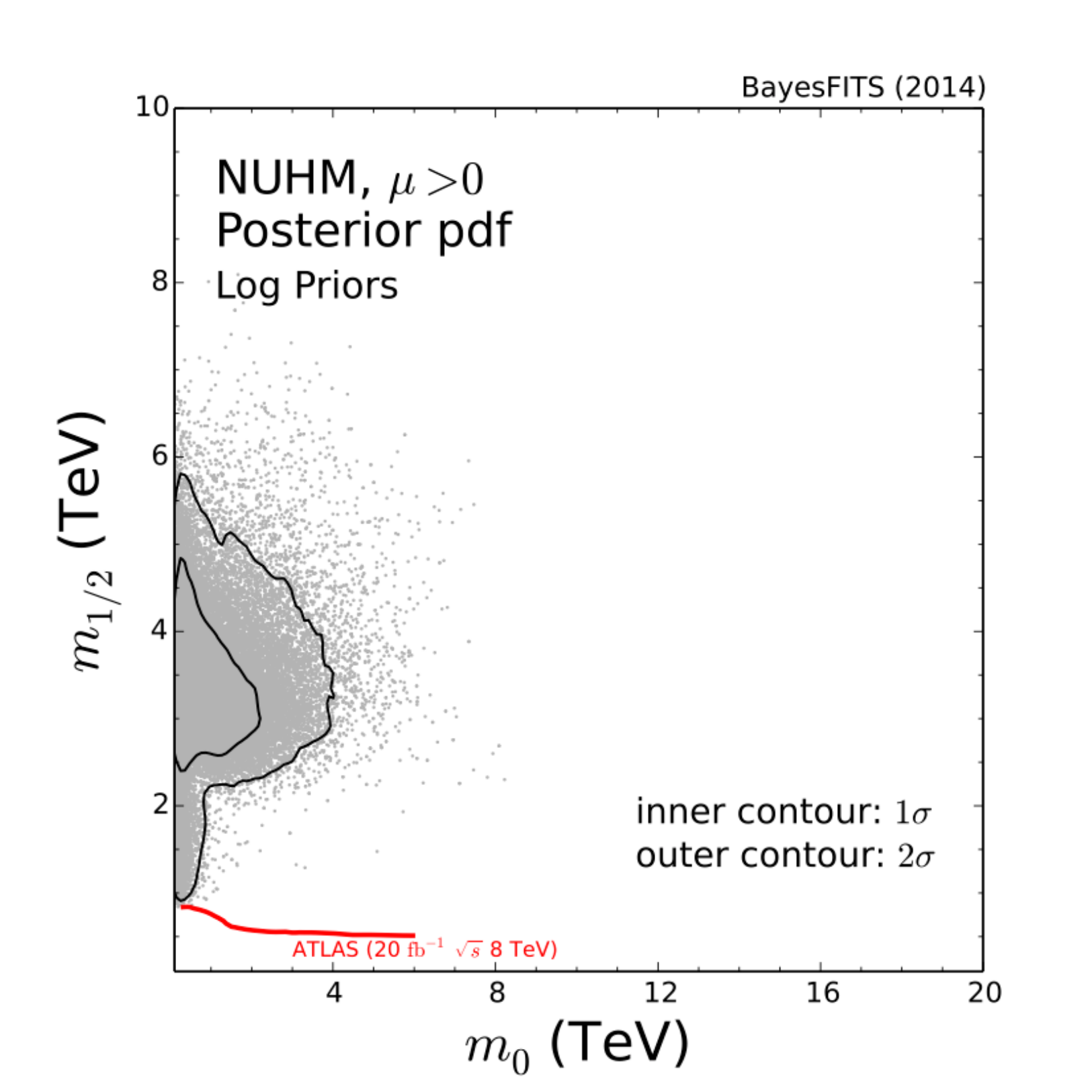}
}%
\hspace{0.01\textwidth}
\subfloat[]{%
\label{fig:b}%
\includegraphics[width=0.47\textwidth]{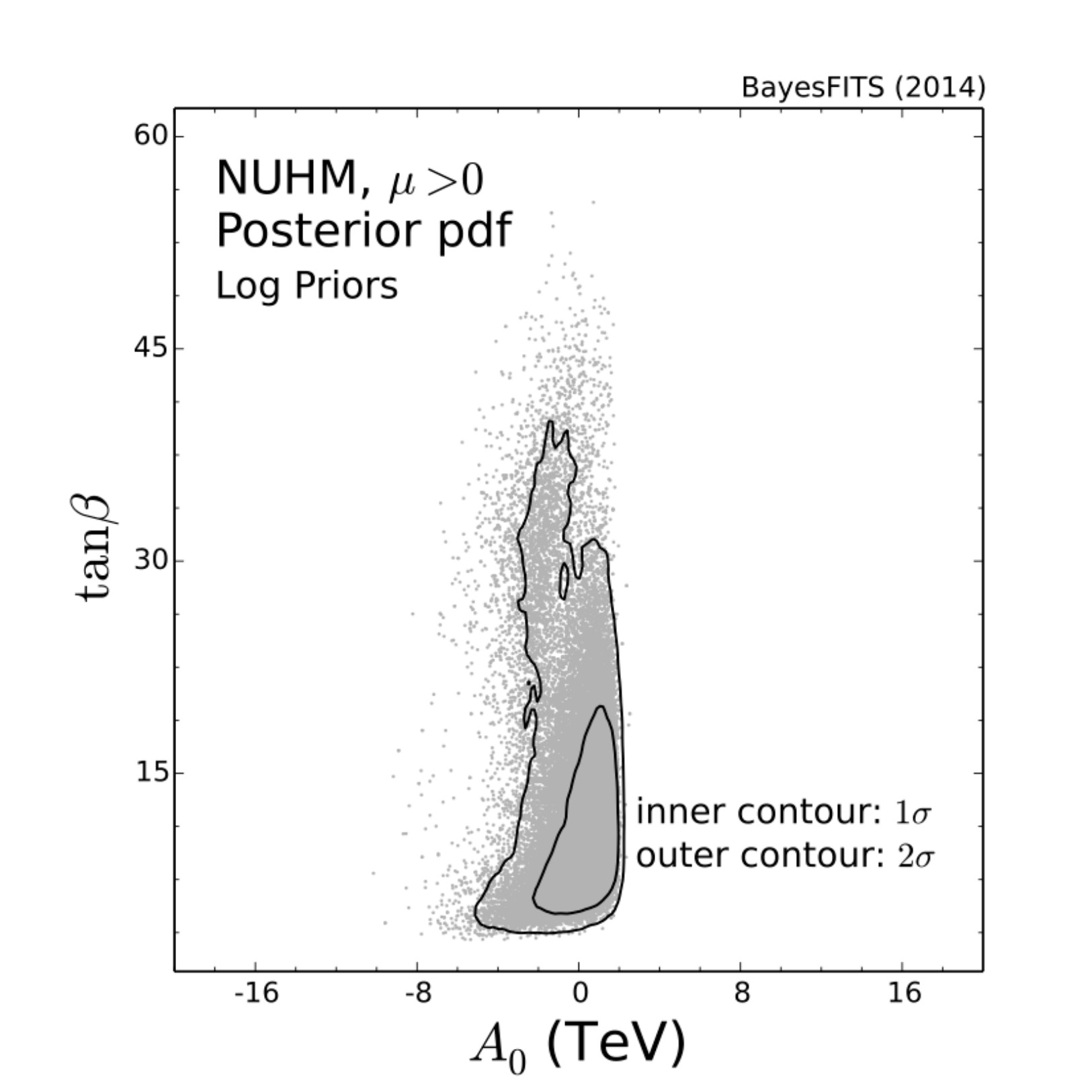}
}%
\\
\subfloat[]{%
\label{fig:c}%
\includegraphics[width=0.47\textwidth]{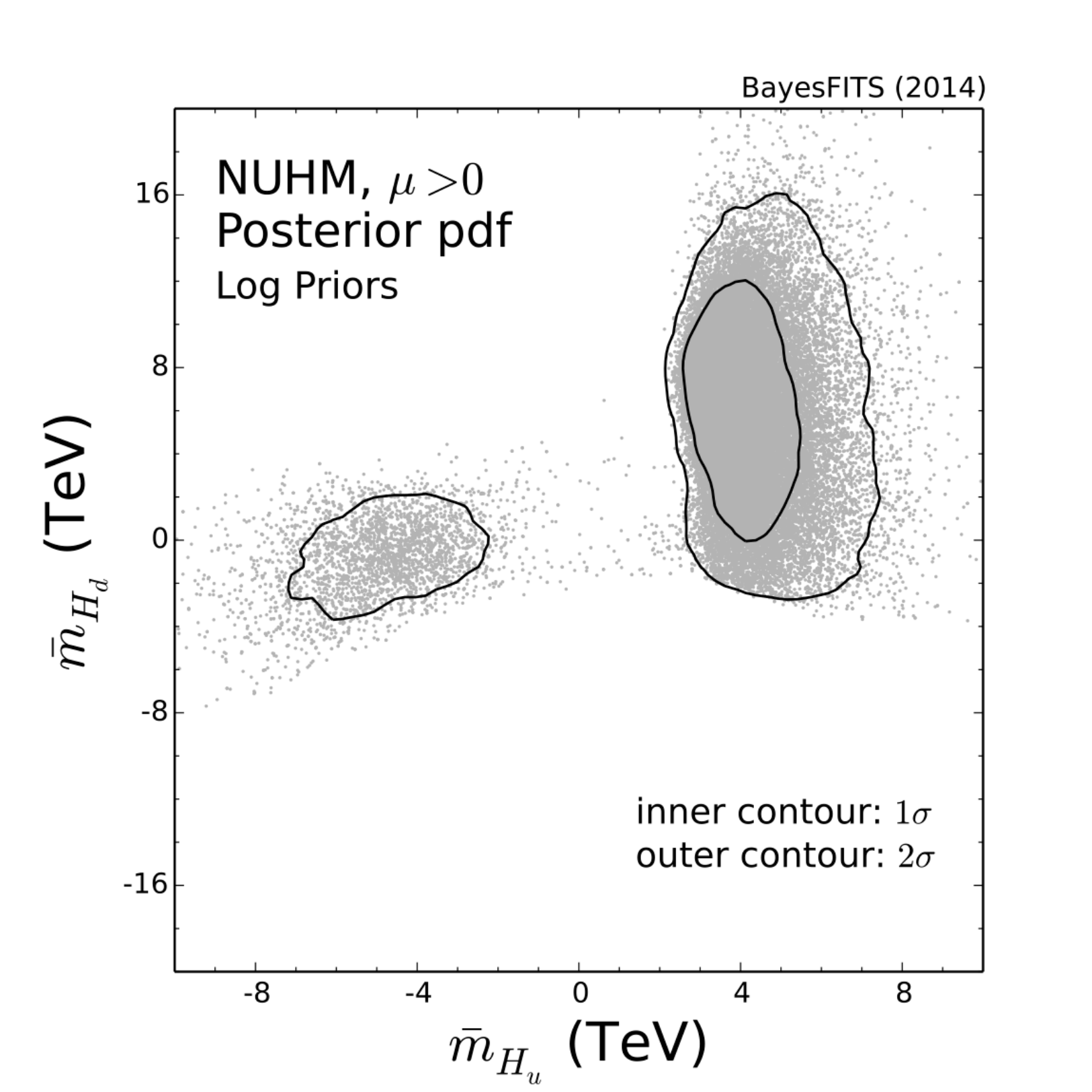}
}%
\hspace{0.01\textwidth}
\subfloat[]{%
\label{fig:d}%
\includegraphics[width=0.47\textwidth]{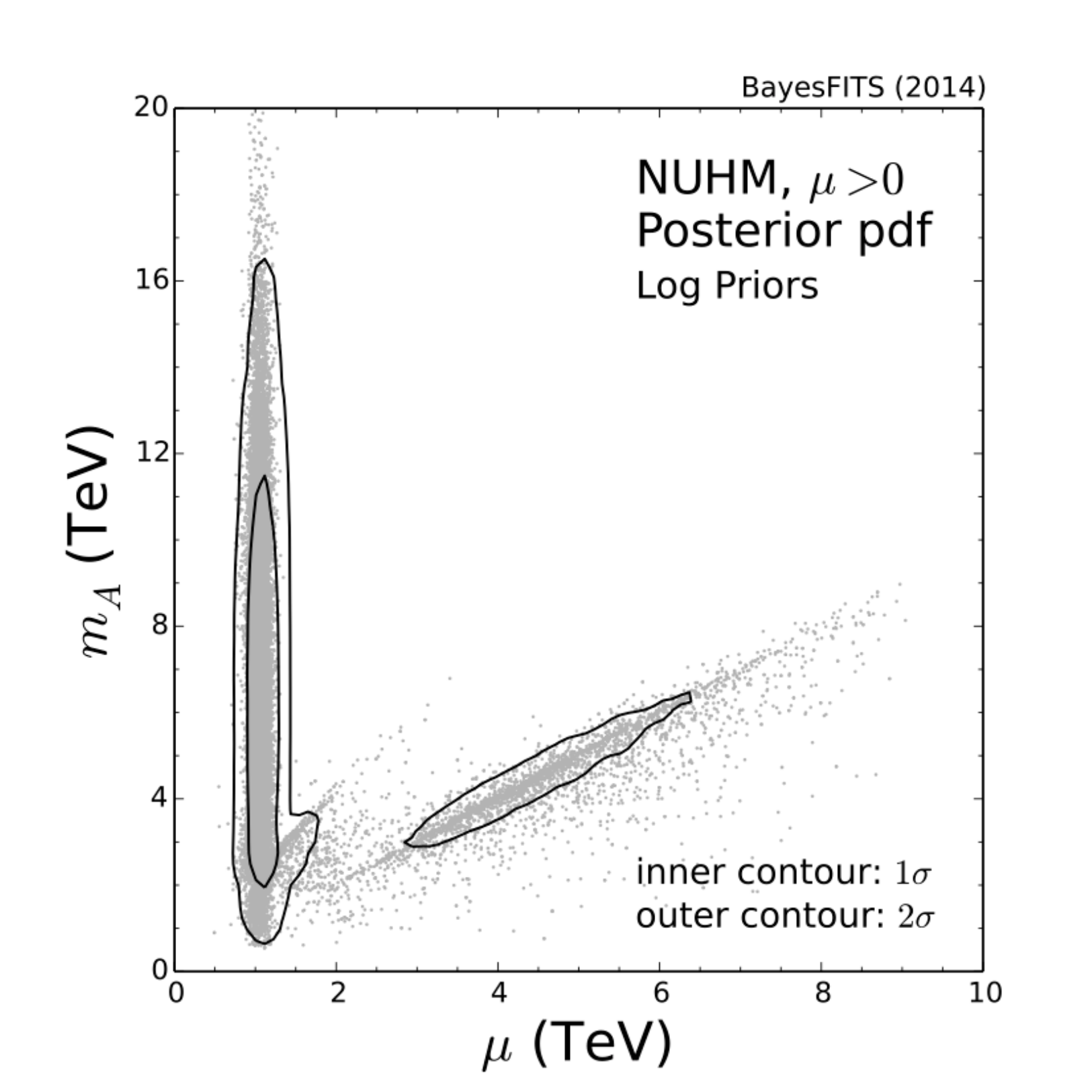}
}%
\caption{\footnotesize Marginalized 2D posterior in \protect\subref{fig:a} the (\mzero, \mhalf), \protect\subref{fig:b} the (\azero, \tanb),  
\protect\subref{fig:c} the (\barmhu, \barmhd), and \protect\subref{fig:d} the ($\mu$, \ma) planes of the NUHM with $\mu>0$. 
68\% and 95\% credible regions are shown by the inner and outer contours, respectively. 
Gray points are distributed according to the posterior probability. The ATLAS 95\%~C.L. exclusion line is shown in red solid for reference.}
\label{fig:nuhm_m0_m12_A0_tanb}
\end{figure}

In \reffig{fig:nuhm_m0_m12_A0_tanb}\subref{fig:a} we show the marginalized 2D posterior 
in the (\mzero, \mhalf) plane. The 68\% and 95\% credible regions are indicated by inner and outer solid contours, respectively.
In this and the following figures we superimpose on the clearly marked credible regions a set of points drawn from the posterior distribution. 
We include those points since, unlike in the CMSSM, the points that satisfy the relic density constraint are not
always found in regions of parameter space well separated along different mechanisms of neutralino annihilation. 
Thus, the distribution and position of the points in the plots highlight details not always   
easy to infer from the contours and help indicate where additional solutions beyond the 95\% credibility regions can be found.


The mechanisms to reduce the relic density are for the most part in common with the CMSSM: stau co-annihilation and $A$-resonance of bino-like neutralinos, 
and annihilation and co-annihilation of $\sim 1\tev$ higgsino-like neutralinos and charginos. 
We shall see, though, that the NUHM also presents some additional ways of obtaining the correct relic density. 
Moreover, all of those mechanisms can be obtained for \mzero\ not too large,
in contrast with the $\sim 1\tev$ higgsino region of the CMSSM, thanks to the additional freedom in the Higgs sector.
Thus the posterior shows a more compact shape in (\mzero, \mhalf) plane, due also to the effect of log prior distributions. 
Had we used flat priors, volume effects would inflate the region of large \mzero\ and \mhalf\ and therefore increase the importance of
the $\sim1\tev$ higgsino region in the NUHM.

The marginalized posterior in the (\azero, \tanb) plane is shown in \reffig{fig:nuhm_m0_m12_A0_tanb}\subref{fig:b}.
As was the case for the CMSSM, the solutions tend to be distributed over the entire \tanb\ range, 
but now favor relatively smaller values of $|\azero|$ than in previous scans.  


In \reffig{fig:nuhm_m0_m12_A0_tanb}\subref{fig:c} we show the marginalized 2D pdf in the (\barmhu, \barmhd) 
plane.\footnote{Here \barmhu\ and \barmhd\ refer to the signed square root of the absolute value of 
\mhusq\ and \mhdsq, respectively. E.g., $\barmhu=\mhusq/\sqrt{|\mhusq|}$.} 
One can see here a previously unexplored 95\% credibility region featuring negative values for \barmhu\ at the GUT scale 
and, for the majority of the points involved, also $\barmhd<0$\,.

Large negative values of \mhusq\ at the GUT scale lead to large negative value for the same parameter at \msusy.
Through the EWSB condition, these points thus feature very large values of $\mu$, up to ranges previously unexplored 
in NUHM analyses. We show the marginalized 2D pdf in the ($\mu$, \ma) plane in \reffig{fig:nuhm_m0_m12_A0_tanb}\subref{fig:d}.
The described solutions can be seen on the lower right end of the plot, for values of $\mu$ that can be as large 
as 9\tev.\footnote{It is clear that these solutions present uncomfortably large values of EW fine tuning due the large $\mu$
parameter, so that they might be unappealing from a theoretical point of view. However, in this paper we are only concerned with the existence of 
viable phenomenological solutions, independently of theoretical considerations. 
We will therefore treat these solutions without further mentioning the fine-tuning issue.} 

\begin{figure}[t]
\centering
\subfloat[]{%
\label{fig:a}%
\includegraphics[width=0.47\textwidth]{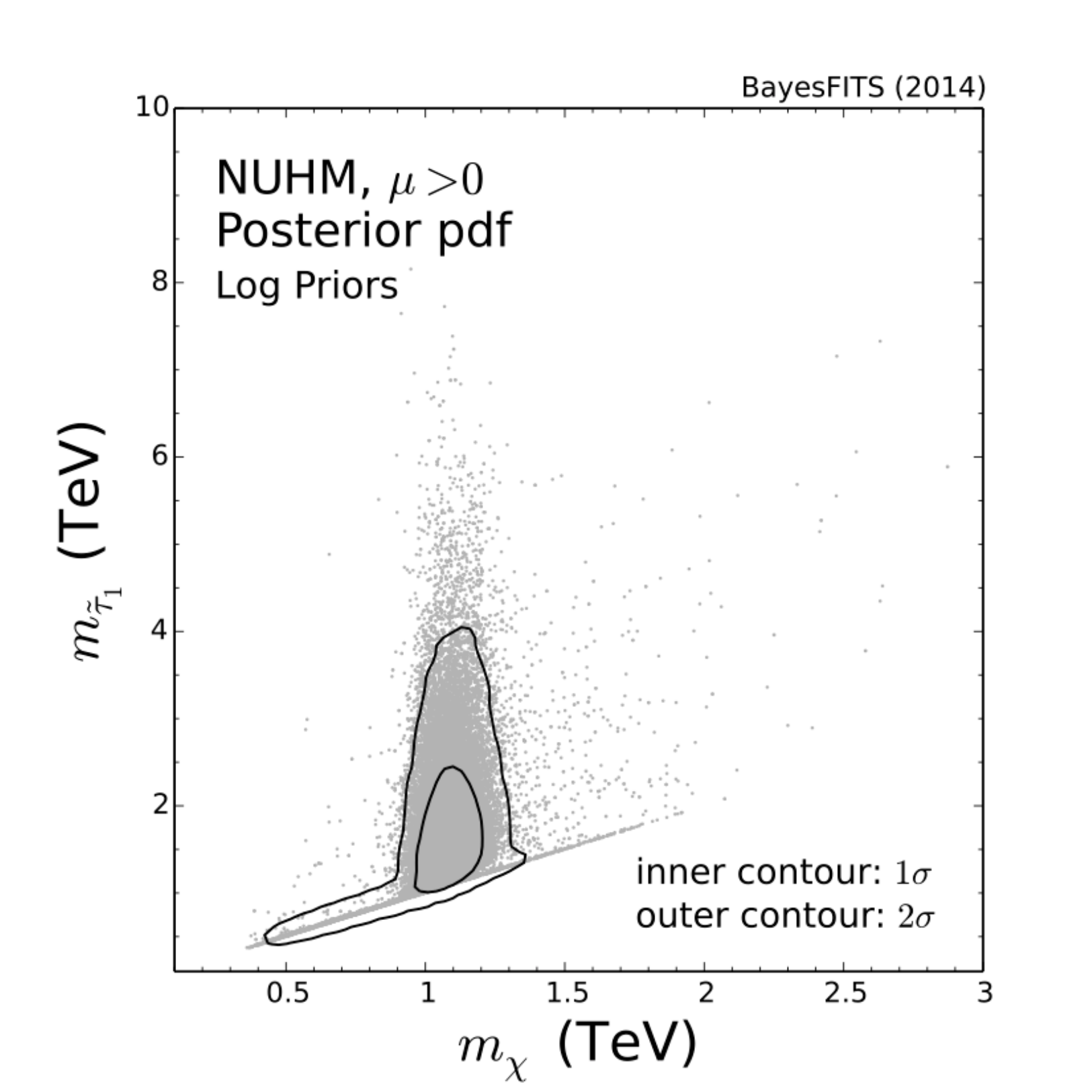}
}%
\hspace{0.01\textwidth}
\subfloat[]{%
\label{fig:b}%
\includegraphics[width=0.47\textwidth]{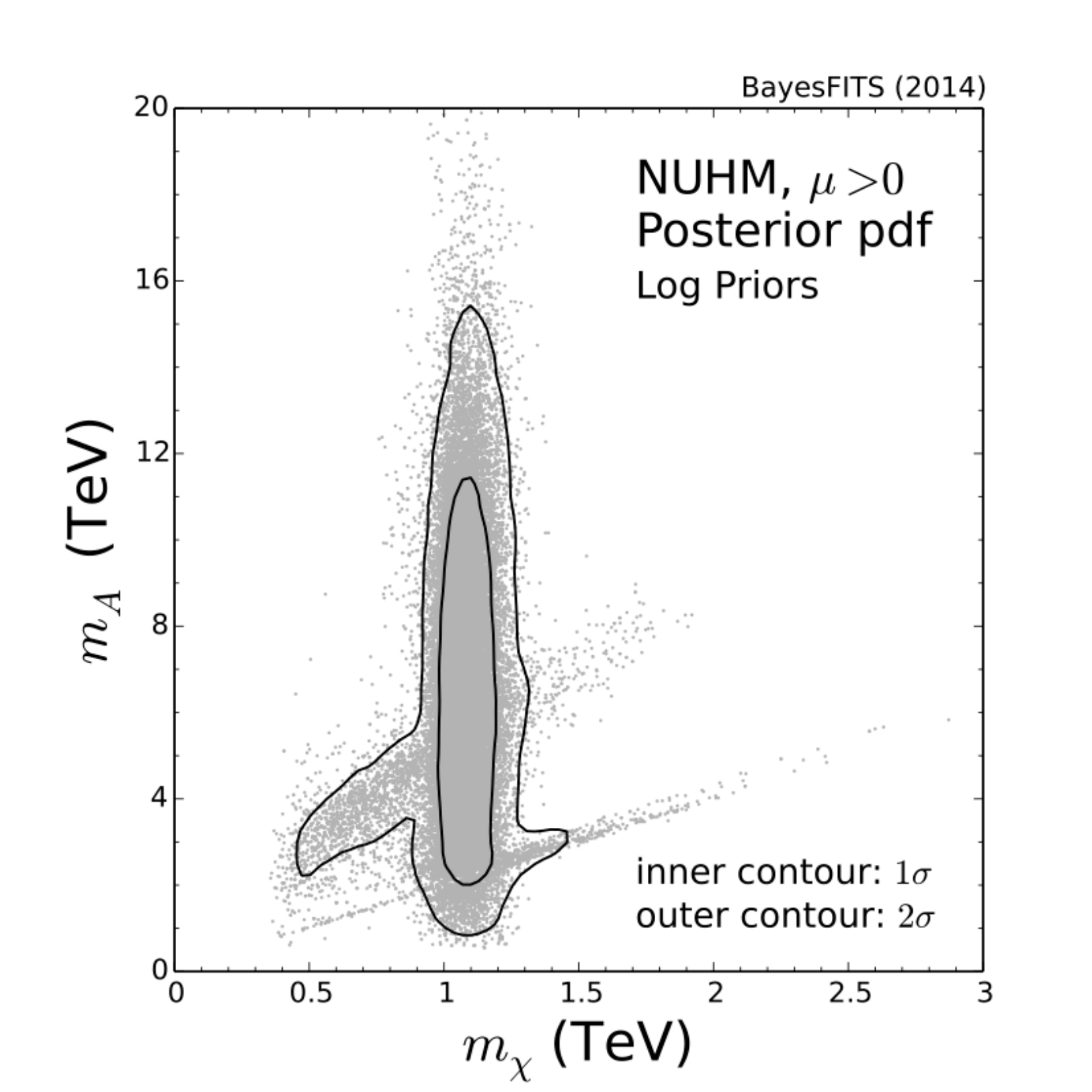}
}%
\caption{\footnotesize Marginalized 2D posterior in \protect\subref{fig:a} the (\mchi, $m_{\stau_1}$) plane and 
\protect\subref{fig:b} the (\mchi, \ma) plane of the NUHM with $\mu>0$. 
68\% and 95\% credible regions are shown by the inner and outer contours, respectively. 
Gray points are distributed according to the posterior probability.}
\label{fig:nuhm_mhu_mhd_mx_ma}
\end{figure}

The correct relic density for these points is obtained through mechanisms of stau-coannihilation.
In fact, values of $\mu$ so large strongly enhance the coupling of the lightest stau to the $H_u$ component
of the lightest Higgs, thus increasing the efficiency of the annihilation channel $\stau \stau \rightarrow h h$.
The $\stau \stau h$ vertex, proportional to $\mu$, leads to a $\mu^4$ enhancement 
of the annihilation cross section, which becomes the dominant mechanism when the lightest stau and neutralino are almost degenerate in mass.
We show in \reffig{fig:nuhm_mhu_mhd_mx_ma}\subref{fig:a} the 2D pdf in the (\mchi, $m_{\stau_1}$) plane. 
One can see that because of the above considerations, stau-coannihilation in the NUHM is efficient up to $\mchi\simeq 2\tev$,
thus significantly extending the range observed in the CMSSM, or in
previous studies performed with more limited ranges of $\mu$\cite{Strege:2012bt,Fowlie:2012im}. 

Figures~\ref{fig:nuhm_m0_m12_A0_tanb}\subref{fig:c} and \ref{fig:nuhm_m0_m12_A0_tanb}\subref{fig:d} show that
the vast majority of the points with high posterior probability feature, at the  GUT scale, positive values of \mhusq\ and \mhdsq, 
and at the weak scale, $\mu\simeq 1\tev$.
Thus, as was the case for the CMSSM, in the NUHM the largest number of solutions belong to the $\sim1\tev$ higgsino region.
The predominance of this region was also shown in Ref.\cite{Strege:2012bt}, where it was the \textit{only} region found, due to their
choice of input parameters and prior ranges. The extended stau-coannihilation region 
shown in Figs.~\ref{fig:nuhm_m0_m12_A0_tanb} and \ref{fig:nuhm_mhu_mhd_mx_ma} is instead a novel finding of this study.
  
Moreover, as was mentioned at the beginning of this section, in the
NUHM there are also a significant number of solutions for which the
neutralino annihilates via the resonance with heavy Higgs bosons, as
already pointed out in\cite{Roszkowski:2009sm,Kowalska:2013hha}.  We show in
\reffig{fig:nuhm_mhu_mhd_mx_ma}\subref{fig:b} the 2D pdf in the
(\mchi, \ma) plane, which shows the presence of many points at
$\ma\approx 2\mchi$. Those points cannot be identified as easily in
other plots: for example, in
\reffig{fig:nuhm_m0_m12_A0_tanb}\subref{fig:c} they are shown as
diffuse points lying between the two high posterior probability modes
described above, and as a subset of points in the right hand mode
restricted to $\barmhd < 4\tev$ (above this value $\mhd^2$ drives \ma\
to be too large to find solutions with $\ma\approx 2\mchi$).  We
point out here that, while in the CMSSM the points of the
$A$-resonance region feature neutralinos with very large bino
composition, and very large \tanb\ values, this is not always the case
in the NUHM. Thus, while in the CMSSM the main annihilation channel
for those points is $\chi\chi\rightarrow b\bar{b}$, many of the
$A$-resonance points with $\mchi>1.2\tev$ in
\reffig{fig:nuhm_mhu_mhd_mx_ma}\subref{fig:b} feature much larger
higgsino fraction and moderate \tanb\ values, so that the correct relic
density is obtained in a combined fashion: with $t\bar{t}$ final
states in addition to $b\bar{b}$, and through resonance with the heavy
scalar $H$, degenerate with $A$, in addition to the pseudoscalar.  As
we shall see in \refsec{sec:dmnuhm} this fact has important
consequences for dark matter direct detection.

Figure~\ref{fig:nuhm_mhu_mhd_mx_ma}\subref{fig:b} neatly shows the different annihilation mechanisms described above: 
the large concentration of points at $\mchi \simeq 1\tev$ gives the $\sim1\tev$ higgsino region; 
at $\ma\approx 2\mchi$ there is the strip of points belonging to the just-described $A/H$-resonance region, 
which induces a lobe in the 95\% credible region of the posterior for $\mchi\gsim1.2\tev$;
and, finally, the stau-coannihilation region, 
which produces a lobe in the posterior with $\mchi \lesssim 1\tev$, but can in fact extend up to $\mchi\lesssim2\tev$.

\begin{figure}[t]
\centering
\includegraphics[width=0.47\textwidth]{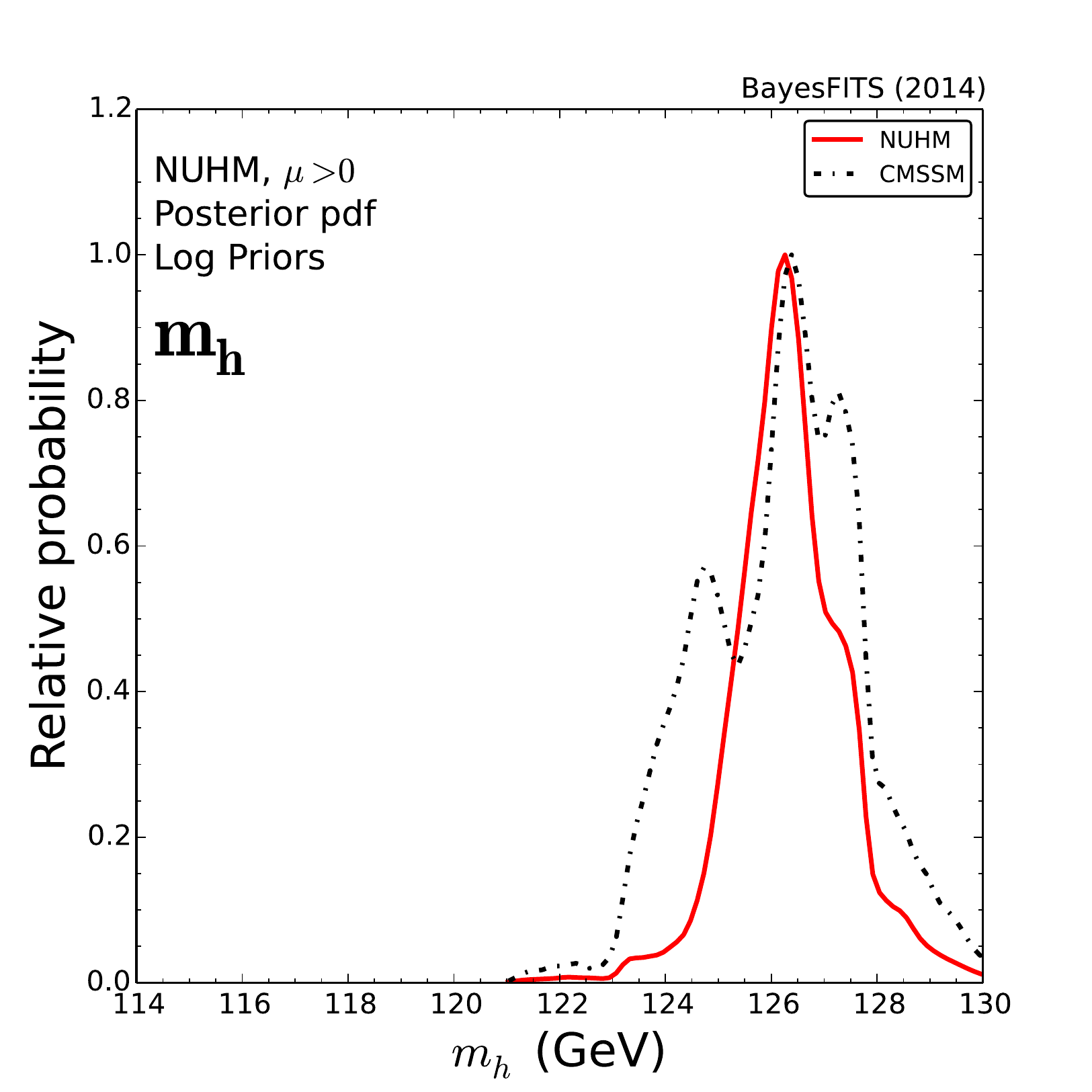}
\caption{\footnotesize Marginalized 1D pdf of $m_h$ for the NUHM with $\mu > 0$. The black dot-dashed line shows the distribution obtained for
  the CMSSM with $\mu>0$ for comparison.}
\label{fig:nuhm_mh}
\end{figure}

We show in \reffig{fig:nuhm_mh} the 1D pdf distribution of the lightest Higgs mass, \mhl. 
The black dot-dashed line shows the distribution obtained for the CMSSM with $\mu>0$ for comparison. 
One can see that the posterior is prevalently uni-modal due to the statistical dominance of the 
$\sim 1\tev$ higgsino region in the NUHM.

\begin{figure}[t]
\centering
\subfloat[]{%
\label{fig:a}%
\includegraphics[width=0.36\textwidth]{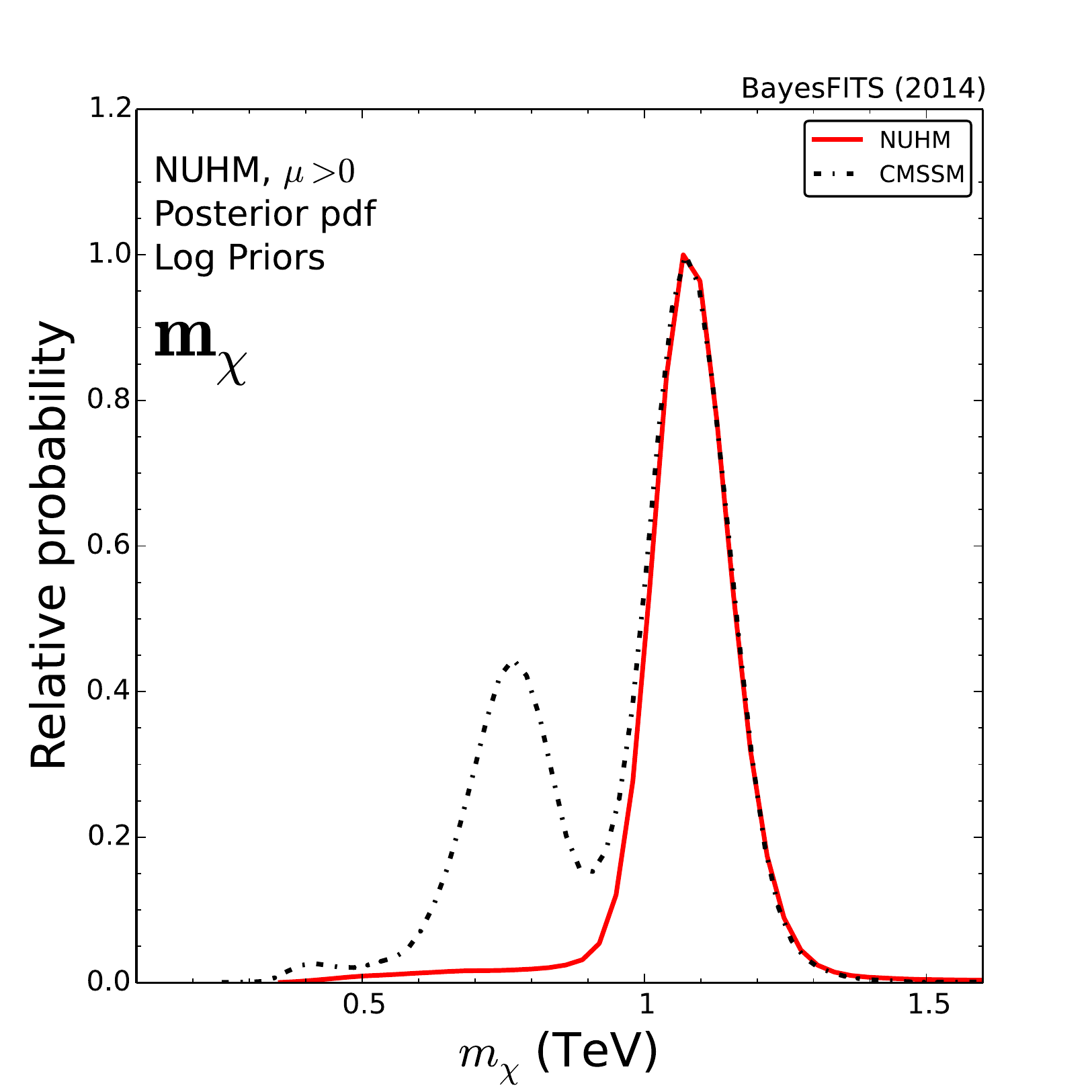}
}%
\hspace{0.07\textwidth}
\subfloat[]{%
\label{fig:b}%
\includegraphics[width=0.36\textwidth]{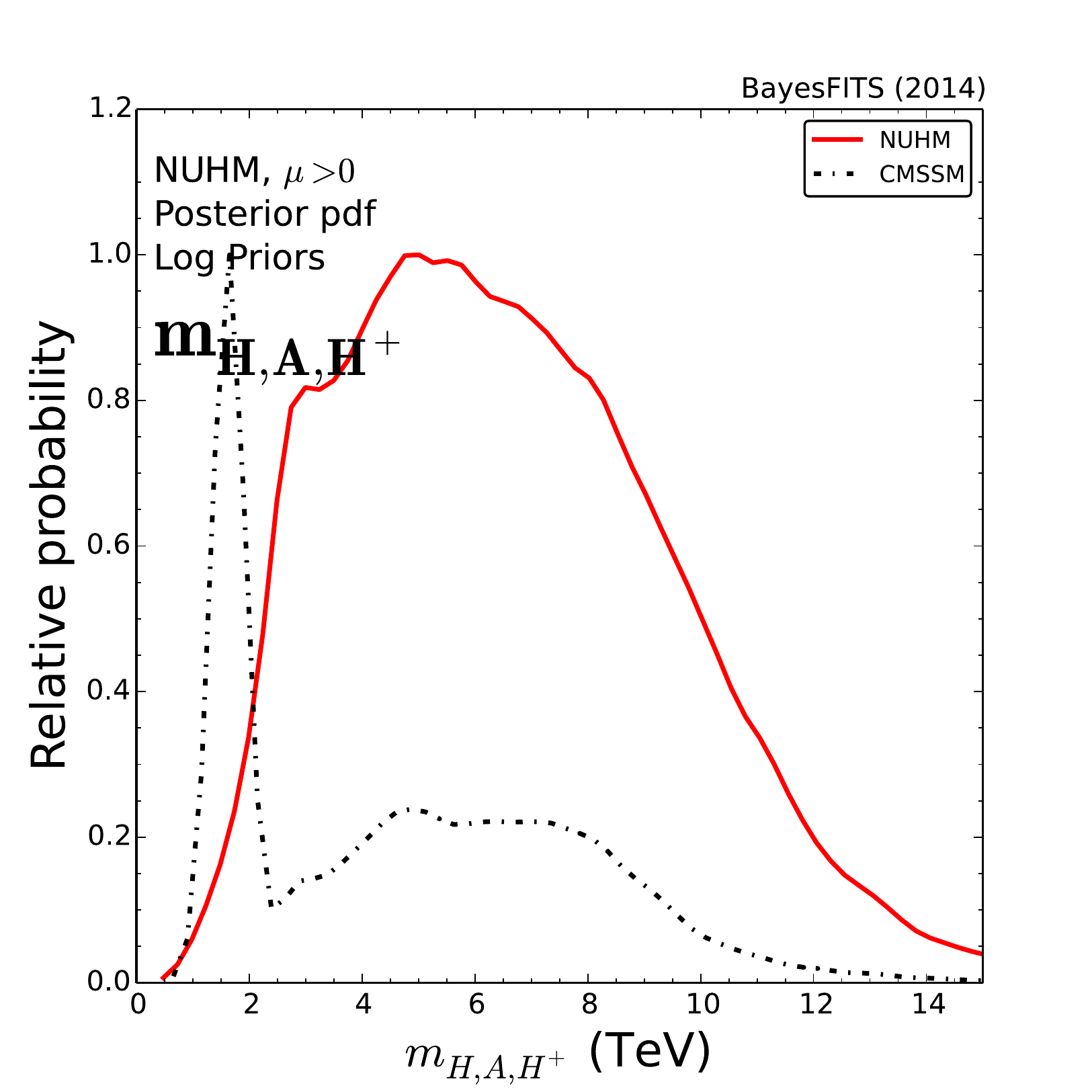}
}%

\subfloat[]{%
\label{fig:c}%
\includegraphics[width=0.36\textwidth]{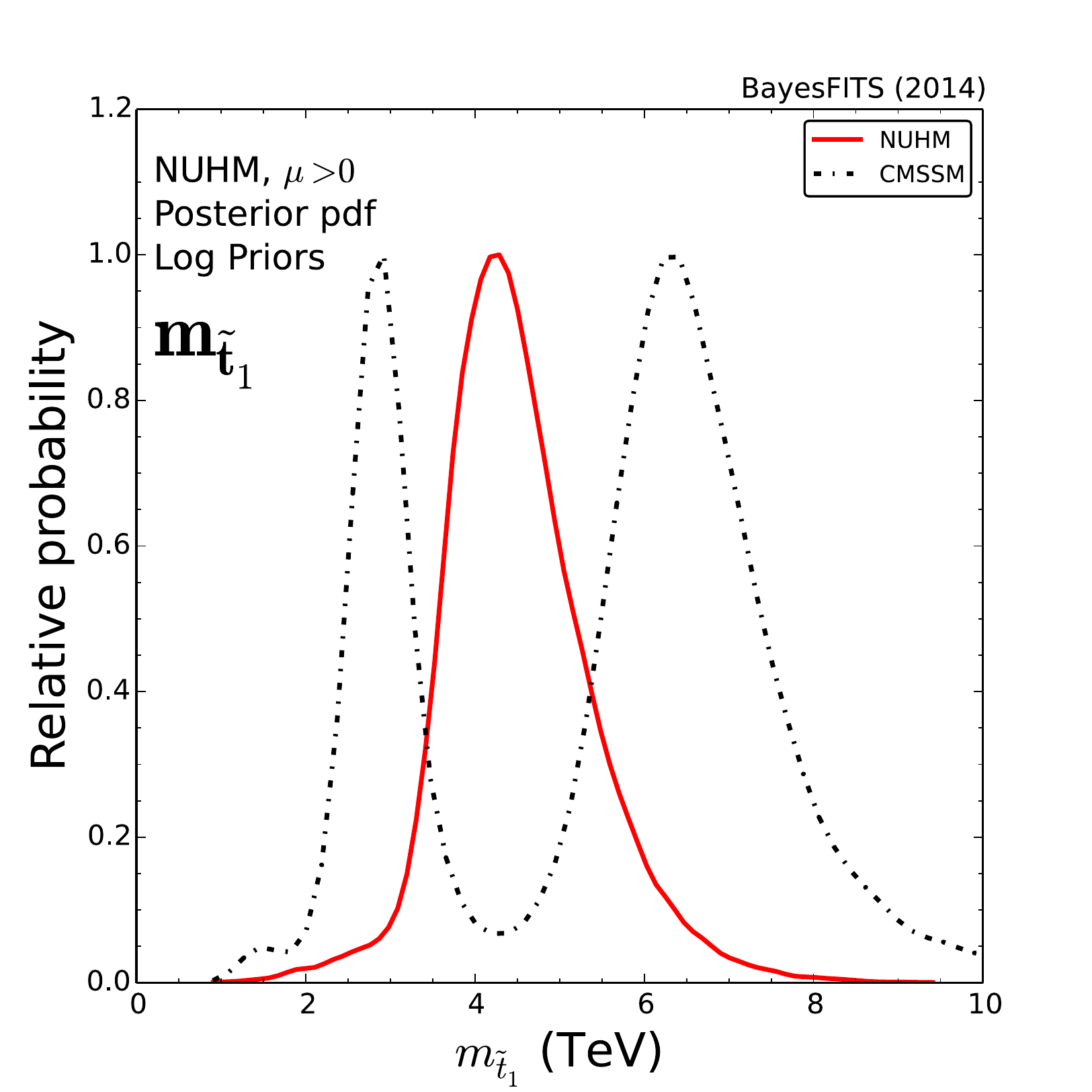}
}%
\hspace{0.07\textwidth}
\subfloat[]{%
\label{fig:d}%
\includegraphics[width=0.36\textwidth]{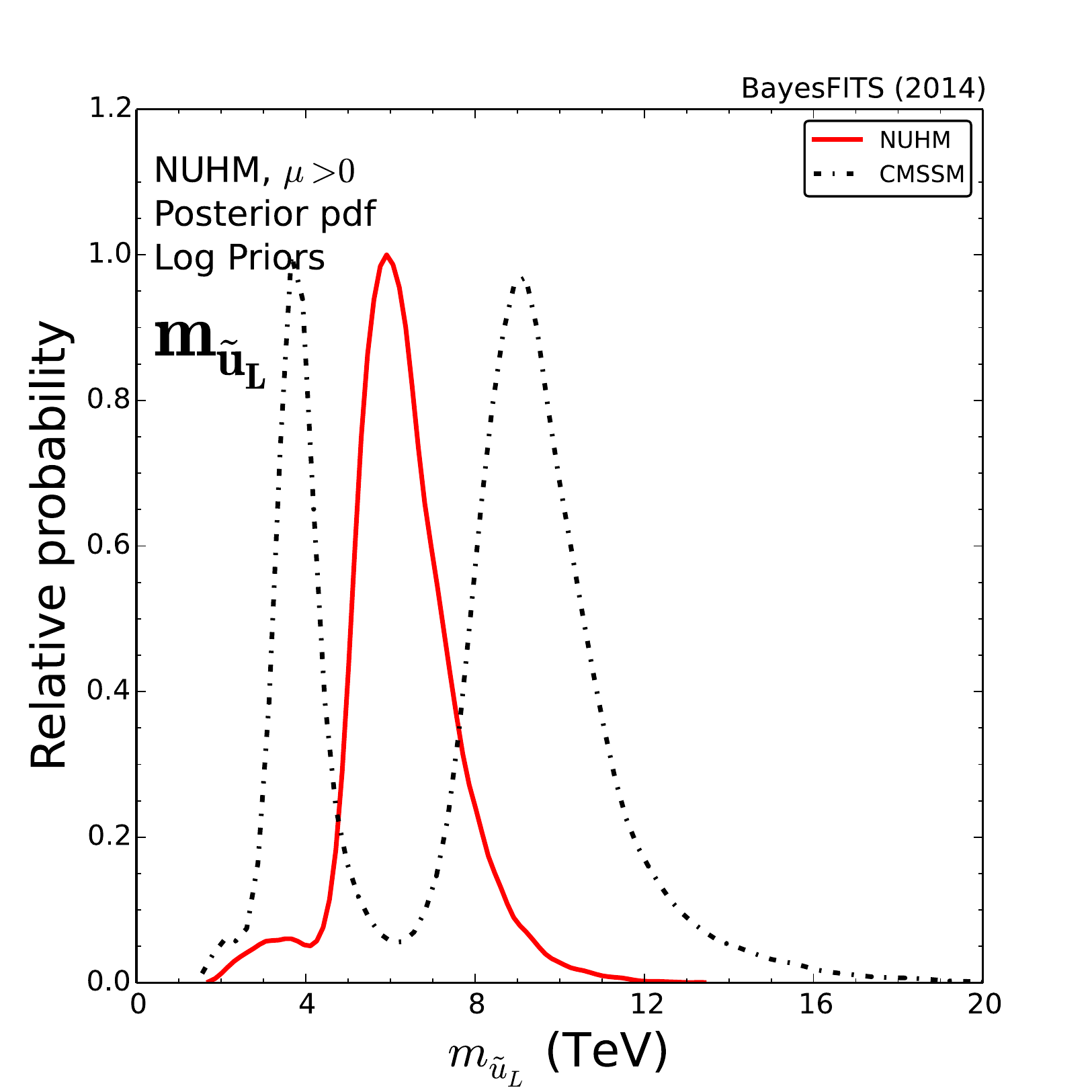}
}%

\subfloat[]{%
\label{fig:e}%
\includegraphics[width=0.36\textwidth]{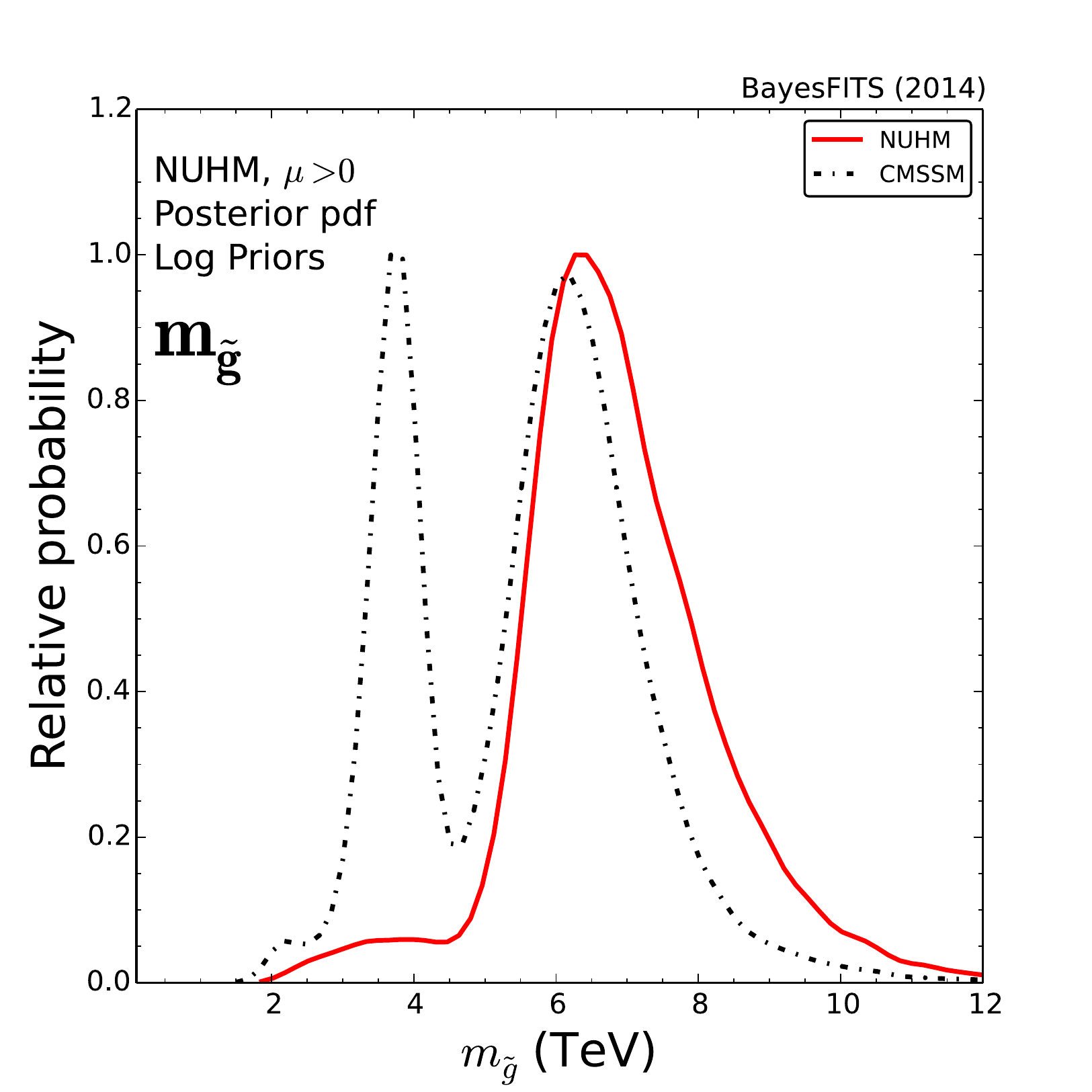}
}%
\hspace{0.07\textwidth}
\subfloat[]{%
\label{fig:f}%
\includegraphics[width=0.36\textwidth]{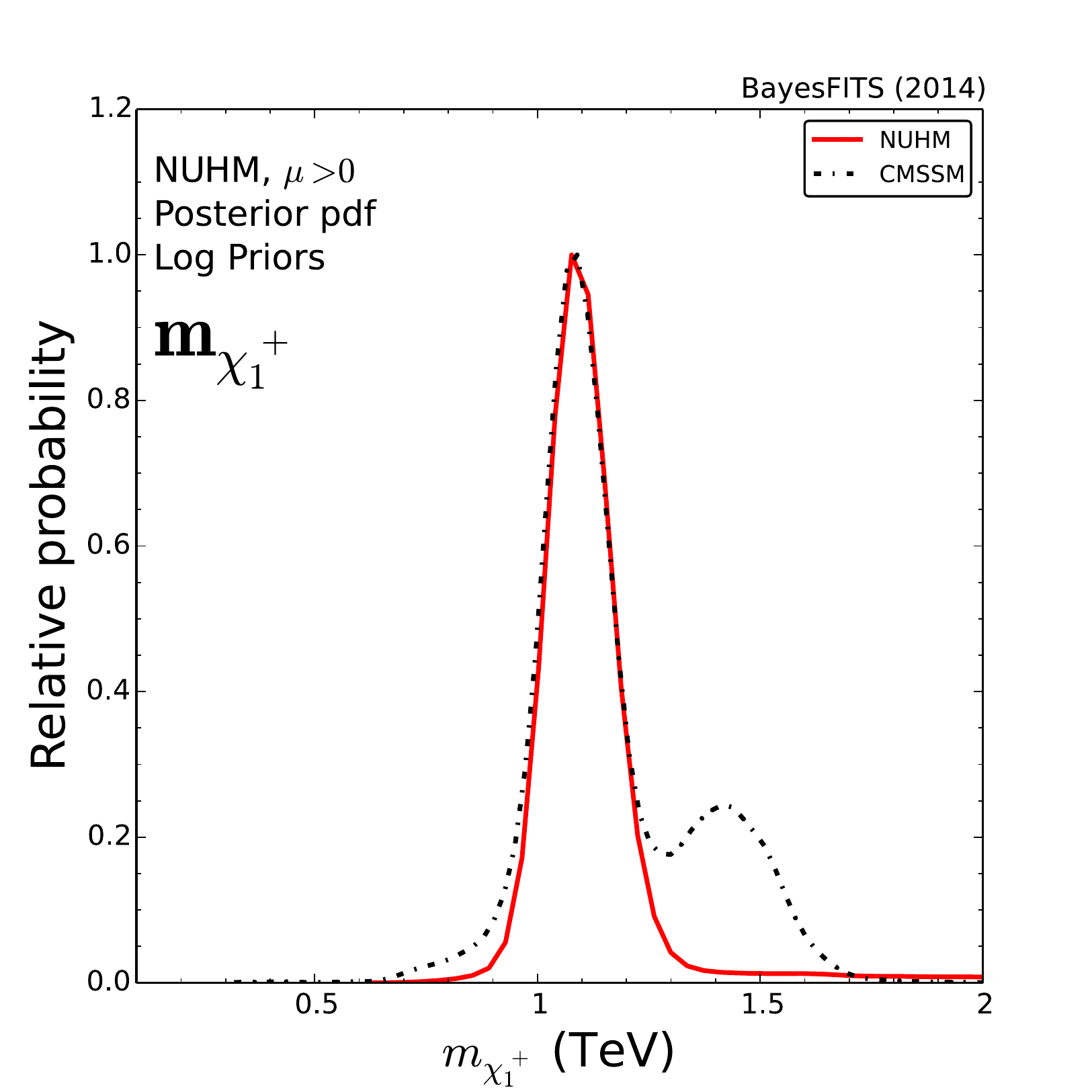}
}%

\caption{\footnotesize Marginalized 1D pdf for the heavy Higgs
  bosons and a selection of superpartner masses in the NUHM with
  $\mu>0$.  Black dot-dashed lines are the distributions obtained for
  the CMSSM with $\mu>0$ for comparison.}
\label{fig:nuhm_1D}
\end{figure}

In \reffig{fig:nuhm_1D} we present the marginalized 1D posterior
distributions for the heavy Higgs bosons and a selection of
superpartner masses.  As a comparison we show the corresponding CMSSM
distributions from \reffig{fig:cmssm_1D} as black dot-dashed lines.
Figure~\ref{fig:nuhm_1D}\subref{fig:a} shows the distribution for
\mchi.  One can see one single peak at $\mchi \simeq 1\tev$,
indicating that in the NUHM the $\sim1\tev$ higgsino region dominates
the posterior probability, with the other two regions significantly
less favored and spread over a broad range of \mchi\ values.

In \reffig{fig:nuhm_1D}\subref{fig:b} we show the distribution of the heavy Higgs masses. 
As mentioned above, the $A/H$-resonance region extends over a large range of \ma\ values, 
and the posterior probability associated with it is much reduced relative to the CMSSM. 
As a consequence, values of $\ma>2-3\tev$, typical of the $\sim1\tev$
higgsino region, are instead favored in the NUHM, thus making the prospects for sensitivity at the LHC ($A\rightarrow\tau^+\tau^-$ direct searches)
much bleaker than in the CMSSM.

Analogous conclusions pertain to the prospects for LHC observation of the other SUSY particles.
In Figs.~\ref{fig:nuhm_1D}\subref{fig:c} and \ref{fig:nuhm_1D}\subref{fig:d} we show 
the distributions for the lightest stop mass and squark masses for the first two generations, respectively. 
In \reffig{fig:nuhm_1D}\subref{fig:e} we show the distribution for the gluino mass. 
The bulk of the squark mass distributions are peaked around mass values significantly smaller 
than in the corresponding $\sim 1\tev$ higgsino region of the CMSSM, as the posterior does not extend as much in \mzero,
but they are still well outside the most optimistic reach for direct detection at the LHC.

One observes some solutions in common with the CMSSM, in the stau-coannihilation region, 
characterized by $\mstopone\lesssim 1.5\tev$, $m_{\tilde{u}_L}\lesssim 3\tev$, or $\mglu\lesssim 3\tev$,
and a neutralino that can be as light as 0.4\tev. Those might begin to be probed at the 
14\tev\ run of the LHC. 
However, as was explained above, the stau-coannihilation region in the NUHM extends significantly with respect to the CMSSM, reaching
quite large \mhalf\ values. 
Thus, it favors heavier gluinos, neutralinos, and scalars, and the statistical weight of 
the parameter space in reach of the LHC is much reduced.  
   
Finally, we show for completeness  in \reffig{fig:nuhm_1D}\subref{fig:f} the 1D pdf for the lightest chargino.
One can see the predominant peak at $m_{\charone}\simeq 1\tev$, encompassing models with higgsino-like \charone, accompanied by a lower tail 
that extends to larger mass values, typical of the wino-dominated charginos.

\subsection{Prospects for dark matter detection}\label{sec:dmnuhm}

\begin{figure}[t]
\centering
\subfloat[]{%
\label{fig:default}%
\includegraphics[width=0.47\textwidth]{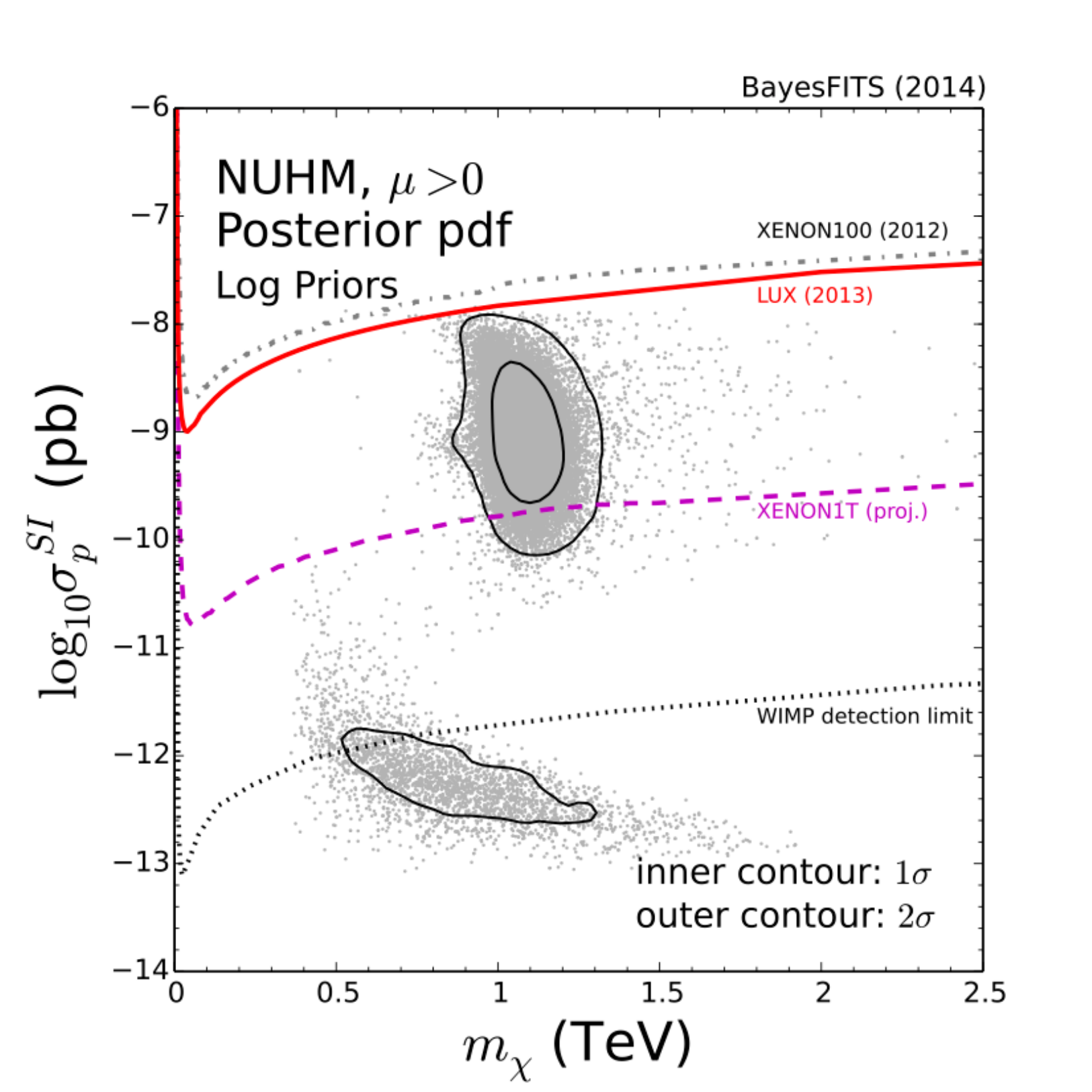}
}%
\hspace{0.01\textwidth}
\subfloat[]{%
\label{fig:default}%
\includegraphics[width=0.47\textwidth]{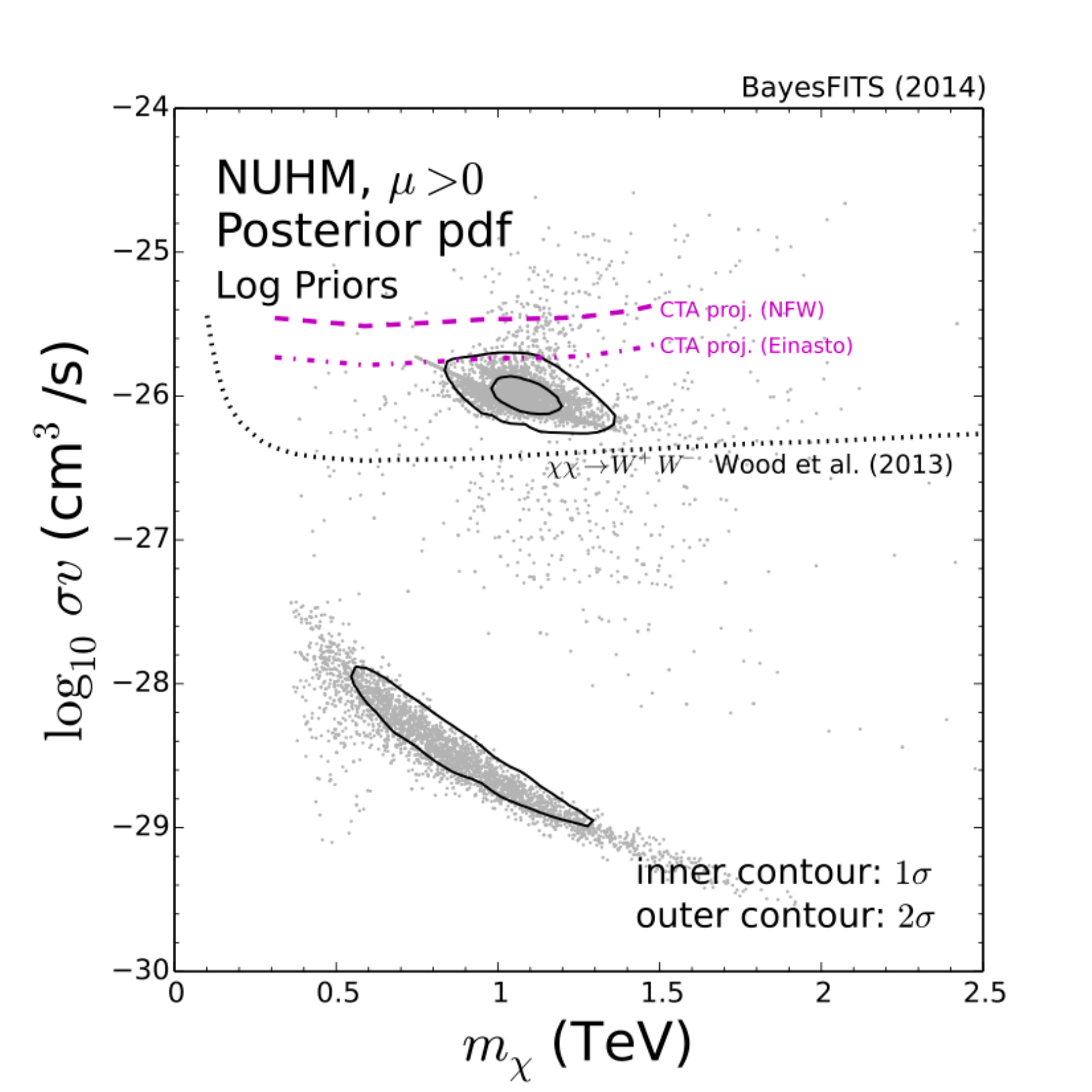}
}%
\caption{\footnotesize \protect\subref{fig:a} Marginalized 2D posterior distribution in the (\mchi, \sigsip) plane of the NUHM with $\mu>0$. 
The solid red line shows the 90\%~C.L. upper bound as given by LUX, here included in the likelihood function. The dot-dashed gray line
shows the 90\%~C.L. 2012 bound of XENON100. The projected sensitivity for 2017 at \xenononet\ is shown in magenta dashed. 
The black dotted line marks the onset of the irreducible neutrino background.
\protect\subref{fig:b} Marginalized 2D posterior distribution for the NUHM with $\mu > 0$ in the (\mchi, $\sigma v$) plane.
The magenta dashed line shows the expected sensitivity of CTA under the assumption of a NFW halo profile. 
The magenta dot-dashed line shows the corresponding sensitivity with Einasto profile.
The thin dotted line shows the projected sensitivity of the CTA expansion\cite{Wood:2013taa}.
}
\label{fig:nuhm_mx_sigsip_sigmav}
\end{figure}

In \reffig{fig:nuhm_mx_sigsip_sigmav}\subref{fig:a} we show the marginalized 2D posterior distribution in the (\mchi, \sigsip) plane. 
As was the case in the CMSSM, shown in \reffig{fig:cmssm_mx_sigmap}\subref{fig:a}, one can easily identify the $\sim 1\tev$ higgsino region 
as the large 68\% and 95\% credible region at $\mchi\simeq 1-1.2\tev$ right below the LUX limit.

The characteristics of this region are largely independent of the model, so that the 
prospects for detection are similar to the CMSSM. 
However, the relative probability of this region is larger in the NUHM, being greater than 90\%, versus approximately 
70\% of the total probability in the same region of the CMSSM. 

On top of this, as was mentioned when discussing \reffig{fig:nuhm_mhu_mhd_mx_ma}\subref{fig:b},  
many of the solutions in the $A/H$-resonance region of the NUHM
feature mixed composition, bino-higgsino neutralinos with $\mchi\gsim1.2\tev$, with consequently enhanced couplings to the nucleus. 
Those points can be seen in \reffig{fig:nuhm_mx_sigsip_sigmav}\subref{fig:a} scattered below the LUX limit, 
well in reach of the \xenononet\ sensitivity, shown with a magenta dashed line,    

On the negative side, one can see that the remaining 95\% credible region, 
the stau-coannihilation region, now extends to much smaller values of \sigsip\ 
and for neutralinos heavier than $\mchi \simeq 0.8\tev$ it lies below the onset of irreducible neutrino 
background\cite{Cabrera:1984rr,Monroe:2007xp} calculated in\cite{Billard:2013qya}, shown with a black dotted line.
For \sigsip\ below the black dotted line the background of atmospheric and diffuse supernova neutrinos becomes important,
so that the sensitivity scales with square root of exposure, making the neutralino more difficult 
to detect via direct detection experiments. 
 
Finally, we show in \reffig{fig:nuhm_mx_sigsip_sigmav}\subref{fig:b} 2D posterior distribution in the (\mchi, $\sigma v$) plane.
As was explained in \refsec{sec:constraints} the estimated CTA sensitivity is 
indicatively shown for the NFW and Einasto halo profile as a magenta dashed and dot-dashed line, respectively.
One must remember that the sensitivity of the individual points in the scan strongly depends on
the annihilation final states, so that the lines in \reffig{fig:nuhm_mx_sigsip_sigmav}\subref{fig:b}
must be taken with care.
However, the indicated sensitivity is robust for final states characterized by gauge bosons,
which are very typical of the $\sim 1\tev$ higgsino region, shown in the figure 
as the predominant 68\% and 95\% posterior region. 

Thus, one can extend to the NUHM the conclusion already stated for the
CMSSM in \refsec{sec:dmcmssm}.  The projected sensitivity in
the minimal configuration studied here\footnote{We use the setup
  considered in\cite{Pierre:2014tra} called Array I consisting of 3
  large size telescopes, 18 medium size telescopes and 56 small size
  telescopes.} seems to fall just short of biting significantly into
the parameter space of the model.  But extended configurations
considered in the literature or, alternatively, an improvement in the
estimated sensitivity, have the potential to deeply probe the bulk of
the model's parameter space, thus yielding a complementary test with
respect to direct detection searches.

%
 
\section{Summary and conclusions}\label{sec:summary} 
 
In this paper we performed a Bayesian analysis of the CMSSM and the NUHM.
We presented the 68\% and 95\% credible regions of the marginalized 2D posterior pdf and the 1D 
distributions of relevant parameters and observables in light of 
the latest experimental constraints and updated numerical tools.

In particular, we updated the results of our previous study\cite{Kowalska:2013hha} by 
\textit{a)} including the corrections to the lightest Higgs mass beyond the 2-loop order 
using $\tt FeynHiggs\, v2.10.0$, which calculates the leading and next-to-leading log corrections in the top/stop sector 
resummed at all orders 
in perturbation theory; \textit{b)} including in the likelihood function the latest constraints 
from direct SUSY searches at the LHC with $\sim20\invfb$ at 8\tev; and 
\textit{c)} including in the likelihood function the 
most recent constraints from direct detection of dark matter at LUX.

We find that the higher-order corrections to the Higgs mass induce modifications to the posterior probability distribution with respect to\cite{Kowalska:2013hha}. 
The correct value of the Higgs boson mass now requires an \msusy\ in general lower,
so that the 95\% credible regions do not extend beyond $\mzero\simeq 12\tev$ in the CMSSM. 
Moreover, regions of the parameter space that in the past struggled to produce a $\sim 126\gev$ Higgs mass, can now do it more easily.
As a consequence, we observe increased statistical relevance of the $A$-resonance region of the CMSSM relative to\cite{Kowalska:2013hha}, 
with approximately 30\% of the total probability. This improves the chances for direct observations of the pseudoscalar Higgs at the LHC 14\tev\ run,
in the $A\rightarrow \tau^+\tau^-$ channel. It also potentially favors 
direct testing in sparticle searches at future, higher-energy proton colliders.  

On the other hand, the bulk of the probability still lies in the $\sim 1\tev$ higgsino region, 
comprising $\sim70\%$ of the total in the CMSSM
and $\sim 90\%$ in the NUHM. Given the almost pure higgsino nature of the dark matter candidate in this region 
the prospect for probing the vast majority of the parameter space of both models through dark matter detection searches is 
enticing.
  
The constraining power of future 1-tonne direct detection experiments like \xenononet\ on the parameter space of the 
CMSSM and the NUHM has been long known, and is confirmed once more in here.
In this study we also showed that indirect detection of dark matter through $\gamma$-rays from the GC 
at CTA is a realistic possibility in the CMSSM and the NUHM.   
We applied the results of Ref.\cite{Pierre:2014tra} for the sensitivity of CTA to 
several annihilation final states to the case of annihilation to multiple final states, as is the MSSM.
We find that the configuration studied in\cite{Pierre:2014tra} falls just short of biting significantly into the parameter space,
but a factor of 5 improvement on the sensitivity can probe $\sim90\%$ of the favored parameter space in both models.

In summary, in both models all the regions favored by the latest constraints show 
good prospects for future observation.
In the CMSSM, the stau-coannihilation region (1\% of the total pdf) will be most likely probed in its entirety through direct SUSY searches 
at the LHC 14\tev\ run; $\sim50\%$ of the $A$-resonance region (corresponding to 15\% of the total pdf) might be probed 
in Higgs searches at the LHC, while it was shown in\cite{Kowalska:2013hha} that the whole region is also very sensitive to 
improvements in the measurement of \brbsmumu; the $\sim 1\tev$ higgsino region (70\% of the pdf)
will be probed via dark matter direct detection in 1-tonne experiments. 
Simultaneously, optimistic but not unrealistic improvements in the projected sensitivity of CTA might
potentially test the whole $\sim 1\tev$ higgsino region and almost all of the $A$-resonance region.

This becomes even more important in the NUHM, where dark matter searches become the privileged instrument to probe 
the parameter space. In particular the largest part of the $\sim 1\tev$ higgsino region and a large number of $A/H$
resonance solutions, that together cover approximately $\sim 90\%$ of the favored parameter space
can be simultaneously probed at 1-tonne detectors and at CTA, if the estimated sensitivity of the latter increases 
as shown in this study.

\begin{center}
\textbf{ACKNOWLEDGMENTS}
\end{center}

  We would like to thank Kamila Kowalska for valuable input on the construction of the LUX likelihood. 
  A.W. would like to thank Jamie Tattersall for email support on using $\tt CheckMATE$. 
  This work has been funded in part by the Welcome Programme
  of the Foundation for Polish Science.
  L.R. is also supported in part by a STFC
  consortium grant of Lancaster, Manchester, and Sheffield Universities. The use
  of the CIS computer cluster at the National Centre for Nuclear Research is gratefully acknowledged.

\appendix
\section{Impact of a recent calculation of the sensitivity of CTA}\label{sec:appendix}

In a recent paper\cite{Roszkowski:2014iqa} we calculated the projected sensitivity of CTA to the WIMP annihilation cross section, 
under the assumption of the Einasto or the NFW DM halo profile. We used a binned likelihood function, and the details of the calculation are presented 
in Appendix~A of\cite{Roszkowski:2014iqa}. Our likelihood function leads to a higher sensitivity than the one obtained in\cite{Pierre:2014tra}, 
which was used in this paper to draw the projected limits shown in Figs.~\ref{fig:cmssm_mx_sigmav_cta}, \ref{fig:cmssm_mx_sigmap}\subref{fig:b}, and \ref{fig:nuhm_mx_sigsip_sigmav}\subref{fig:b}. The result of our calculations for several individual final states in the MSSM, 
assuming the Einasto profile and 500 hours of observation, can be found in Fig.~14 of\cite{Roszkowski:2014iqa}.

\begin{figure}[t]
\centering
\subfloat[]{%
\label{fig:a}%
\includegraphics[width=0.47\textwidth]{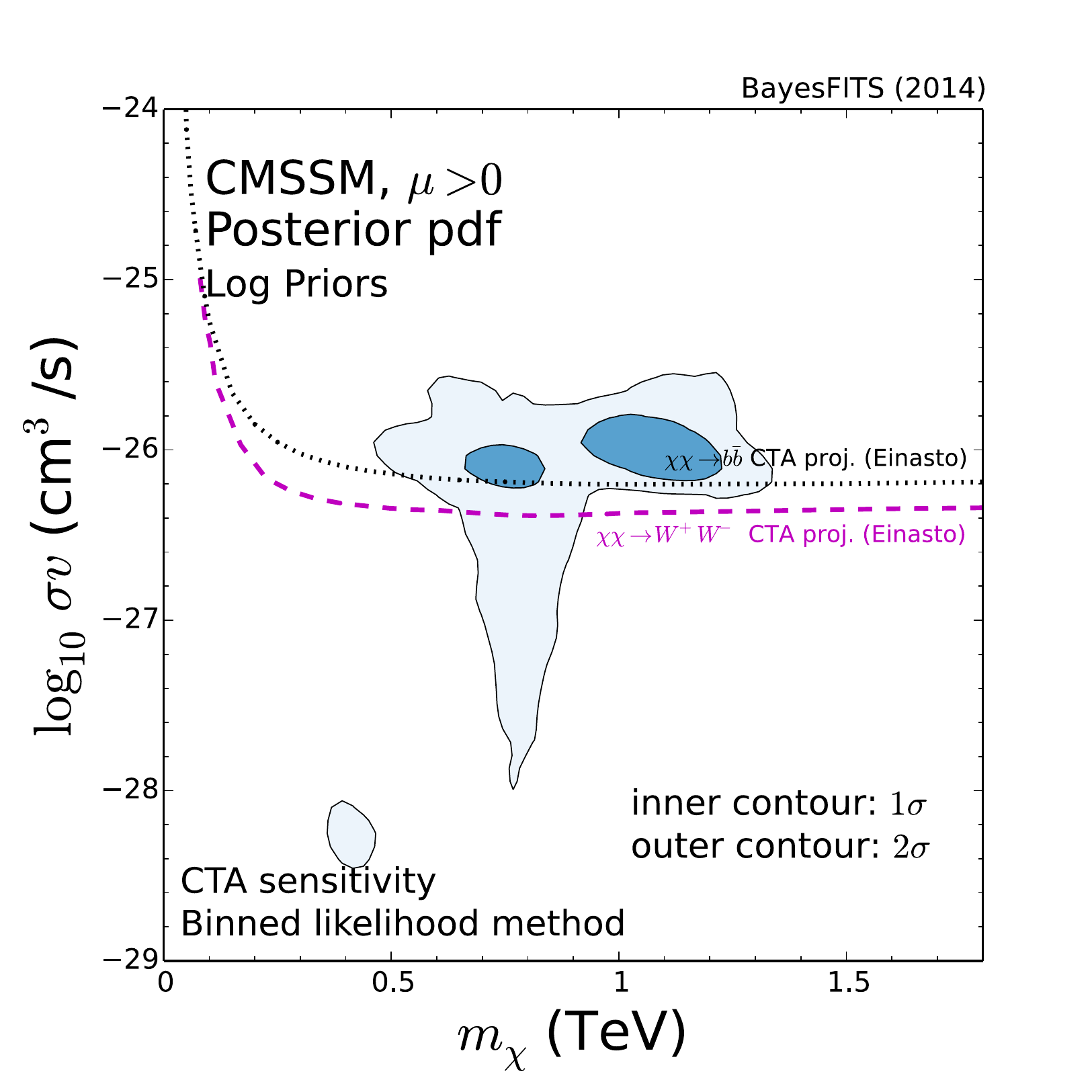}
}%
\hspace{0.01\textwidth}
\subfloat[]{%
\label{fig:b}%
\includegraphics[width=0.47\textwidth]{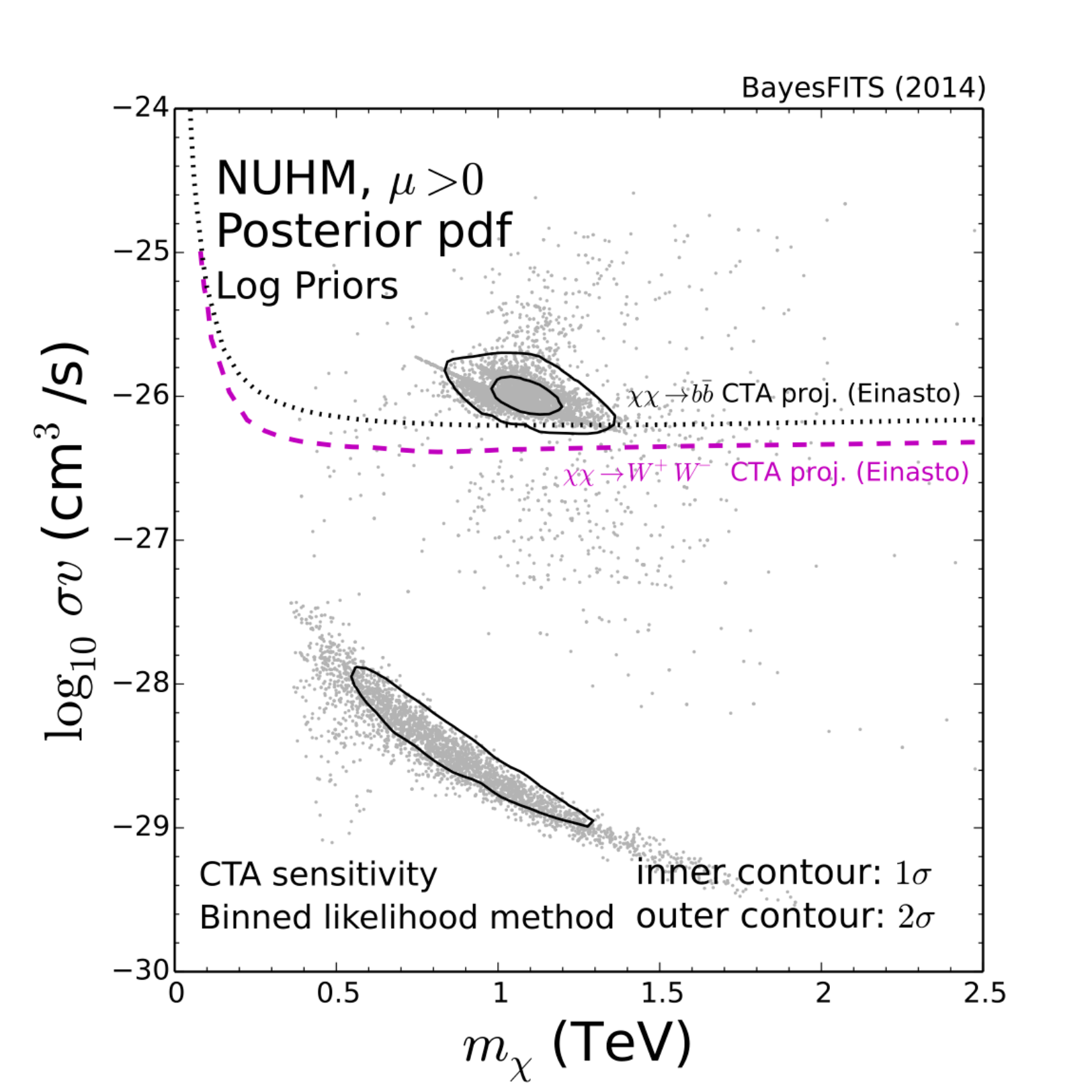}
}%
\caption{\footnotesize \protect\subref{fig:a} Marginalized 2D posterior distribution for the CMSSM with $\mu > 0$ 
in the (\mchi, $\sigma v$) plane. \protect\subref{fig:b} Marginalized 2D posterior distribution for the NUHM with $\mu > 0$ 
in the (\mchi, $\sigma v$) plane. In both panels the magenta dashed line shows the expected sensitivity of CTA 
for the $W^+W^-$ final state under the Einasto profile assumption, calculated in Ref.\cite{Roszkowski:2014iqa}
using the binned likelihood method. 
The black dotted line shows the corresponding sensitivity for the $b\bar{b}$ final state. 
This figure updates Figs.~\ref{fig:cmssm_mx_sigmap}\protect\subref{fig:b} and \ref{fig:nuhm_mx_sigsip_sigmav}\protect\subref{fig:b}.
}
\label{fig:new_cta}
\end{figure}

In \reffig{fig:new_cta}\subref{fig:a} we show the new $W^+W^-$ and $b\bar{b}$ projected sensitivities, obtained using the binned likelihood method,
in the (\mchi, \sigv) plane of the CMSSM. 
The $W^+W^-$ final state bound provides a rough estimate of the exclusion reach for the $\sim1\tev$ higgsino region, whereas
the $b\bar{b}$ bound applies to the $A$-resonance region. In \reffig{fig:new_cta}\subref{fig:b} we show the same for the NUHM.
Figure~\ref{fig:new_cta} shows that the binned likelihood method leads to significantly improved sensitivity relative to that of Ref.\cite{Pierre:2014tra}, 
which was presented in Figs.~\ref{fig:cmssm_mx_sigmap}\subref{fig:b} and \ref{fig:nuhm_mx_sigsip_sigmav}\subref{fig:b}.
  
In the remainder of this appendix we show that, in light of the results of\cite{Roszkowski:2014iqa}, 
under reasonable assumptions the full extent of the 95\% credible region of the CMSSM will be 
probed by a limited number of upcoming experiments. These are, by increasing mass scale:

1. CMS and ATLAS direct SUSY searches at the LHC Run II;

2. Precise measurement of \brbsmumu\ at LHCb in Run II;

3. Direct searches for dark matter at \xenononet\ and other tonne-scale experiments;

4. Indirect detection of dark matter at CTA through $\gamma$ rays from the GC.

\begin{figure}[t]
\centering
\subfloat[]{%
\label{fig:a}%
\includegraphics[width=0.47\textwidth]{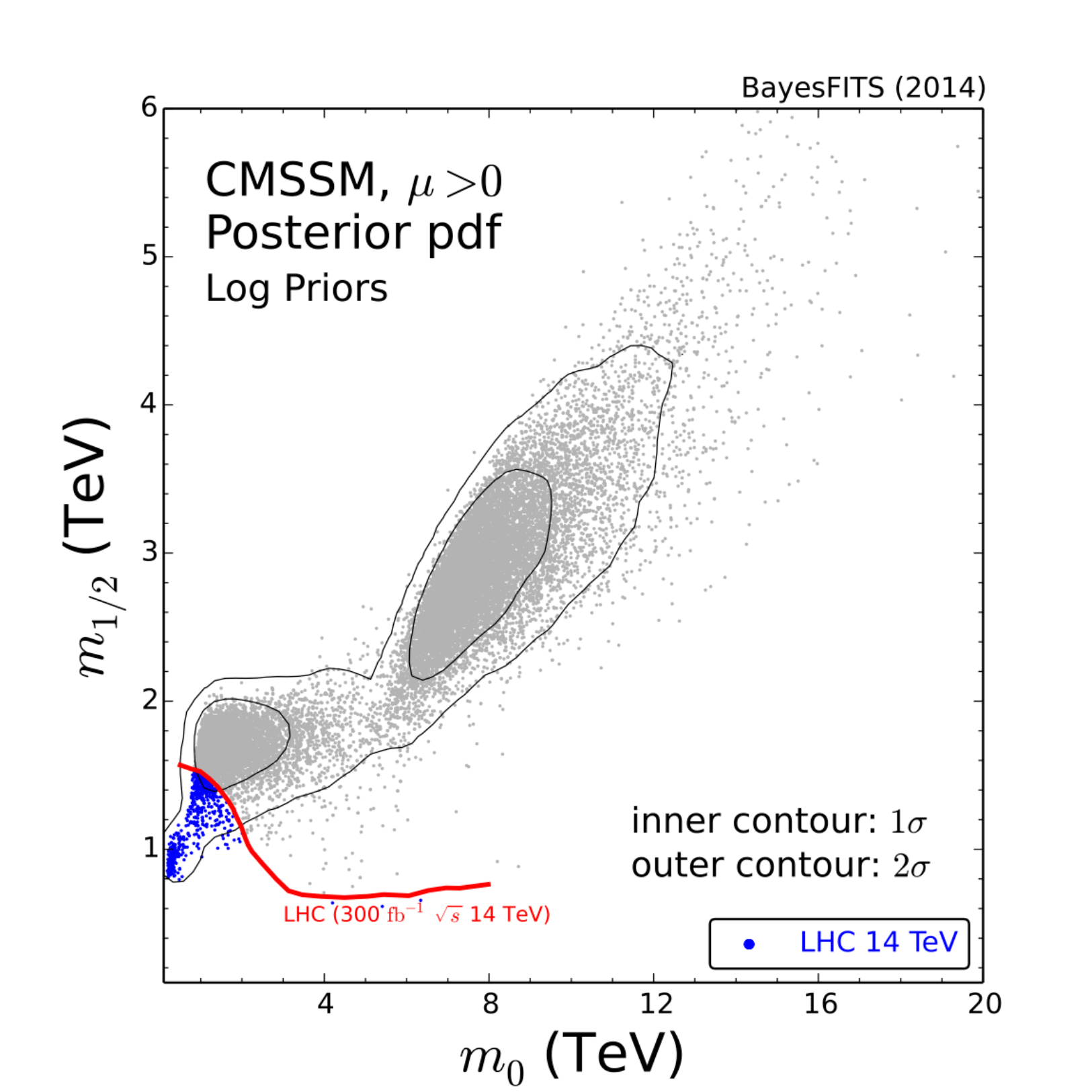}
}%
\hspace{0.01\textwidth}
\subfloat[]{%
\label{fig:b}%
\includegraphics[width=0.47\textwidth]{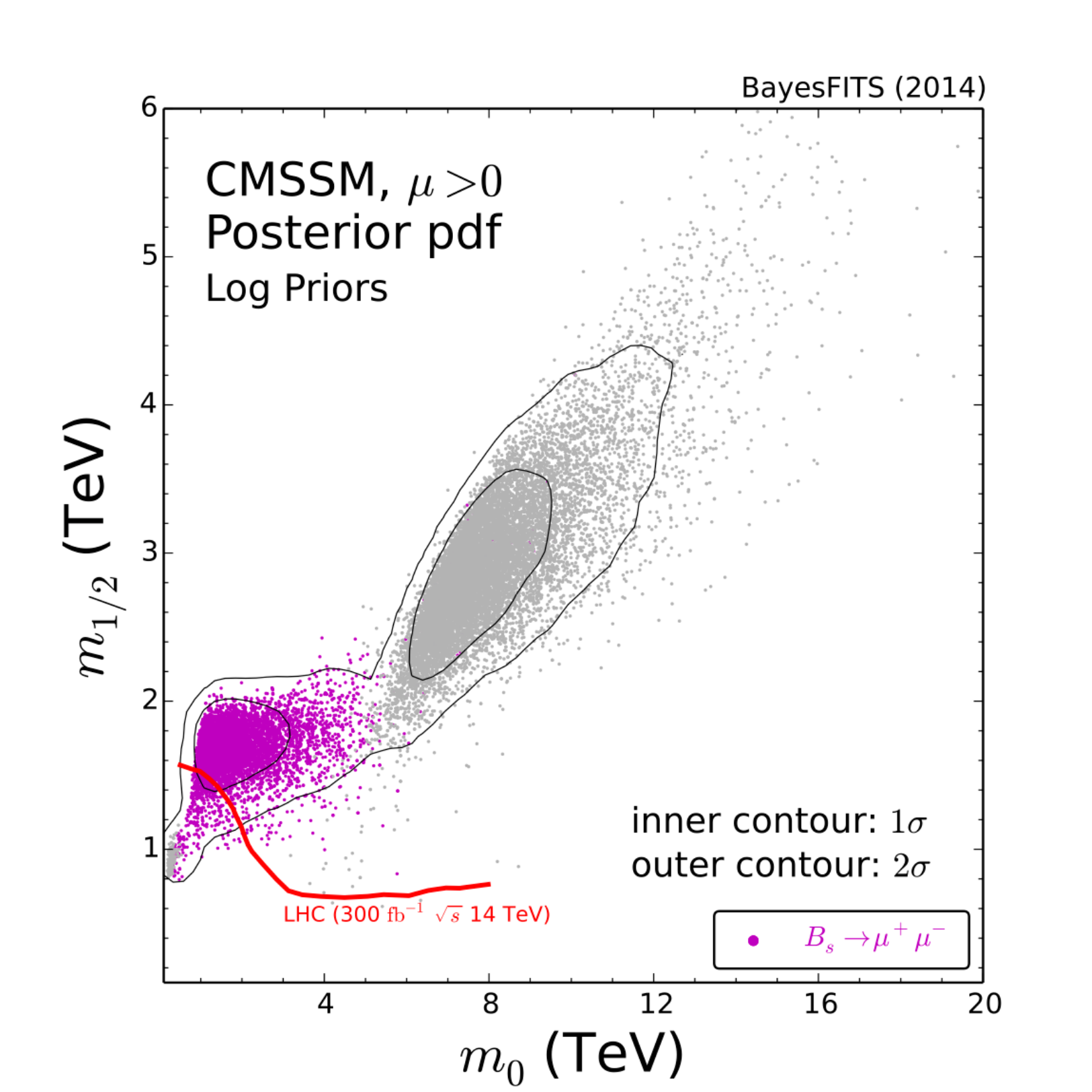}
}%
\\
\subfloat[]{%
\label{fig:c}%
\includegraphics[width=0.47\textwidth]{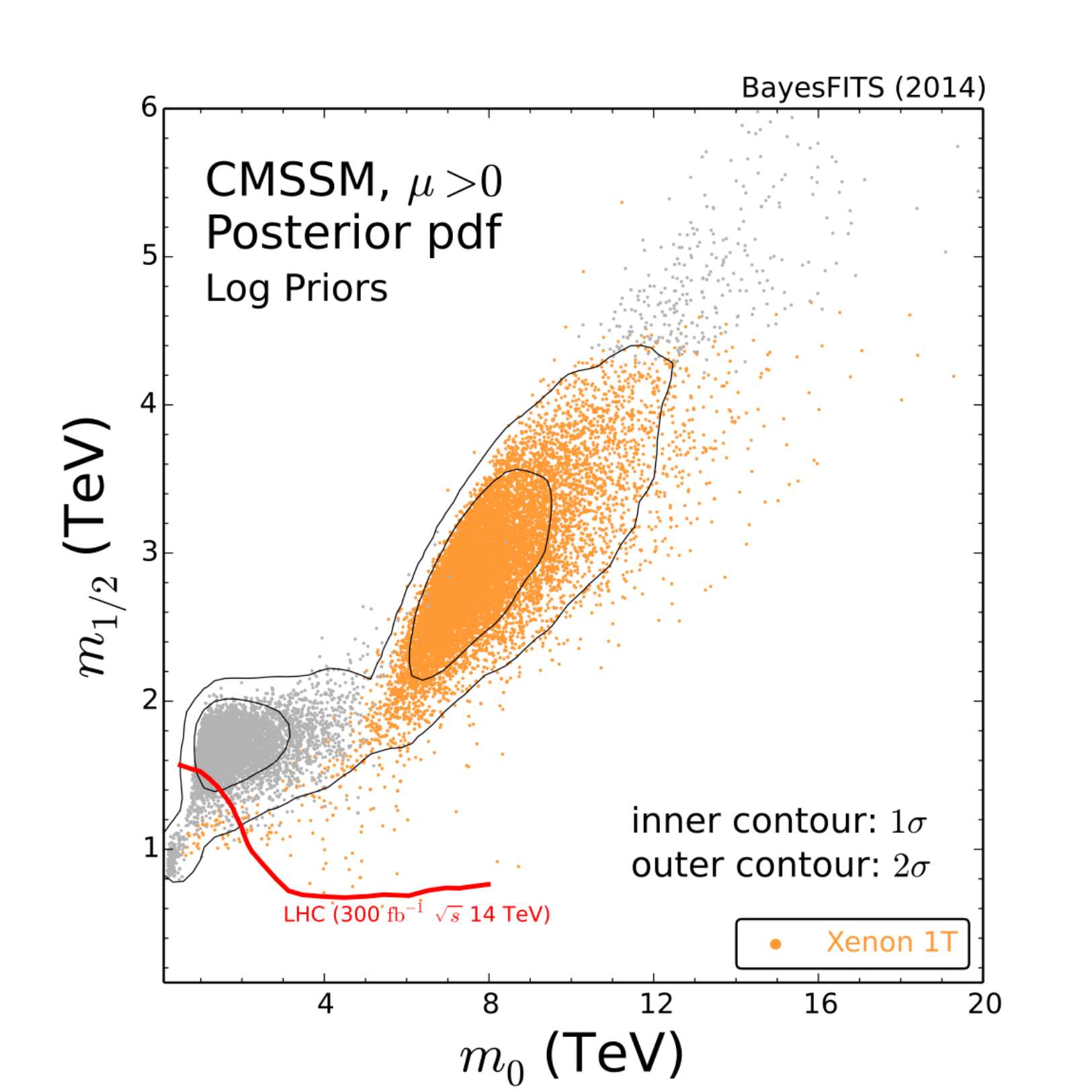}
}%
\hspace{0.01\textwidth}
\subfloat[]{%
\label{fig:d}%
\includegraphics[width=0.47\textwidth]{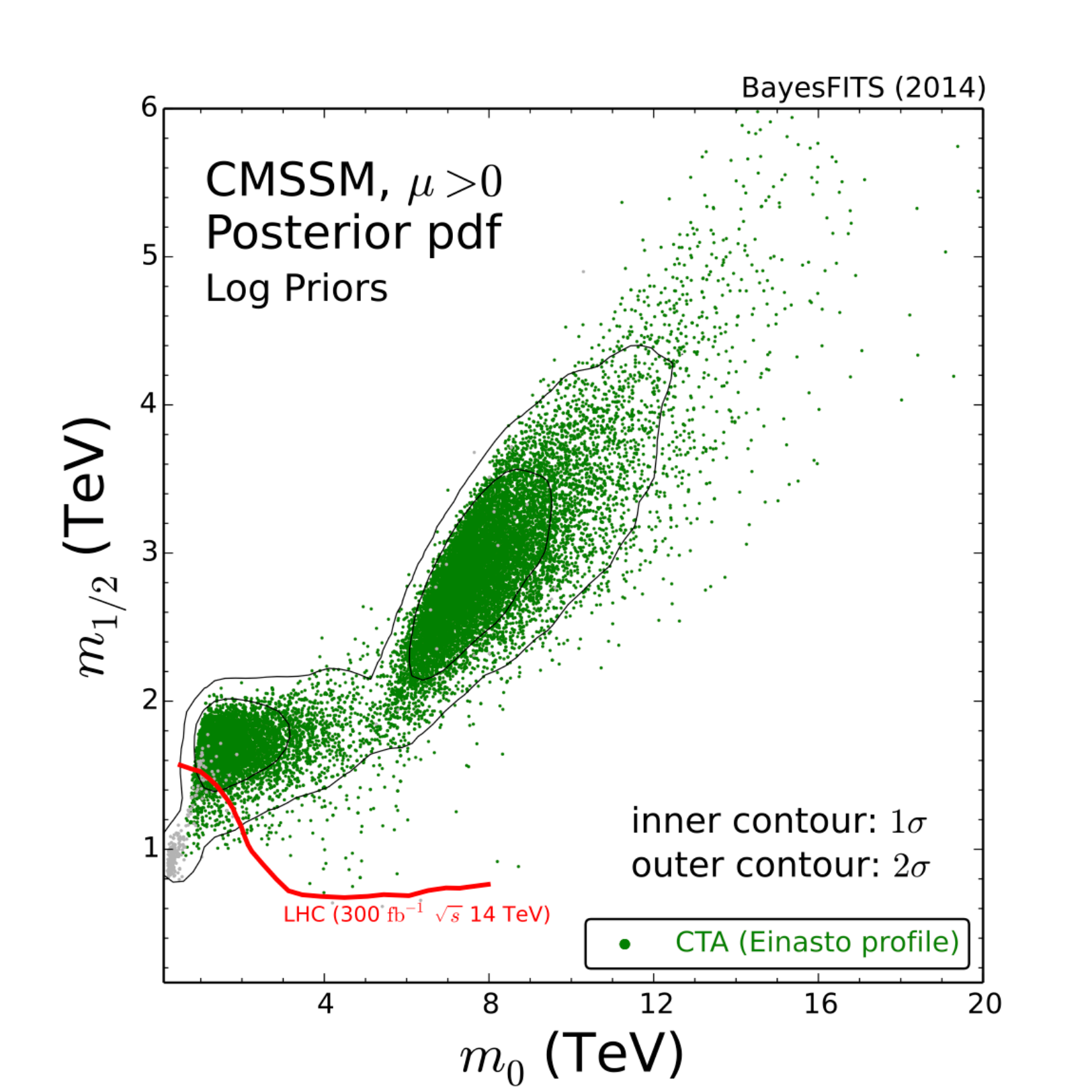}
}%
\caption{\footnotesize \protect\subref{fig:a} Marginalized 2D posterior in the (\mzero, \mhalf) plane of the CMSSM with $\mu>0$. 
68\% and 95\% credible regions are shown by the inner and outer contours, respectively. 
Points are distributed according to the posterior probability. 
The projected LHC Run II 95\%~C.L. exclusion line from Ref.\cite{Baer:2013fva} 
is shown in red solid for reference. Blue points lie within sensitivity for direct SUSY searches, gray points are unconstrained. 
\protect\subref{fig:b} Same as \protect\subref{fig:a} except that points in magenta are sensitive to future measurements of \brbsmumu, 
as described in Ref.\cite{Kowalska:2013hha}. \protect\subref{fig:c} Same as \protect\subref{fig:a} except that points in orange lie 
within sensitivity of tonne-scale underground DM detectors. \protect\subref{fig:d} Same as \protect\subref{fig:a} except that points in green lie 
within sensitivity of CTA to $\gamma$ rays from DM annihilations, as calculated in Ref.\cite{Roszkowski:2014iqa}.}
\label{fig:cmssm_future}
\end{figure}

The impact of each experiment's projected sensitivity on the parameter space of the CMSSM is shown in the four panels of \reffig{fig:cmssm_future}.
In \reffig{fig:cmssm_future}\subref{fig:a} we show a sample of points distributed according to the posterior probability, 
whose 68\% and 95\% credible regions are shown by the inner and outer contours, respectively.
The points in blue lie within the projected sensitivity 
of searches for squarks and gluinos at the LHC Run II calculated in Ref.\cite{Baer:2013fva}, whereas the gray points remain unconstrained.
The expected limit from the 0-1 lepton + jets + missing energy searches with 300\invfb at 14\tev\ is shown as a red solid line. 
As is well known, Run II will be sensitive to most of the remaining part of the stau-coannihilation region, presently not excluded by the LHC 8\tev\ run.

As was explained in Ref.\cite{Kowalska:2013hha}, one will exclude the 95\% credible posterior region corresponding to 
the $A$-resonance region of the CMSSM if the measurement of \brbsmumu\ eventually converges to its SM expectation with an error of $\sim5\%$, 
and at the same time the theoretical uncertainties reach approximately the same precision. 
In \reffig{fig:cmssm_future}\subref{fig:b} we show in magenta the points that will be excluded under the projection
considered in Ref.\cite{Kowalska:2013hha}: $\brbsmumu_{proj} =  (3.50 \pm 0.17 \pm 0.17) \times 10^{-9}$, 
with experimental and theoretical uncertainties added in quadrature.
As was mentioned in \refsec{sec:CMSSM}, it should be noted that
there will be considerable overlap between the constraint from \brbsmumu\ and searches for a heavy Higgs decaying to tau pairs.  

The projected 90\%~C.L. sensitivity for \xenononet\ in the (\mchi, \sigsip) plane of the CMSSM was 
shown in \reffig{fig:cmssm_mx_sigmap}\subref{fig:a} as a dashed magenta line.
This corresponds to the excluded points in the (\mzero, \mhalf) plane that are presented in orange in \reffig{fig:cmssm_future}\subref{fig:c}. 
\xenononet\ and other tonne-scale detectors will probe the $\sim1\tev$ higgsino region in its near entirety, with the exception of the points
at large \mzero\ and \mhalf.

We finally come to CTA's sensitivity to $\gamma$ rays from the GC. 
We applied the binned likelihood function constructed in\cite{Roszkowski:2014iqa} to the points of the CMSSM, under the assumption of the Einasto profile for the 
DM distribution in the GC. The points of the (\mzero, \mhalf) plane excluded at the 95\%~C.L. by the likelihood function with 500 hours of observation 
are shown in green in \reffig{fig:cmssm_future}\subref{fig:d}.   
CTA is going to provide a strong constraint in the $\sim1\tev$ higgsino region, as well as on the majority of the $A$-resonance region's points. 
The sensitivity of CTA reaches the part of the $\sim1\tev$ higgsino region that lies outside the sensitivity of tonne-scale underground DM detectors 
and provides complementary probe to the rest of the parameter space with the exception of the stau-coannihilation region.

\begin{figure}[t]
\centering
\includegraphics[width=0.75\textwidth]{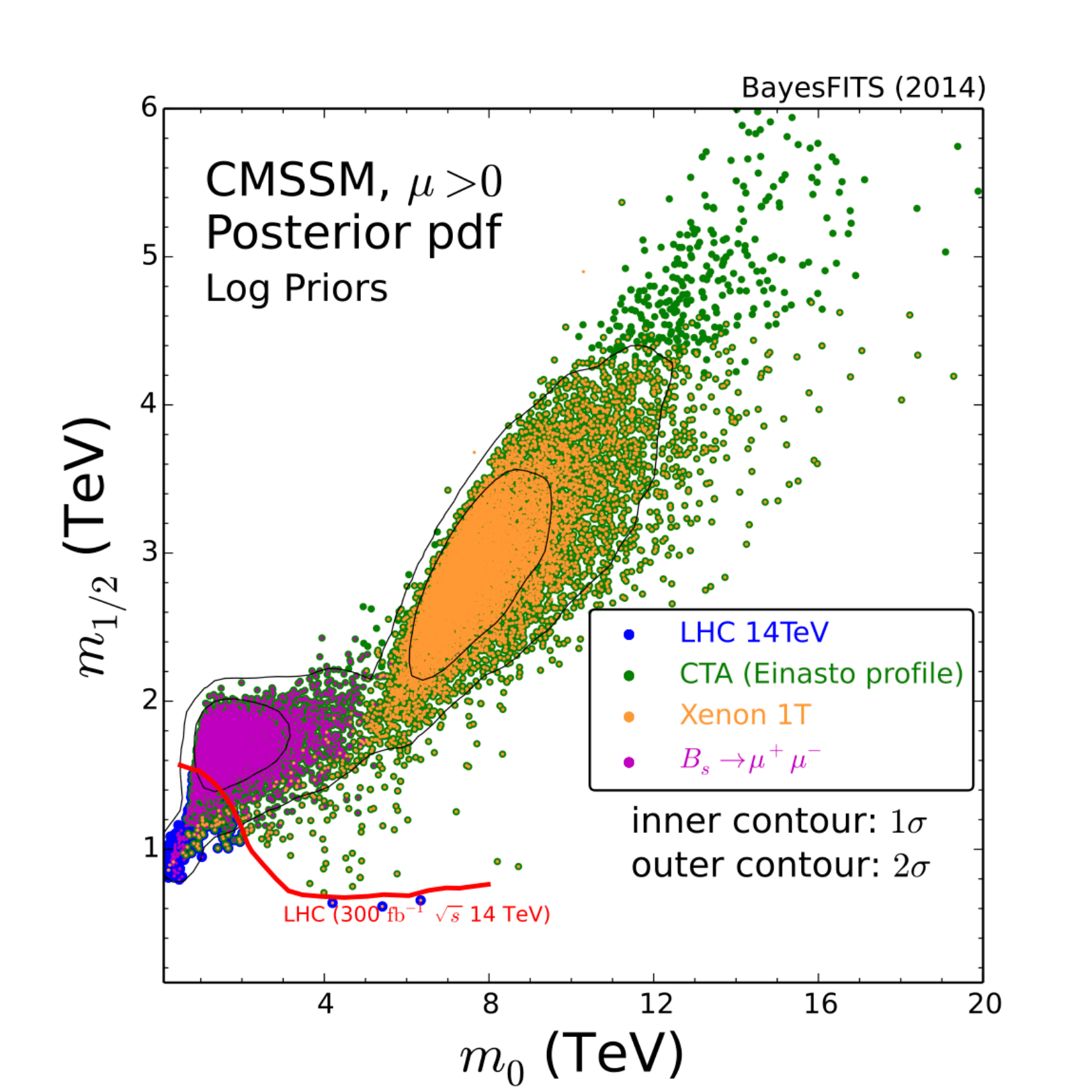}
   \caption{\footnotesize Marginalized 2D posterior in the (\mzero, \mhalf) plane of the CMSSM with $\mu>0$. 
68\% and 95\% credible regions are shown by the inner and outer contours, respectively. 
Points are distributed according to the posterior probability. The projected LHC Run II 95\%~C.L. exclusion line is shown in red solid for reference.
Colored points show the future sensitivity to direct SUSY searches in blue, measurement of \brbsmumu\ in magenta, tonne-scale underground detectors in orange,
and CTA in green.}
\label{fig:cmssm_future_all}
\end{figure}

To conclude, in \reffig{fig:cmssm_future_all} we show the combination of the constraints presented in the panels of \reffig{fig:cmssm_future}. 
We want to point out that there is no single gray point left in the figure. Since the posterior sample shown contains approximately 20000
points, this implies that even regions well beyond the 95\% credible level will be constrained.
In addition, \reffig{fig:cmssm_future_all} highlights the considerable amount of overlapping between the parameter space probed by CTA and that of other experiments,
so that for most of the parameter space detection/exclusion will not rely on a single measurement.

In this regard, we want to point out that if one adopts a flatter profile than the Einasto for the DM distribution in the halo, 
the above projection for CTA will inevitably be weakened. 
We have checked that under the NFW profile assumption most points belonging to the $A$-resonance region 
and those included in the 68\% credible part of the $\sim 1\tev$ higgsino region will be excluded. However,  
most points belonging to the 95\% part of the $\sim 1\tev$ higgsino region, 
and those characterized by even larger \mzero\ and \mhalf\ will be beyond reach.  
Even so, CTA is going to provide an invaluable instrument for complementary testing of the regions of parameter 
space shown in Figs.~\ref{fig:cmssm_future}\subref{fig:b} and \ref{fig:cmssm_future}\subref{fig:c}, that will be probed at the LHC and in tonne-scale
underground detectors.  

\bibliographystyle{JHEP}

\bibliography{cmssm}

\end{document}